\documentclass[fleqn,usenatbib]{mnras}
\pdfoutput=1
\usepackage{newtxtext,newtxmath}

\usepackage[T1]{fontenc}

\DeclareRobustCommand{\VAN}[3]{#2}
\let\VANthebibliography\thebibliography
\def\thebibliography{\DeclareRobustCommand{\VAN}[3]{##3}\VANthebibliography}

\usepackage[english]{babel}
\usepackage{graphicx}
\usepackage{graphicx,amsmath,bm,hyperref,subfig,booktabs,multirow}
\usepackage{textcomp}
\usepackage{gensymb}
\usepackage[euler]{textgreek}
\usepackage{multicol,lipsum, babel}
\usepackage{natbib}
\bibpunct{(}{)}{;}{a}{}{,} 
\usepackage{comment}
\hypersetup{colorlinks=true,citecolor=blue,
linkcolor=blue,filecolor=magenta,urlcolor=blue}
\usepackage{graphicx}	
\usepackage{amsmath}	
\usepackage{pdflscape}

\title[Gaia view of SMC's stellar sub-structure]{Gaia view of a stellar sub-structure in front of the Small Magellanic Cloud}
 
\author[Omkumar et al.]{
Abinaya O. Omkumar,$^{1}$\thanks{E-mail: abinayakumar97@gmail.com}
Smitha Subramanian,$^{1}$\thanks{E-mail: smitha.subramanian@iiap.res.in}
Florian Niederhofer,$^{2}$
Jonathan Diaz,$^{3}$
\newauthor{Maria-Rosa L. Cioni,$^{2}$
Dalal El Youssoufi,$^{2}$
Kenji Bekki,$^{3}$
Richard de Grijs,$^{4,5,6}$}
\newauthor{Jacco Th. van Loon$^{7}$}\\
$^{1}$Indian Institute of Astrophysics, Koramangala II Block, Bangalore-560034, India\\
$^{2}$Leibniz-Institut f\"ur Astrophysik Potsdam (AIP), An der Sternwarte 16. D-14482 Potsdam, Germany\\
$^{3}$ICRAR, M468, The University of Western Australia 35 Stirling Highway,Crawley Western Australia, 6009, Australia\\
$^{4}$Department of Physics and Astronomy, Macquarie University, Balaclava Road, Sydney, NSW 2109, Australia\\
$^{5}$Research Centre for Astronomy, Astrophysics and Astrophotonics, Macquarie University, Balaclava Road, Sydney, NSW 2109, Australia\\
$^{6}$International Space Science Institute--Beijing, 1 Nanertiao, Zhongguancun, Hai Dian District, Beijing 100190, China\\
$^{7}$Lennard-Jones Laboratories, Keele University, ST5 5BG, UK\\
}

\date{Accepted XXX. Received YYY; in original form ZZZ}

\pubyear{2020}

\begin{document}
\label{firstpage}
\pagerange{\pageref{firstpage}--\pageref{lastpage}}
\maketitle

\begin{abstract}
Recent observational studies identified a foreground stellar sub-structure traced by red clump (RC) stars ($\sim$ 12 kpc in front of the main body) in the eastern regions of the Small Magellanic Cloud (SMC) and suggested that it formed during the formation of the Magellanic Bridge (MB), due to the tidal interaction of the Magellanic Clouds. Previous studies investigated this feature only up to 4$\rlap{.}^{\circ}$0 from the centre of the SMC due to the limited spatial coverage of the data and hence could not find a physical connection with the MB. To determine the spatial extent and properties of this foreground population, we analysed data from the \textit{Gaia} data release 2 (DR2) of a $\sim$314 deg$^2$ region centred on the SMC, which cover the entire SMC and a significant portion of the MB. We find that the foreground population is present only between 2$\rlap{.}^{\circ}$5 to $\sim$ 5$^\circ$--6$^\circ$ from the centre of the SMC in the eastern regions, towards the MB and hence does not fully overlap with the MB in the plane of the sky. The foreground stellar population is found to be kinematically distinct from the stellar population of the main body with $\sim$ 35 km s$^{-1}$ slower tangential velocity and moving to the North-West relative to the main body. Though the observed properties are not fully consistent with the simulations, a comparison indicates that the foreground stellar structure is most likely a tidally stripped counterpart of the gaseous MB and might have formed from the inner disc (dominated by stars) of the SMC. A chemical and 3D kinematic study of the RC stars along with improved simulations, including both tidal and hydro-dynamical effects, are required to understand the offset between the foreground structure and MB.
\end{abstract}



\section{Introduction}
{\indent The Magellanic System, one of the nearest examples of an interacting system of galaxies, comprises two dwarf galaxies, the Large Magellanic Cloud (LMC) and the Small Magellanic Cloud (SMC), a bridge of gas and stars connecting these galaxies known as the Magellanic Bridge (MB), a leading stream of gas known as the Leading Arm (LA) and a trailing stream of gas known as the Magellanic Stream (MS). The Magellanic Clouds (MCs) are located at a distance of 50$\pm$2 kpc (LMC -- \citealp{degrijs2014lmc}) and 62$\pm$1 kpc (SMC -- \citealp{degrijs2015smc}). The MB, MS and the LA are prominent features in HI maps \citep{putman2003}.\\ 
\indent Simulations of the Magellanic System  \citep{Besla2012,Diaz2012}, based on the revised proper motion estimates of the MCs (\citealp{kallivayalil2006,vieira2010}) are able to explain the formation of many of the observed gaseous features around the MCs as a result of their mutual interactions. According to these models, the MS was formed $\sim$ 1.5 Gyr ago and the MB was formed $\sim$ 100-300 Myr ago, mainly from the material stripped from the SMC. However, the MS has also been suggested to have material stripped from the LMC (\citealp{nidever2008, hammer2015, Ritcher2017}). Based on the relative motions of the MCs and the recent proper motion measurements of stars within the MB region \citep{gaiadr1,gaia2018}, the tidal interaction event which formed the MB is suggested to have happened $\sim$ 150 Myr ago (\citealp{schmidt2019}; \citealp{zivick2018,zivick2019}). Though the tidal interactions must have played a dominant role, the ram-pressure effects due to the Milky Way halo could have also altered the present shape of the gaseous features of the Magellanic System (\citealp{hammer2015,salem2015, Tepper2019} and \citealp{Wang2019}). Simulations predict that the dominant nature of the interaction is tidal, resulting in stellar  sub-structures along with gaseous features around the MCs. Stellar sub-structures formed during the formation of the MB and MS are expected to have stars older than 150 Myr and 1.5 Gyr respectively. No conclusive evidence for a stellar counterpart (consisting of stars older than 1.5 Gyr) to the MS has been found so far. A young ($\sim$ 117 Myr-old) metal-poor star cluster has been discovered recently in the vicinity of the LA (\citealp{princewhelan2019, nidever2019}). These studies suggest that this young stellar population could have formed during the interaction of the LA with the Milky Way disc.\\
\indent Several studies have focused on the MB region, in search of stellar populations. The MB contains stellar populations of few Myrs (\citealp{demers1998,harris2007,chen2014,skowron2014} and references therein) which might have formed from the gas stripped during the interaction. \cite{demers1998} and \cite{harris2007} did not find intermediate-age/old (age > 2 Gyr) stellar populations in the fields centred on the HI ridge line of the MB. Despite later studies \citep{nidever2011,bagheri2013,skowron2014,noel2013,noel2015,jacyszyn2017} uncovering the presence of intermediate-age/old stellar populations in the central and western regions of the MB, the interpretations as to their origin differed. While \cite{noel2013,noel2015} and \cite{carrera2017} supported a tidal origin for these intermediate-age stars in the MB, \cite{jacyszyn2017} and \cite{wagner2017} suggested them as part of the overlapping stellar halos of the MCs.\\
\indent \cite{belokurov2017} reported the existence of a stellar tidal bridge from the study of RR Lyrae stars in the Gaia data release 1 (DR1) \citep{gaiadr1} and found that this old stellar bridge is not aligned with the gaseous MB but is shifted by $\sim$ 5$^\circ$ from the bridge traced by young main sequence stars (the latter is well aligned with the gaseous bridge). They suggested that this offset is due to the ram-pressure effect of the Milky Way halo on the gas stripped during the tidal interaction between the MCs. However, \cite{jacyszyn2019} found a smooth distribution of RR Lyrae stars, similar to that of two extended overlapping structures, instead of a bridge-like distribution. Thus the identification of stellar sub-structures in the low density environment around the MCs and to provide an observational proof of the tidal origin are not trivial.\\ 
\indent As the SMC is less massive than the LMC and simulations predict stripping of stars and gas from it, the effect of tidal interaction is expected to have left an imprint on the SMC structure. Earlier studies of the SMC \citep{Subramanian2009,subramanian2012} concentrated on the inner 2$^\circ$-radius region and found that the old/intermediate-age stellar populations have a smooth and ellipsoidal distribution, with no signatures of interactions. The study by \citet{Nidever2013} identified an interesting feature, the presence of a foreground population of red clump (RC) stars in four distinct 0.36 deg$^2$ fields at a radius of 4$^\circ$ from the SMC centre to the East (in the direction of the MB and the LMC) suggesting a tidal origin. \citet{Subramanian2017} studied this feature using the data from the VISTA (Visible and Infrared Survey Telescope for Astronomy) survey of the MCs (VMC) \citep{cioni2011} in the YJK$_s$ near-infrared (NIR) bands. VMC data being continuous and homogeneous, allowed them to trace this feature over 2$\rlap{.}^{\circ}$5--4$\rlap{.}^{\circ}$0 to the East. Both studies suggested that the foreground RC stars represent a stellar population stripped from the SMC during the tidal interaction between the MCs around 300 Myr ago, which formed the MB. The study by \cite{Subramanian2017} was based on a subset of VMC data. Tatton et al. (2020) studied the entire VMC data and confirmed the presence of this feature across the entire eastern SMC up to at least 4$\rlap{.}^{\circ}$0 from the center. A spectroscopic study by \cite{dobbie2014} of Red Giant Branch (RGB) stars beyond 3$^\circ$ from the SMC centre, in the eastern regions, supports this interpretation. However, due to the limited spatial coverage of the data, \citet{Nidever2013}, \citet{Subramanian2017} and Tatton et al. (2020) 
could not probe this feature beyond 4$\rlap{.}^{\circ}$0 from the SMC centre and assess its physical connection with the MB.\\
\indent In this study we use \textit{Gaia} data release 2 (DR2) data \citep{gaia2018}, which covers the entire Magellanic System, to determine the extent (the physical connection with the MB) and properties of this stellar sub-structure in front of the SMC. In addition to \textit{Gaia} DR2 data, we analyse results from the simulations of \citet{Diaz2012} to identify the possible origin of this feature. We present a detailed analysis of the RC stars within a circular region with radius r $\le$ 10$^\circ$ from the SMC centre ($\sim$ 314 deg$^2$), which covers the entire SMC and MB regions adjacent to the SMC. The RC stars are more massive and metal-rich counterparts of the horizontal-branch stars. They have an age range of 2--9 Gyr and a mass range of 1--2.2 M$_{\sun}$ \citep{girardisalaris2001,girardi2016}. As they start their core helium-burning phase at an almost-fixed core mass they have fixed absolute magnitudes. Hence they are useful probes to study the 3D structure of their host galaxies. \\
 \indent The structure of this paper is as follows. Sections \ref{section2} and \ref{section3} explain the selection criteria applied to the \textit{Gaia} DR2 data and the analysis respectively. In Section \ref{section4} we discuss the bimodality in RC magnitude distribution and distance effect. Section \ref{section5} presents the 3D structure and in Section \ref{section6} we discuss the kinematics of the dual RC population. In Section \ref{section7} we compare our results with the simulations and in Section \ref{section8} we provide a summary and conclusions.} 
\section{Data}
\label{section2}
{\subsection{Gaia DR2 data selection}
\label{section2.1}
\medskip
The photometric and astrometric data from the \textit{Gaia} DR2 (\citealp{gaia2018, lindegren2018, evans2018}) is used in this study. The \textit{G} (330 - 1050 nm), \textit{G$_{BP}$} (330 - 680 nm) and \textit{G$_{RP}$} (630 - 1050 nm) bands are the three pass bands in Gaia with mean wavelengths of 673 nm, 532 nm and 797 nm respectively \citep{jordi2010}. Fig. \ref{fig:cart_selection} shows \textit{Gaia} DR2 sources towards the Magellanic System, in Cartesian coordinates (zenithal equidistant projection). \textit{X} and \textit{Y} are defined as in \cite{vandermarel2001} centred on the MB (\textalpha$_0$ : 03\textsuperscript{h}08\textsuperscript{m}, \textdelta$_0$ : $-72^\circ$).  Different components of the Magellanic System are shown. In the present work we study in detail the region marked by a circle of 10$\degree$ radius, centered on the optical centre of the SMC, which covers the entire SMC region and a significant region of the MB.\\ 
\begin{figure}
    \centering
    \hspace*{-2em}
    \includegraphics[scale=0.4]{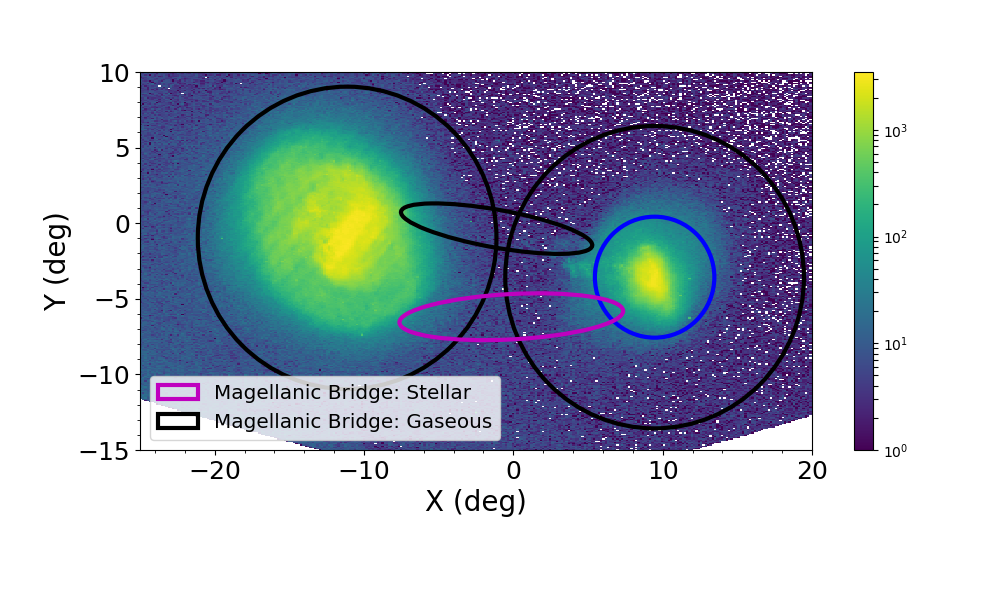}
    \vspace{-1cm}
    \caption{Cartesian plot of the LMC and SMC with the MB (\textalpha$_0$ : 03\textsuperscript{h}08\textsuperscript{m}, \textdelta$_0$ : $-72^\circ$) 
    as the centre of \textit{Gaia} DR2 sources. X and Y are defined as in \protect\cite{vandermarel2001} centred on the MB. The East and North are towards left and up respectively. The blue and the black circles around the SMC mark 4$\degree$ and 10$\degree$ radial regions from the optical centre of SMC ($\alpha_{S}$ = 00\textsuperscript{h}52\textsuperscript{m}12\textsuperscript{s}.5 and $\delta_{S}$ = $-72^\circ$49\arcmin43\arcsec ; J2000 \protect\citealp{deVaucouleurs1972}) respectively. The black circle around the LMC shows a 10$\degree$ radial region from its optical centre coordinates ($\alpha_{L}$ = 05\textsuperscript{h}23\textsuperscript{m}35\textsuperscript{s} and $\delta_{L}$ = $-69^\circ$45\arcmin22\arcsec ; J2000 \protect\citealp{deVaucouleurs1972}).
    The black and the magenta ellipses show the approximate locations of the HI (gaseous) bridge region of the MB, also traced by young stars, and the old stellar bridge identified by \protect\cite{belokurov2017} respectively. The colour bar represents the stellar density (number of stars per bin of size 16.2 arcmin$^2$, in units of arcmin$^{-2}$).}
    \label{fig:cart_selection}
\end{figure}
\indent In order to select stars in the SMC, we applied various selection criteria. We set an initial cut-off in parallax (parallax $\leq$ 0.2 mas ; \citealp{Luri2018}), which corresponds to a distance of 5 kpc, to reduce the contribution from Milky Way foreground stars.  A 5-sigma cut was applied to the flux signal-to-noises in all three bands \textit{(G, G$_{BP}$, G$_{RP}$)}, resulting in magnitude errors $\le$ 0.2 mag. Fig. \ref{fig:cart_selection} shows that the stellar density in the central regions of the MCs is very high. The astrometric data obtained from \textit{Gaia} DR2 suffer from small-scale systematic variations due to crowding (refer to \citealt{lindegren2018, Vasiliev2018b} for more details). Astrometric excess noise is a quality indicator provided by \textit{Gaia} DR2 to assess the reliability of the astrometric data. This parameter is expressed in units of mas, and can be used to select sources that are reliable and consistent with the five parameters astrometric solution. We selected sources which have astrometric excess noise values $\le$ 1.3 mas. \textit{Gaia} DR2 data provide proper motion measurements in the RA ($\mu_{\alpha}$) and Dec ($\mu_{\delta}$) directions. We applied a cut to these values based on the expected range in proper motion values ($-$3 mas yr$^{-1}$ $\le$  $\mu_{\alpha}$  $\le$ $+$3 mas yr$^{-1}$ and $-$3 mas yr$^{-1}$ $\le$  $\mu_{\delta}$ $\le$ $+$3 mas yr$^{-1}$) predicted by simulations \citep{Diaz2012} for stellar tidal features around the MCs. This cut-off applied to the proper motion values further reduces the Milky Way contamination in the data.

\subsection{Spatial distribution and sub-regions}
\label{section2.2}
The spatial distribution of stars inside the 10$\degree$ radial region of the SMC is shown in the XY plane in Fig. \ref{fig:Cartesian Plot}. We divided the observed region of the SMC into several sub-regions. Initially, we divided the 10$^\circ$-radius region into 20 annular sub-regions, each with a width of 0$\rlap{.}^{\circ}$5, shown as the black concentric circles in Fig. \ref{fig:Cartesian Plot}. Then each annular region is further divided into four sectors, viz. North East (NE, Y $>$ 0 \& X $<$ 0), North West (NW, Y $>$ 0 \& X $>$ 0), South East (SE, Y $<$ 0 \& X $<$ 0) and South West (SW, Y $<$ 0 \& X $>$ 0). Due to the low number of stars in the outer sub-regions (beyond 5$^\circ$) it is difficult to identify and analyse the properties of RC stars in 0$\rlap{.}^{\circ}$5 annular sub-regions. There we merge some of the sub-regions (sector-wise) and analyse the data in the merged regions. This is discussed in Section \ref{section3.2}. 
\begin{figure}
    \centering
    \includegraphics[scale=0.55]{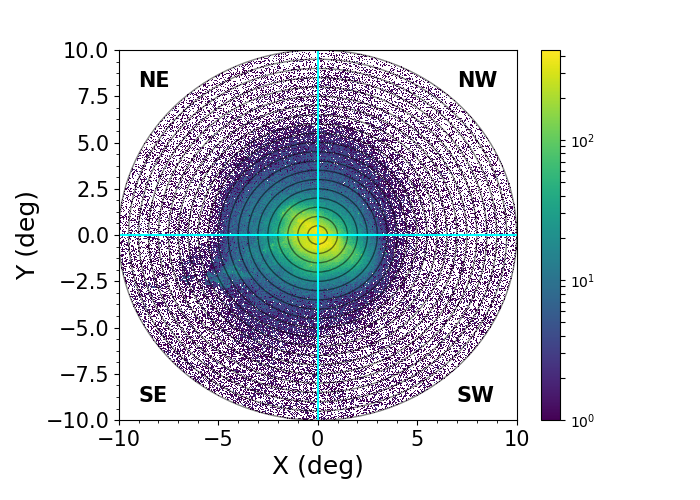}
    \caption{Cartesian plot of the 10$\degree$ region of the SMC with its optical centre ($\alpha_{S}$ = 00\textsuperscript{h}52\textsuperscript{m}12\textsuperscript{s}.5 and $\delta_{S}$ = $-72^\circ$49\arcmin43\arcsec ; J2000 \protect\citealp{deVaucouleurs1972}) as origin using \textit{Gaia} DR2 data. The X and Y are defined as in \protect\cite{vandermarel2001}. The black concentric circles in the plot show the 0$\rlap{.}^{\circ}$5 radial sub-regions. The blue to yellow colour bar indicates the increase in stellar density. The cyan lines show the division of NE, NW, SE and SW sub-regions. The colour bar has the same units as that in Fig. \ref{fig:cart_selection}.}
    \label{fig:Cartesian Plot}
\end{figure}}
\section{Analysis}
\label{section3}
RC stars in different sub-regions are identified using the ${G_0}$ vs. \textit{$(G_{BP}$} - \textit{$G_{RP})_{0}$} colour-magnitude diagram (CMD) while their magnitude distributions are analysed to find the spatial extent of the foreground stellar sub-structure in eastern sub-regions. Below we describe the details of the extinction correction and the analysis. 

\begin{figure*}
    \captionsetup[subfigure]{labelformat=empty}
    \centering
    \hspace*{-2.2em}
    \subfloat[]{\includegraphics[width=0.28\textwidth]{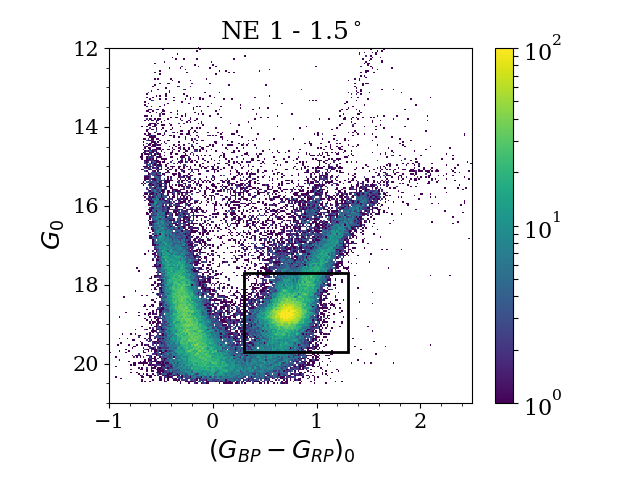}}
    \hspace*{-2.2em}
    \subfloat[]{\includegraphics[width=0.28\textwidth]{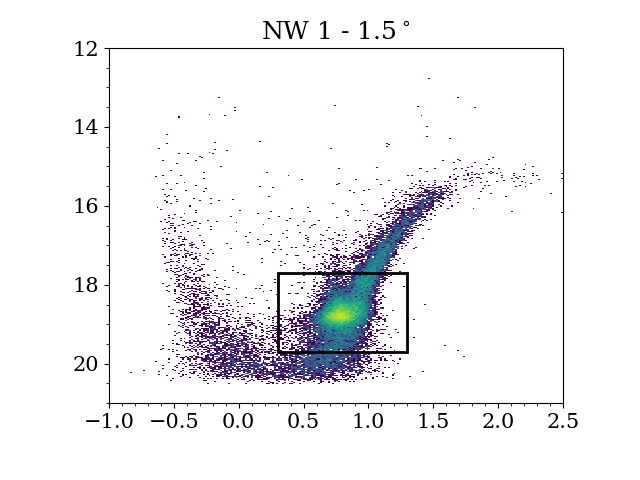}}
    \hspace*{-2.2em}
    \subfloat[]{\includegraphics[width=0.28\textwidth]{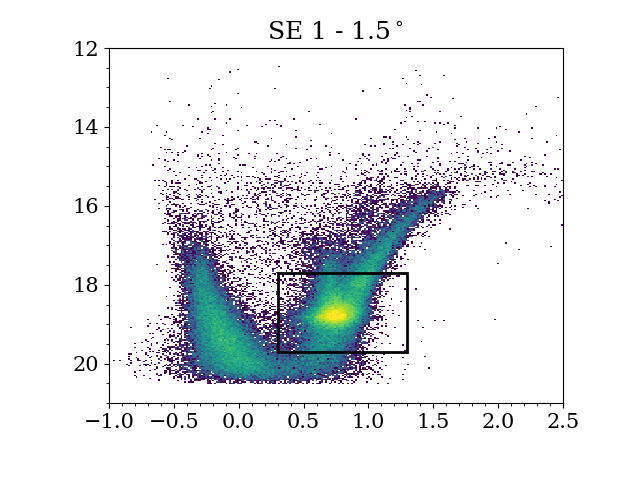}}
    \hspace*{-2.2em}
    \subfloat[]{\includegraphics[width=0.28\textwidth]{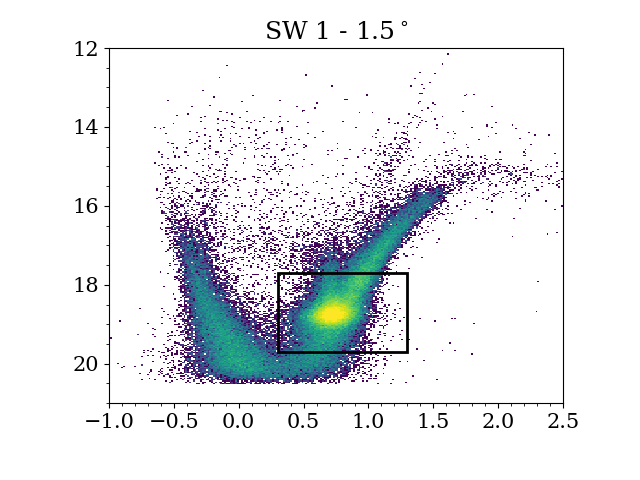}}\\
    \vspace{-0.9cm}
    \hspace*{-2.2em}
    \subfloat[]{\includegraphics[width=0.28\textwidth]{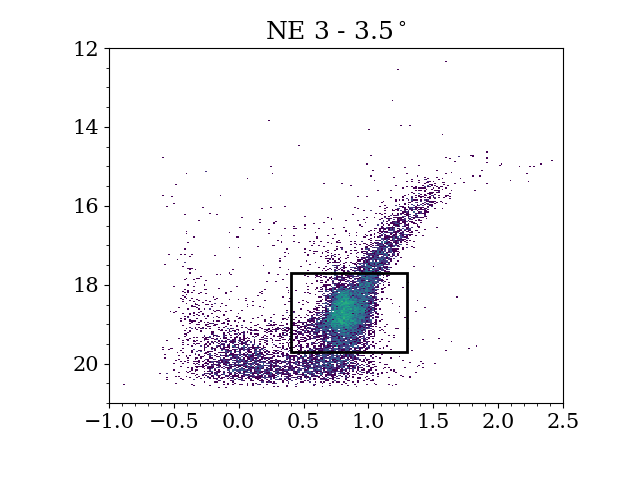}}
    \hspace*{-2.2em}
    \subfloat[]{\includegraphics[width=0.28\textwidth]{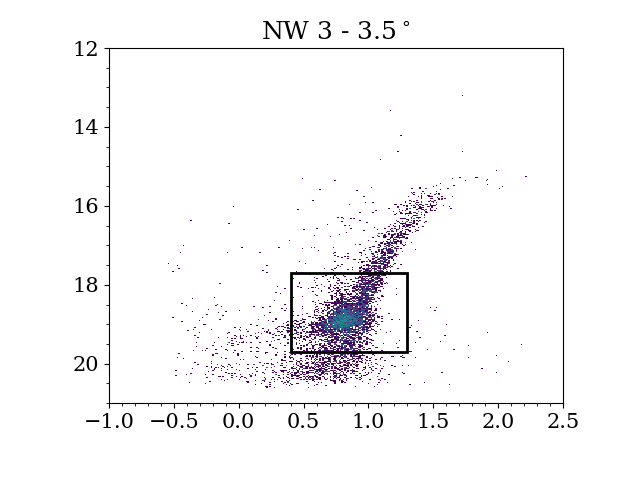}}
    \hspace*{-2.2em}
    \subfloat[]{\includegraphics[width=0.28\textwidth]{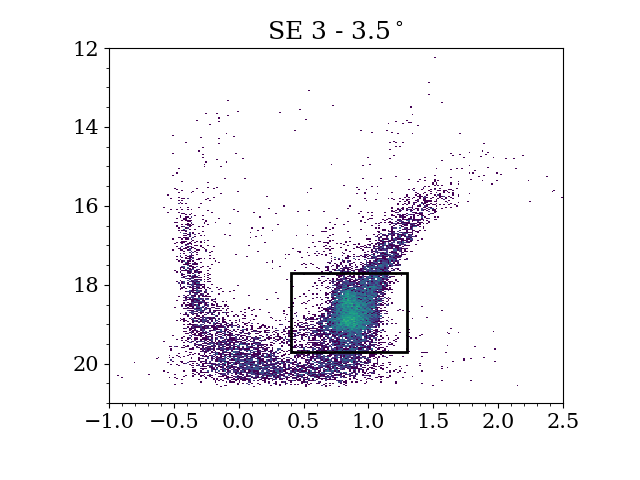}}
    \hspace*{-2.2em}
    \subfloat[]{\includegraphics[width=0.28\textwidth]{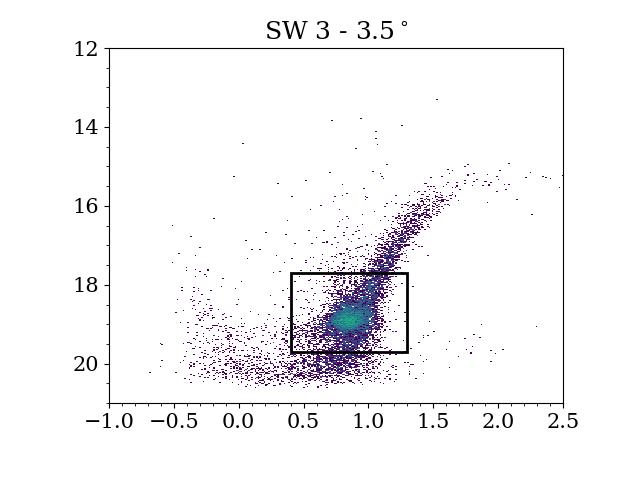}}\\
    \vspace{-0.9cm}
    \hspace*{-2.2em}
    \subfloat[]{\includegraphics[width=0.28\textwidth]{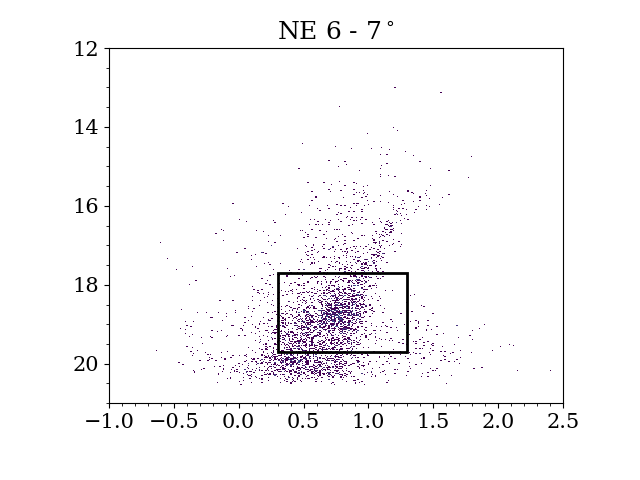}}
    \hspace*{-2.2em}
    \subfloat[]{\includegraphics[width=0.28\textwidth]{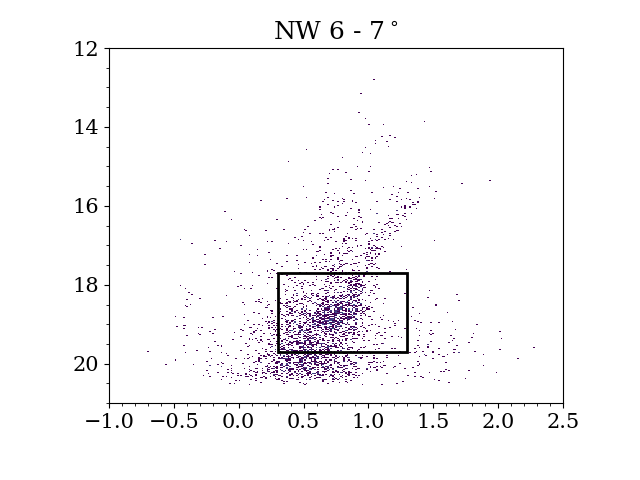}}
    \hspace*{-2.2em}
    \subfloat[]{\includegraphics[width=0.28\textwidth]{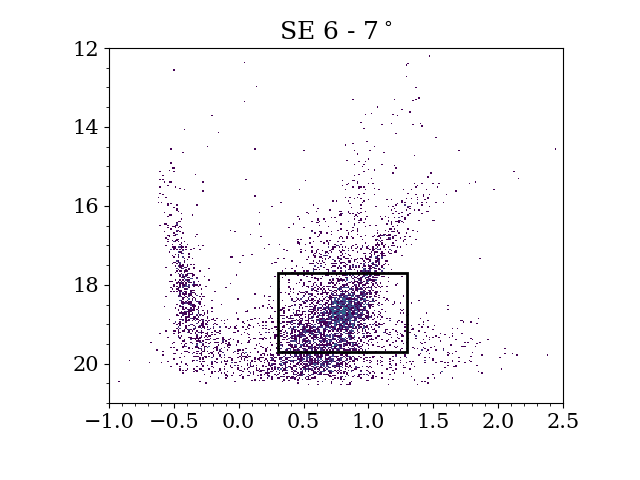}}
    \hspace*{-2.2em}
    \subfloat[]{\includegraphics[width=0.28\textwidth]{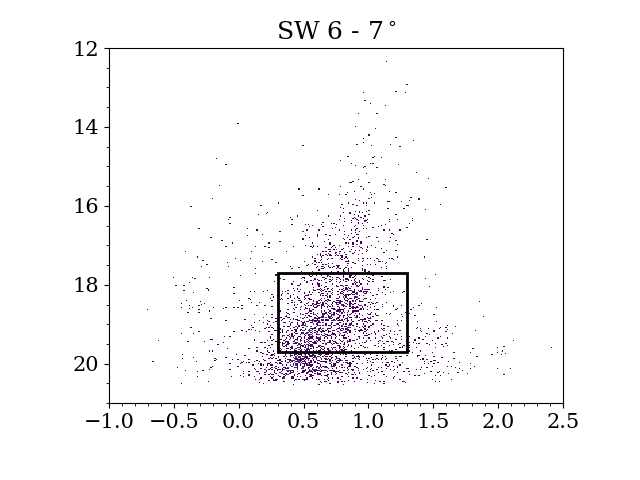}}\\
    \caption{Hess diagrams representing the stellar density in the observed CMD of \textit{Gaia} DR2 sources enclosed in sub-regions within the 1--1$\rlap{.}^{\circ}$5 (first row), 3--3$\rlap{.}^{\circ}$5 (second row) and 6--7$^\circ$ (third row) radial region from the SMC centre. The colour bar from blue to yellow represents the increase in stellar density. In black are the rectangular boxes to select RC stars within each sub-region. The axis labels and colorbar are the same for all plots and are shown for the top left panel only.}
    \label{fig:CMD}
\end{figure*}

\subsection{Extinction correction}
\label{section3.1}
\indent We need to correct for the effect of interstellar extinction before we analyse the properties of the RC stars. \citet{Rubele2018} provide an extinction map for the regions within $\sim$ 3$^\circ$ from the SMC centre and we use those values to correct \textit{Gaia} DR2 data for the extinction effect. They used a synthetic CMD technique to retrieve the star-formation history, metallicities, distances and extinction values of each sub-region in their study. We applied the extinction values corresponding to the sub-region in which the stars in our sample fall. For stars belonging to regions beyond their extinction map we used the extinction values of the nearest sub-region. We converted the extinction values in the visual band (A$_V$) to extinction values in the \textit{Gaia} bands (A$_G$, A$_{G_{BP}}$, A$_{G_{RP}}$) using constant multiplicative factors (0.859, 1.068, and 0.652, respectively) provided by \citet{chen2019}, which were derived using the extinction law from \citet{Cardelli1989}. In the analysis that follows, we use extinction corrected magnitudes of the \textit{G}, \textit{G$_{BP}$} \& \textit{G$_{RP}$} bands. We note that applying the extinction values of inner regions, derived for regions at 3$^\circ$, to the outer regions may lead to over-subtraction of extinction in these  regions. This can affect the distance estimation and is discussed in Section \ref{section5}. 
\subsection{Hess diagrams and identification of RC stars in different sub-regions}
\label{section3.2}
We constructed Hess diagrams (stellar density plots) of the ${G_0}$ vs. $(G_{BP} - G_{RP})_0$ CMD for all sub-regions, with bin sizes of 0.01 mag in $(G_{BP} - G_{RP})_0$ colour and 0.04 mag in ${G_0}$. As the stellar density in the outer sub-regions (beyond 5$^\circ$) is low compared to the inner regions, it is difficult to clearly identify the RC feature in the Hess diagrams. Hence, instead of analysing a 0$\rlap{.}^{\circ}$5 annular sub-regions we carried out the same analysis for 1$\degree$ sub-regions between 5--7$^\circ$ radius. Beyond 7$\degree$, the number density of sources is even smaller, so we merged the 7--10$^\circ$ annuli into one, but retaining the four sectors and performed the analysis. \\
\indent Fig. \ref{fig:CMD} shows the Hess diagrams for the sub-regions (NE, NW, SE and SW) within 1--1$\rlap{.}^{\circ}$5 (first row), 3--3$\rlap{.}^{\circ}$5 (second row) and 6--7$^\circ$ (third row) 
from the SMC centre. Hess diagrams for all the other sub-regions are shown in 
Figs. \ref{fig:CMD0-2.5}, \ref{fig:CMD2.5-5} and \ref{fig:CMD5-10}. The black box in the Hess diagrams represents the RC region and the box size is $\sim$1 mag in $(G_{BP} - G_{RP})_0$ and $\sim$2 mag in ${G_0}$. The exact magnitude and colour range of the RC box in different sub-regions are defined based on the visual inspection of each Hess diagram, ensuring that the entire RC feature is included in the selection box. In the 7--10$^\circ$ region, the size of the selection box is smaller ($\sim$0.5 mag in $(G_{BP} - G_{RP})_0$ and $\sim$1.2 mag in ${G_0}$) to reduce the contamination from Milky Way stars. A vertical extension in the form of a double RC feature is visible in the eastern sub-regions of the SMC between $\sim$ 2$\rlap{.}^{\circ}$5 to 5$\rlap{.}^{\circ}$0. The colour of the two clumps as seen in the Hess diagrams is similar. Such a vertical extension of the RC feature is not visible in the other sub-regions. In the next sub-section we analyse the magnitude distribution of the RC stars in different sub-regions to better describe these variations.
\begin{figure*}
    \captionsetup[subfigure]{labelformat=empty}
    \centering
    \hspace*{-1.6em}
    \vspace{-0.8cm}
    \subfloat[]{\includegraphics[width=0.26\textwidth]{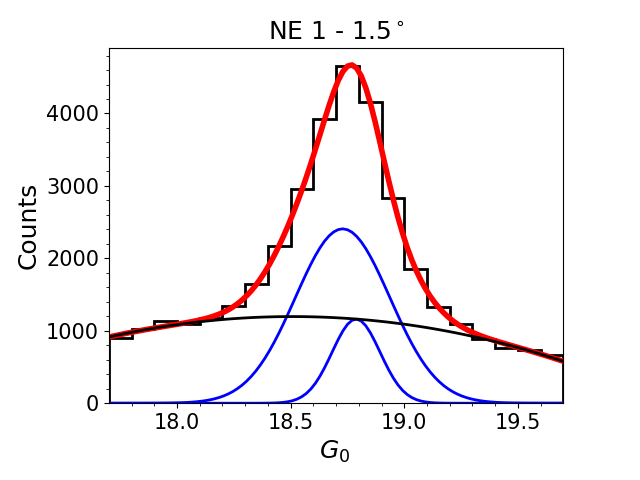}}
    \hspace*{-1.6em}
    \subfloat[]{\includegraphics[width=0.26\textwidth]{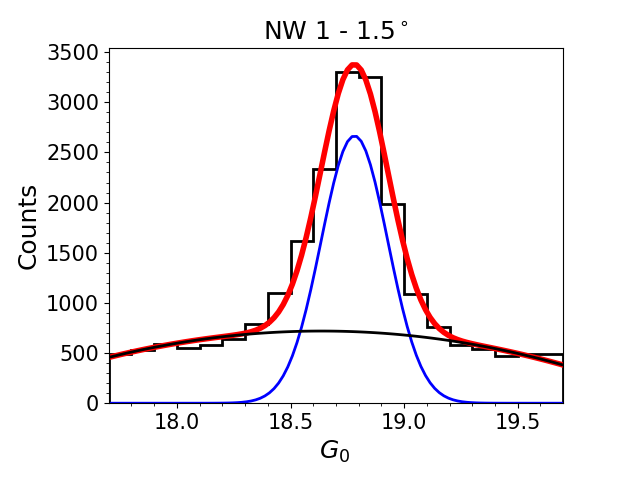}}
    \hspace*{-1.6em}
    \subfloat[]{\includegraphics[width=0.26\textwidth]{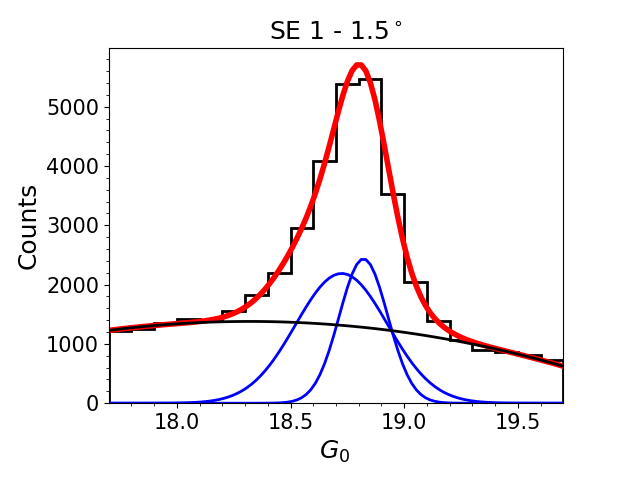}}
    \hspace*{-1.6em}
    \subfloat[]{\includegraphics[width=0.26\textwidth]{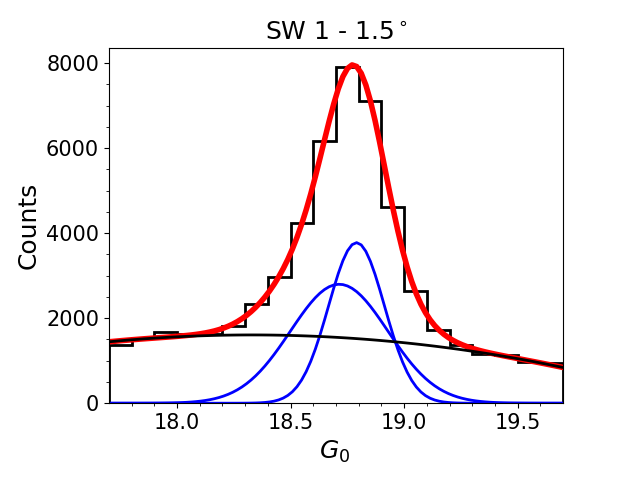}}\\
    \hspace*{-1.6em}
    \vspace{-0.8cm}
    \subfloat[]{\includegraphics[width=0.26\textwidth]{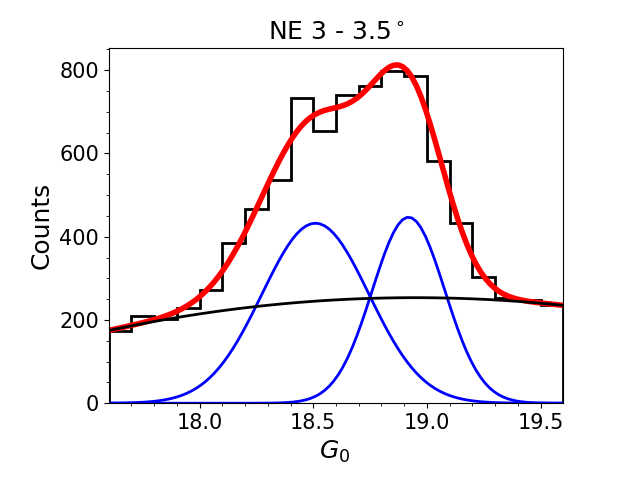}}
    \hspace*{-1.6em}
    \subfloat[]{\includegraphics[width=0.26\textwidth]{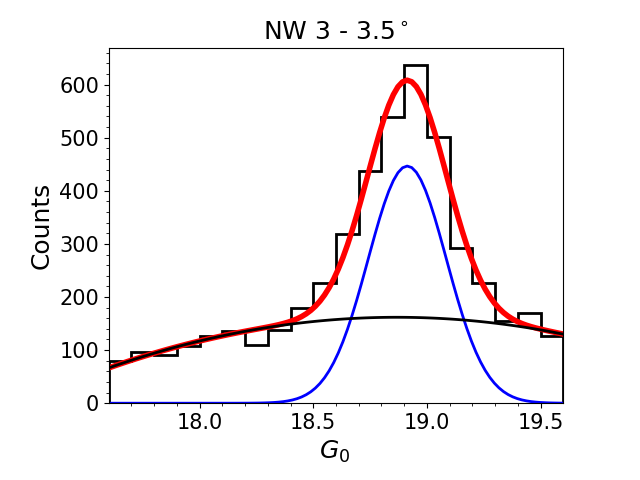}}
    \hspace*{-1.6em}
    \subfloat[]{\includegraphics[width=0.26\textwidth]{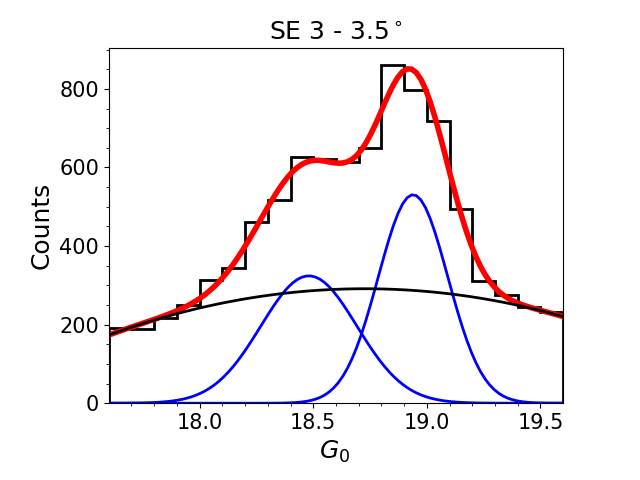}}
    \hspace*{-1.6em}
    \subfloat[]{\includegraphics[width=0.26\textwidth]{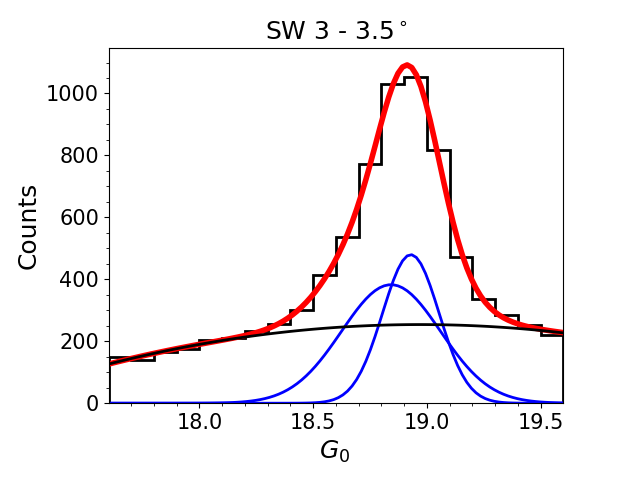}}\\
    \hspace*{-1.1em}
    \vspace{-1cm}
    \subfloat[]{\includegraphics[width=0.26\textwidth]{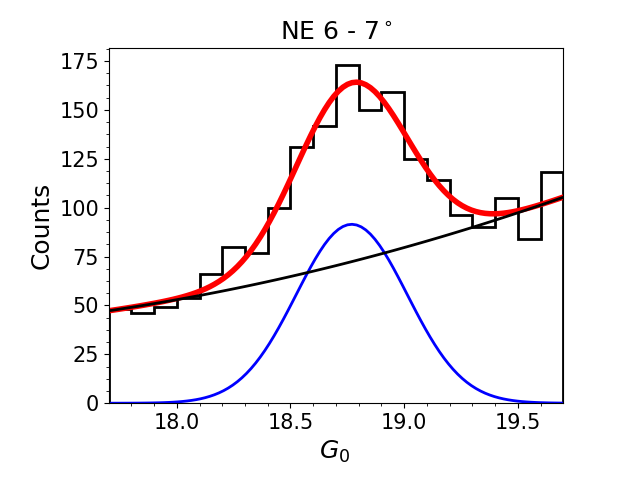}}
    \hspace*{-1.1em}
    \subfloat[]{\includegraphics[width=0.26\textwidth]{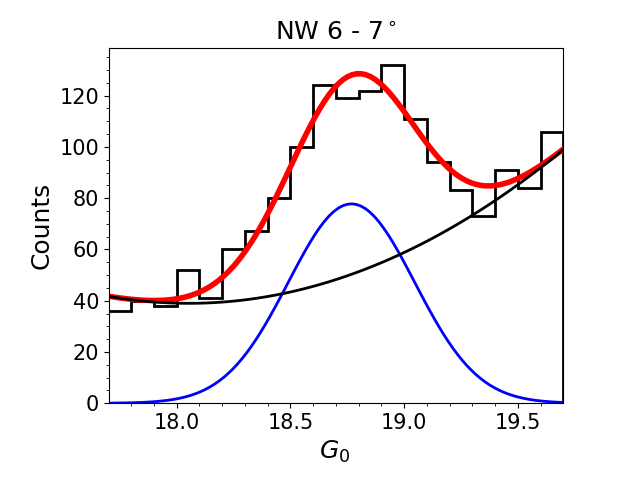}}
    \hspace*{-1.1em}
    \subfloat[]{\includegraphics[width=0.26\textwidth]{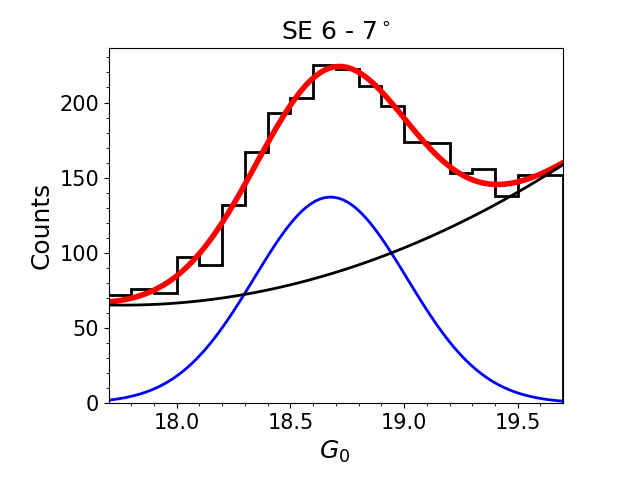}}\\
    \hspace*{-1.1em}
    \vspace{0.5cm}
    \caption{Magnitude distributions of RC stars in the 1--1$\rlap{.}^{\circ}$5 (first row), 3--3$\rlap{.}^{\circ}$5 (second row) and 6--7$^\circ$ (third row) sub-regions and their best fits. Blue, black and red lines indicate the Gaussian function, the quadratic polynomial and the total fit respectively.}
    \label{fig:magfit}
\end{figure*}

\begin{table*}
\centering
\caption{Gaussian fit parameters for the magnitude distributions of RC stars in the NE, NW, SE and SW regions.}
\label{tab:mag_ne}
\resizebox{\textwidth}{!}{%
\begin{tabular}{|c|c|c|c|c|c|c|c|c|c|c|}
\hline
\multirow{2}{*}{\begin{tabular}[c]{@{}c@{}}Radius\\ (deg)\end{tabular}} & \multicolumn{5}{c|}{NORTH EAST} & \multicolumn{5}{c|}{NORTH WEST} \\ \cline{2-11} 
 & Peak1 (mag)& Sigma1 (mag)& Peak2 (mag)& Sigma2 (mag)& $\chi^2$ & Peak1 (mag)& Sigma1 (mag)& Peak2 (mag)& Sigma2 (mag)& $\chi^2$ \\ \hline
0--0.5 & 18.947 $\pm$ 0.014 & 0.102 $\pm$ 0.008 & 18.863 $\pm$ 0.017 & 0.196 $\pm$ 0.011 & 1.48 & 18.835 $\pm$ 0.004 & 0.150 $\pm$ 0.003 & --  & --  & 1.52 \\ 
0.5--1 & 19.011 $\pm$ 0.010 & 0.127 $\pm$ 0.006 & 18.932 $\pm$ 0.023 & 0.215 $\pm$ 0.020 & 3.07 & 19.006 $\pm$ 0.007 & 0.104 $\pm$ 0.004 & 18.921 $\pm$ 0.014 & 0.195 $\pm$ 0.009 & 3.18 \\ 
1--1.5 & 18.788 $\pm$ 0.007 & 0.106 $\pm$ 0.005 & 18.729 $\pm$ 0.006 & 0.208 $\pm$ 0.008 & 1.52 & 18.781 $\pm$ 0.005 & 0.147 $\pm$ 0.004 & --  & --  & 8.13 \\ 
1.5--2 & 18.840 $\pm$ 0.014 & 0.148 $\pm$ 0.010 & 18.725 $\pm$ 0.046 & 0.240 $\pm$ 0.029 & 2.06 & 18.935 $\pm$ 0.012 & 0.108 $\pm$ 0.007 & 18.850 $\pm$ 0.026 & 0.191 $\pm$ 0.014 & 3.27 \\ 
2--2.5 & 18.927 $\pm$ 0.009 & 0.204 $\pm$ 0.010 & --  & --  & 5.58 & 19.038 $\pm$ 0.004 & 0.167 $\pm$ 0.003 & --  & --  & 2.27 \\ 
2.5--3 & 18.922 $\pm$ 0.016 & 0.190 $\pm$ 0.008 & 18.499 $\pm$ 0.029 & 0.189 $\pm$ 0.015 & 0.52 & 18.975 $\pm$ 0.005 & 0.114 $\pm$ 0.003 & 18.904 $\pm$ 0.010 & 0.229 $\pm$ 0.012 & 0.36 \\ 
3--3.5 & 18.919 $\pm$ 0.036 & 0.160 $\pm$ 0.016 & 18.510 $\pm$ 0.066 & 0.235 $\pm$ 0.054 & 1.88 & 18.914 $\pm$ 0.008 & 0.172 $\pm$ 0.007 & --  & --  & 2.39 \\ 
3.5--4 & 18.782 $\pm$ 0.022 & 0.183 $\pm$ 0.013 & 18.359 $\pm$ 0.027 & 0.180 $\pm$ 0.013 & 0.67 & 18.841 $\pm$ 0.009 & 0.183 $\pm$ 0.009 & --  & --  & 2.28 \\ 
4--4.5 & 18.825 $\pm$ 0.018 & 0.159 $\pm$ 0.010 & 18.362 $\pm$ 0.019 & 0.209 $\pm$ 0.016 & 0.85 & 18.814 $\pm$ 0.006 & 0.187 $\pm$ 0.006 & --  & --  & 1.00 \\ 
4.5--5 & 18.582 $\pm$ 0.018 & 0.332 $\pm$ 0.085 & --  & --  & 1.03 & 18.806 $\pm$ 0.004 & 0.207 $\pm$ 0.005 & --  & --  & 0.30 \\ 
5--6 & 18.697 $\pm$ 0.015 & 0.199 $\pm$ 0.018 & --  & --  & 2.07 & 18.816 $\pm$ 0.014 & 0.230 $\pm$ 0.019 & --  & --  & 2.95 \\ 
6--7 & 18.769 $\pm$ 0.019 & 0.244 $\pm$ 0.036 & --  & --  & 0.90 & 18.769 $\pm$ 0.021 & 0.278 $\pm$ 0.053 & --  & --  & 0.76 \\ 
7--10 & 18.856 $\pm$ 0.047 & 0.193 $\pm$ 0.048 & --  & --  & 0.81 & 18.939 $\pm$ 0.026 & 0.105 $\pm$ 0.014 & --  & --  & 1.99 \\ \hline
\multirow{2}{*}{\begin{tabular}[c]{@{}c@{}}Radius\\ (deg)\end{tabular}} & \multicolumn{5}{c|}{SOUTH EAST} & \multicolumn{5}{c|}{SOUTH WEST} \\ \cline{2-11} 
 & Peak1 (mag) & Sigma1 (mag)& Peak2 (mag)& Sigma2 (mag)& $\chi^2$ & Peak1 (mag)& Sigma1 (mag)& Peak2 (mag)& Sigma2 (mag)& $\chi^2$ \\ \hline
0--0.5 & 18.852 $\pm$ 0.004 & 0.160 $\pm$ 0.003 & --  & --  & 2.11 & 18.871 $\pm$ 0.006 & 0.199 $\pm$ 0.006 & --  & --  & 1.07 \\ 
0.5--1 & 18.985 $\pm$ 0.005 & 0.157 $\pm$ 0.005 & --  & --  & 11.72 & 18.987 $\pm$ 0.003 & 0.182 $\pm$ 0.003 & --  & --  & 2.65 \\ 
1--1.5 & 18.821 $\pm$ 0.005 & 0.107 $\pm$ 0.003 & 18.725 $\pm$ 0.014 & 0.202 $\pm$ 0.009 & 2.22 & 18.790 $\pm$ 0.006 & 0.123 $\pm$ 0.004 & 18.714 $\pm$ 0.015 & 0.214 $\pm$ 0.014 & 2.98 \\ 
1.5--2 & 18.918 $\pm$ 0.006 & 0.123 $\pm$ 0.004 & 18.799 $\pm$ 0.024 & 0.230 $\pm$ 0.017 & 1.83 & 18.860 $\pm$ 0.003 & 0.158 $\pm$ 0.003 &--&--& 7.36 \\ 
2--2.5 & 19.046 $\pm$ 0.007 & 0.132 $\pm$ 0.005 & 18.828 $\pm$ 0.041 & 0.337 $\pm$ 0.064 & 2.21 & 19.003 $\pm$ 0.004 & 0.161 $\pm$ 0.003 &--&--& 6.49 \\ 
2.5--3 & 18.960 $\pm$ 0.009 & 0.166 $\pm$ 0.006 & 18.472 $\pm$ 0.028 & 0.214 $\pm$ 0.026 & 0.82 & 18.943 $\pm$ 0.005 & 0.118 $\pm$ 0.004 & 18.865 $\pm$ 0.015 & 0.230 $\pm$ 0.016 & 1.11 \\ 
3--3.5 & 18.939 $\pm$ 0.019 & 0.153 $\pm$ 0.009 & 18.481 $\pm$ 0.038 & 0.211 $\pm$ 0.036 & 1.51 & 18.929 $\pm$ 0.009 & 0.124 $\pm$ 0.006 & 18.842 $\pm$ 0.028 & 0.218 $\pm$ 0.018 & 0.67 \\ 
3.5--4 & 18.852 $\pm$ 0.038 & 0.191 $\pm$ 0.024 & 18.372 $\pm$ 0.057 & 0.219 $\pm$ 0.039 & 1.55 & 18.803 $\pm$ 0.007 & 0.202 $\pm$ 0.008 & --  & --  & 0.62 \\ 
4--4.5 & 18.852 $\pm$ 0.027 & 0.194 $\pm$ 0.015 & 18.387 $\pm$ 0.030 & 0.206 $\pm$ 0.020 & 0.54 & 18.817 $\pm$ 0.015 & 0.243 $\pm$ 0.023 & --  & --  & 1.29 \\ 
4.5--5 & 18.880 $\pm$ 0.097 & 0.202 $\pm$ 0.050 & 18.413 $\pm$ 0.117 & 0.252 $\pm$ 0.092 & 1.58 & 18.832 $\pm$ 0.028 & 0.276 $\pm$ 0.051 & --  & --  & 1.25 \\ 
5--6 & 18.554 $\pm$ 0.050 & 0.295 $\pm$ 0.131 & --  & --  & 4.31 & 18.941 $\pm$ 0.057 & 0.311 $\pm$ 0.116 & --  & --  & 1.33 \\ 
6--7 & 18.677 $\pm$ 0.014 & 0.336 $\pm$ 0.071 & --  & --  & 0.52 & --  & --  & --  & --  & -- \\ 
7--10 & 18.734 $\pm$ 0.021 & 0.203 $\pm$ 0.034 & --  & --  & 0.57 & --  & --  & --  & --  & -- \\ \hline
\end{tabular}%
}
\end{table*}

\subsection{Magnitude distributions of RC stars}
\label{section3.3}
\begin{figure*}
    \centering
   \subfloat{\includegraphics[width=0.45\textwidth]{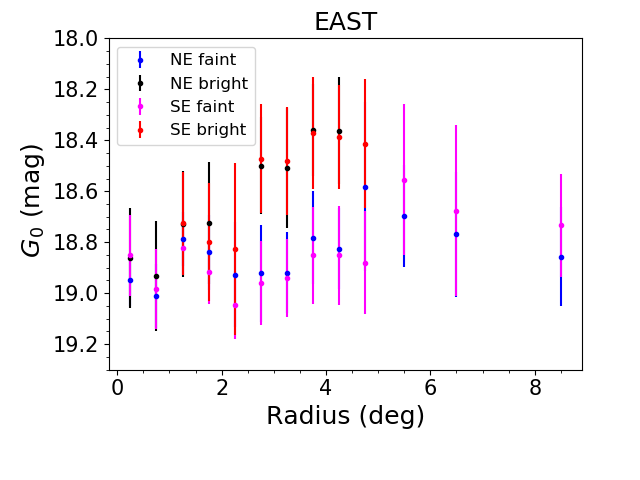}}
    \subfloat{\includegraphics[width=0.45\textwidth]{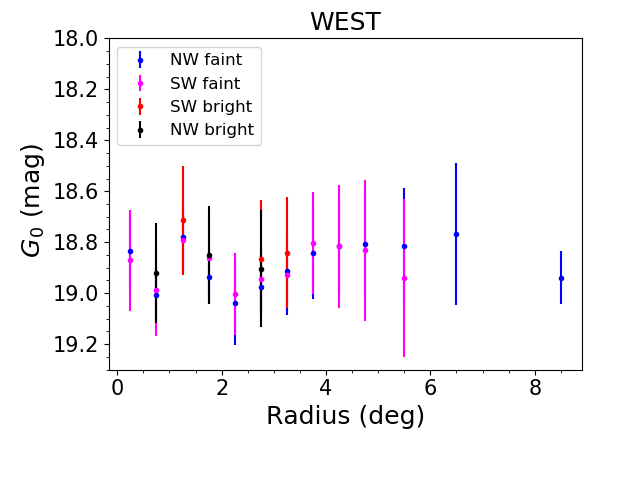}}
    \vspace{-0.5cm}
    \caption{Peak magnitude vs. radius in eastern (NE and SE; left) and western (NW and SW; right) regions. Blue (magenta) points indicate the peak magnitude of faint RC stars in the  northern (southern) sub-regions and black (red) points correspond to the peak magnitudes of bright RC stars in the northern (southern) sub-regions.}
    \label{fig:eastandwest}
\end{figure*}

We created histograms of the ${G_0}$ magnitude distributions of the selected RC stars in each of the sub-regions with a bin size of 0.1 mag. The observed magnitude distribution is initially fit using a single Gaussian function to account for the RC distribution and a quadratic polynomial to account for the RGB stars in the RC selection box. An additional Gaussian is added only if the reduced $\chi^2$ of the fit improves by at least 25$\%$ compared to the reduced $\chi^2$ value of the fit with a single Gaussian function and the width of the second Gaussian is more than the bin size of the distribution. The magnitude distributions of RC stars along with the best fit profiles and multiple components are shown in Figs. \ref{fig:magfit}, \ref{fig:magfit0-2.5}, \ref{fig:magfit2.5-5} and \ref{fig:magfit5-10} for different sub-regions. 
The fits to the distributions are performed using the curvefit function in Python-Scipy \citep{scipy2020}, which employs the non-linear least squares method. Many sub-regions show bimodality in the RC magnitude distribution. The fit parameters are tabulated along with the fit errors in Table \ref{tab:mag_ne}. Due to the very low number of stars in the 6--7$\degree$ and 7--10$\degree$ SW sub-regions, reasonable fits to the magnitude distributions were not obtained.\\
\indent The left and right panels of Fig. \ref{fig:eastandwest} show the RC peak magnitudes in the eastern and western sub-regions respectively, as a function of radius. The faint RC peak magnitude in the northern (southern) sub-regions are indicated by blue (magenta) points and the peak magnitude of bright RC stars in the northern (southern) sub-regions are indicated by black (red) points respectively. The error bar corresponding to each point represents the observed dispersion (width) of the respective Gaussian component. The left panel of Fig. \ref{fig:eastandwest} shows that there is $\sim$ 0.45 mag difference between the bright and the faint RC peak magnitudes in the eastern sub-regions between 2$\rlap{.}^{\circ}$5--5$^\circ$ radius and the difference is more than the dispersion of the respective Gaussian components (as indicated by the error bars). This suggests the presence of a dual RC population in the eastern sub-regions between 2$\rlap{.}^{\circ}$5--5$^\circ$ radius (2$\rlap{.}^{\circ}$5--4$\rlap{.}^{\circ}$5 in NE and 2$\rlap{.}^{\circ}$5--5$^\circ$ in SE). For all the other sub-regions which show double Gaussian components, the peak magnitudes are not significantly different. In all the sub-regions between 5--10$\degree$ (including those in the East), only a single RC is present (see Figs. \ref{fig:magfit5-10} and \ref{fig:eastandwest}). The single peak RC magnitudes in the NE 4$\rlap{.}^{\circ}$5--5$^\circ$ and SE 5--6$^\circ$ sub-regions are closer to the bright RC peak found in the NE 2$\rlap{.}^{\circ}$5--4$\rlap{.}^{\circ}$5 and SE 2$\rlap{.}^{\circ}$5--5$^\circ$ sub-regions. This could be due to the presence of two overlapping RC populations in these sub-regions. We also note that these sub-regions have large dispersion, but we were not able to fit two Gaussian components to reduce the dispersion like in other regions.\\ 
\indent To investigate whether there is any change in the peak magnitudes of RC stars due to the way in which we have divided the observed region, we sub-divided the SE sector and performed the same analysis. The SE sectors (3--3$\rlap{.}^{\circ}$5, 3$\rlap{.}^{\circ}$5--4$^\circ$, 5--7$^\circ$ and 7--10$^\circ$) are further divided diagonally into two (upper: $\phi$ $\leq$ 45$\degree$ and lower: $\phi$ > 45$\degree$). From the obtained fit parameters, we find that there is no significant difference in the peak magnitudes of upper and lower sectors and also with peak values obtained for the combined sector. As a typical example, the peak and dispersion in magnitude for SE 7--10$^\circ$ upper and lower are 18.737$\pm$0.114 mag and 18.741$\pm$0.270 mag whereas 18.734$\pm$0.203 mag is for the entire sector.\\
\indent We note that in some of the CMDs (notably in SE) the vertical extension of RC stars goes beyond the upper edge of our current selection box. As we can see from the Hess diagrams, the number density of this feature beyond our selection box is very low. Extending the selection box to cover the entire extension simultaneously increases the contribution from RGB stars (which are larger in number than the extended tail of the bright RC feature) in the brighter magnitude range. However it is important to verify the effect of excluding some stars at the brighter end, on the magnitude distribution and best-fitting parameters. To include the entire RC feature in some of the SE sub-regions, we slightly extended the magnitude range of the RC selection box. Then we performed the same analysis and obtained the best-fitting parameters. As a typical example, the parameters (peak and dispersion) for the 2$\rlap{.}^{\circ}$5--3$^\circ$ sub-region with the extended RC selection box are 18.962 $\pm$ 0.167 mag (faint) and 18.471 $\pm$ 0.223 mag (bright), respectively. Comparison of the best-fitting parameters, based on the extended RC selection box, and the parameters in Table \ref{tab:mag_ne} shows that there is no significant difference. Thus the final results are not affected by excluding a few stars at the brighter end of the vertical extension of the RC star distribution in some sub-regions.\\
\indent Also note that one of the applied selection criteria, of flux signal-to-noise $\ge$ 5 in all three bands, can reduce the number of stars at fainter magnitudes (mostly \textit{G} $\ge$ 18.7 mag) and may affect the RC magnitude distribution. However, the number of sources removed from the RC selection box, based on this criterion, is negligible ($\sim$1.24\% of sources in the central crowded sub-regions and $\sim$ 0.05\% in the outer sub-regions) and hence this is highly unlikely to affect the RC magnitude distributions and derived parameters.

\begin{figure}
    \centering
    \hspace*{-1em}
    \subfloat{\includegraphics[width=0.26\textwidth]{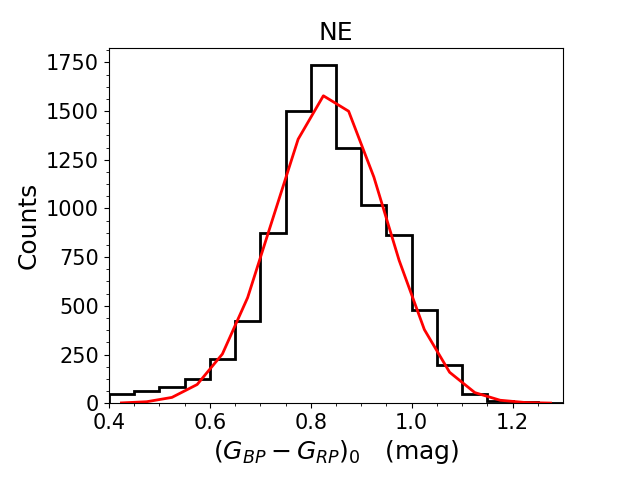}}
    \hspace*{-1.1em}
    \subfloat{\includegraphics[width=0.26\textwidth]{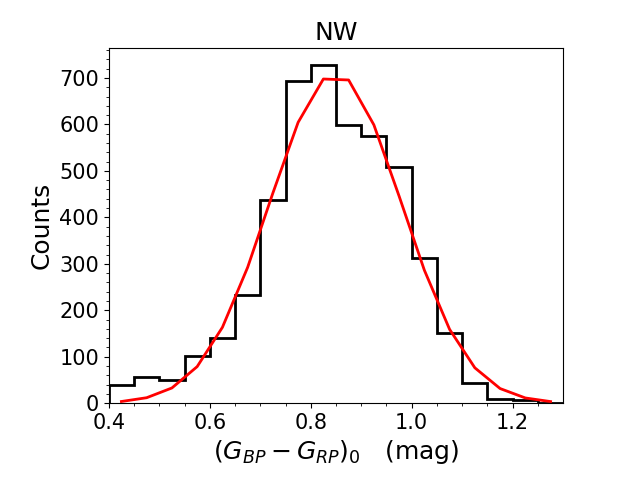}}\\
    \vspace{-0.5cm}
    \hspace*{-1em}
    \subfloat{\includegraphics[width=0.26\textwidth]{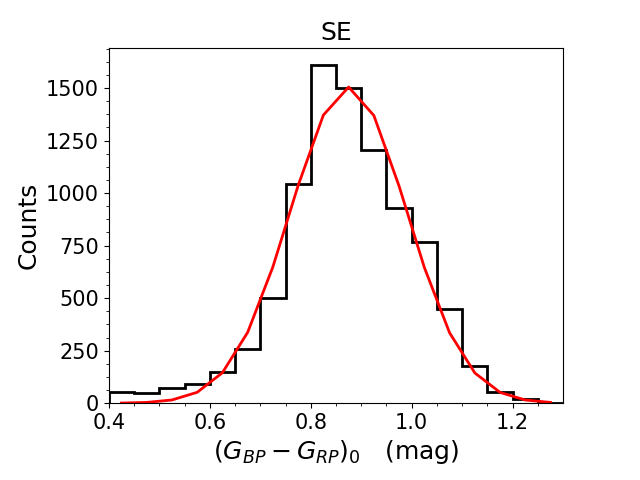}}
    \hspace*{-1.1em}
    \subfloat{\includegraphics[width=0.26\textwidth]{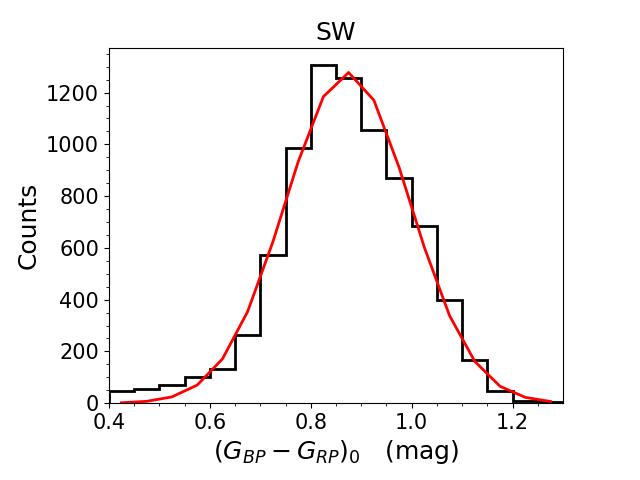}}\\
    \caption{The $(G_{BP} - G_{RP})_0$ colour distributions of RC stars in 3--3$\rlap{.}^{\circ}$5 sub-regions with the best fit marked in red.}
    \label{rad7}
\end{figure}
\begin{table*}
\centering
\caption{Gaussian fit parameters for the colour distributions of RC stars.}
\label{tab:color}
\resizebox{\textwidth}{!}{%
\begin{tabular}{|c|c|c|c|c|c|c|c|c|c|c|c|c|c|c|c|c|}
\hline
\multirow{2}{*}{\begin{tabular}[c]{@{}c@{}}Radius\\ (deg)\end{tabular}} & \multicolumn{2}{c|}{NORTHEAST} & \multicolumn{2}{c|}{NORTHWEST} & \multicolumn{2}{c|}{SOUTHEAST} & \multicolumn{2}{c|}{SOUTHWEST} \\ \cline{2-9} 
 & {PEAK $\pm$ Error} & {SIGMA $\pm$ Error} & {PEAK $\pm$ Error} & {SIGMA $\pm$ Error} & {PEAK $\pm$ Error} & {SIGMA $\pm$ Error} & {PEAK $\pm$ Error} & {SIGMA $\pm$ Error} \\ \hline
0--0.5 & 0.665 $\pm$ 0.005 & 0.146 $\pm$ 0.003 & 0.665 $\pm$ 0.005 & 0.136 $\pm$ 0.003 & 0.649 $\pm$ 0.005 & 0.145 $\pm$ 0.003 & 0.626 $\pm$ 0.006 & 0.163 $\pm$ 0.004 \\ 
0.5--1 & 0.764 $\pm$ 0.004 & 0.137 $\pm$ 0.002 & 0.849 $\pm$ 0.003 & 0.125 $\pm$ 0.002 & 0.774 $\pm$ 0.004 & 0.131 $\pm$ 0.002 & 0.768 $\pm$ 0.004 & 0.140 $\pm$ 0.002 \\ 
1--1.5 & 0.725 $\pm$ 0.004 & 0.129 $\pm$ 0.002 & 0.804 $\pm$ 0.004 & 0.112 $\pm$ 0.002 & 0.751 $\pm$ 0.004 & 0.121 $\pm$ 0.002 & 0.736 $\pm$ 0.004 & 0.119 $\pm$ 0.002 \\ 
1.5--2 & 0.820 $\pm$ 0.005 & 0.117 $\pm$ 0.002 & 0.857 $\pm$ 0.003 & 0.124 $\pm$ 0.002 & 0.845 $\pm$ 0.004 & 0.126 $\pm$ 0.002 & 0.826 $\pm$ 0.004 & 0.124 $\pm$ 0.002 \\ 
2--2.5 & 0.912 $\pm$ 0.005 & 0.120 $\pm$ 0.002 & 0.914 $\pm$ 0.007 & 0.134 $\pm$ 0.004 & 0.934 $\pm$ 0.005 & 0.121 $\pm$ 0.002 & 0.917 $\pm$ 0.005 & 0.124 $\pm$ 0.002 \\ 
2.5--3 & 0.870 $\pm$ 0.004 & 0.113 $\pm$ 0.002 & 0.874 $\pm$ 0.005 & 0.124 $\pm$ 0.003 & 0.886 $\pm$ 0.005 & 0.116 $\pm$ 0.002 & 0.885 $\pm$ 0.004 & 0.117 $\pm$ 0.002 \\ 
3--3.5 & 0.837 $\pm$ 0.005 & 0.111 $\pm$ 0.002 & 0.849 $\pm$ 0.006 & 0.131 $\pm$ 0.003 & 0.875 $\pm$ 0.006 & 0.115 $\pm$ 0.003 & 0.873 $\pm$ 0.004 & 0.124 $\pm$ 0.002 \\ 
3.5--4 & 0.802 $\pm$ 0.005 & 0.107 $\pm$ 0.002 & 0.804 $\pm$ 0.005 & 0.129 $\pm$ 0.003 & 0.843 $\pm$ 0.004 & 0.110 $\pm$ 0.002 & 0.832 $\pm$ 0.004 & 0.130 $\pm$ 0.002 \\ 
4--4.5 & 0.793 $\pm$ 0.005 & 0.107 $\pm$ 0.002 & 0.793 $\pm$ 0.005 & 0.136 $\pm$ 0.002 & 0.837 $\pm$ 0.005 & 0.112 $\pm$ 0.002 & 0.834 $\pm$ 0.005 & 0.130 $\pm$ 0.003 \\ 
4.5--5 & 0.799 $\pm$ 0.005 & 0.114 $\pm$ 0.002 & 0.784 $\pm$ 0.006 & 0.142 $\pm$ 0.003 & 0.833 $\pm$ 0.005 & 0.120 $\pm$ 0.003 & 0.804 $\pm$ 0.009 & 0.162 $\pm$ 0.006 \\ 
5--6 & 0.789 $\pm$ 0.007 & 0.124 $\pm$ 0.004 & 0.770 $\pm$ 0.008 & 0.152 $\pm$ 0.005 & 0.826 $\pm$ 0.006 & 0.125 $\pm$ 0.003 & 0.732 $\pm$ 0.014 & 0.220 $\pm$ 0.013 \\ 
6--7 & 0.719 $\pm$ 0.013 & 0.198 $\pm$ 0.011 & 0.690 $\pm$ 0.013 & 0.214 $\pm$ 0.012 & 0.777 $\pm$ 0.013 & 0.186 $\pm$ 0.010 & 0.685 $\pm$ 0.010 & 0.247 $\pm$ 0.011 \\ 
7--10 & 0.716 $\pm$ 0.007 & 0.165 $\pm$ 0.006 & 0.694 $\pm$ 0.010 & 0.177 $\pm$ 0.009 & 0.793 $\pm$ 0.013 & 0.214 $\pm$ 0.017 & 0.736 $\pm$ 0.017 & 0.268 $\pm$ 0.037\\ \hline
\end{tabular}%
}
\end{table*}
\section{Bimodality in RC magnitude distribution and distance effect} 
\label{section4}
As described in Section \ref{section3.3}, the magnitude distributions of the RC stars in the eastern sub-regions, 2$\rlap{.}^{\circ}$5--5$^\circ$ from the centre show clear bimodality. Since RC stars are standard candles \citep{girardi2016} the natural explanation for this bimodality could be a distance effect. The average magnitude difference between the two peaks (faint and bright RC) is 0.45$\pm$0.09 mag. This translates to a difference in distance of 12$\pm$2 kpc, if we assume the faint clump is at the distance of the main body of the SMC. Apart from the distance effect, other possible effects that can contribute to the observed bimodality in the RC magnitude distribution are  extinction effects and RC population effects. \cite{Subramanian2017} discussed and analysed all these effects in detail while analysing the dual RC in the eastern SMC (between 2$\rlap{.}^{\circ}$5--4$^\circ$ from the centre). They found that these effects cannot explain the observed bimodality and suggested that the main cause for the dual RC feature is a distance effect and obtained a value similar to our results. In their study they also modelled the observed CMD as a linear combination of stellar partial models, assuming a single and a double distance separately. The observed CMD was well fit using the models including a double distance, hence supporting a distance effect for the observed bimodality in the RC magnitude distribution. \\
\indent \cite{Subramanian2017} used the NIR data from the VMC survey for their study. Extinction has a minimal effect in NIR bands, but in our present study we use the optical data from \textit{Gaia} DR2. Hence we address the effect of extinction in the observed bimodality of the RC magnitude distribution. A magnitude difference of 0.45 mag in G can be due to a dust layer between two populations. But an extinction of 0.45 mag in m$_G$ corresponds to a colour difference of $\sim$ 0.21 mag in $(G_{BP} - G_{RP})$ colour. The Hess diagrams shown in Fig. \ref{fig:CMD2.5-5} do not show such a colour difference between the brighter and fainter RC. In order to quantify this, we analysed the colour distributions (with a bin size of 0.05 mag in colour) of the stars in the RC selection boxes; they do not show any signatures of bimodality and are well fit by a single Gaussian function. The observed colour distributions and the best fit Gaussian profiles (in red) for the 3--3.5$^\circ$ sub-regions (where magnitude distributions show bimodality in eastern sub-regions) are shown in Fig. \ref{rad7}. The best fit Gaussian parameters for all the sub-regions are given in Table \ref{tab:color}. From the table we can see that in the eastern sub-regions (2$\rlap{.}^{\circ}$5--5$^\circ$) the dispersion value, which is a measure of internal extinction, is also less than 0.21 mag. Also note that the width of the RC colour distribution has contributions from photometric errors and population effects along with internal extinction. So a value of $\sim$ 0.1 mag for the dispersion in colour is an upper limit for the internal extinction. This analysis shows that the effect of extinction cannot explain the observed bimodality in the RC magnitude distribution.\\
\indent Based on the study by \cite{Subramanian2017} and our analysis of the effect extinction has on the observed bimodality in the RC magnitude distribution, we suggest that the magnitude difference between the faint and bright RC is most likely due to the presence of stellar populations at two different distances.  
\begin{figure*}
    \centering
    \subfloat{\includegraphics[scale=0.55]{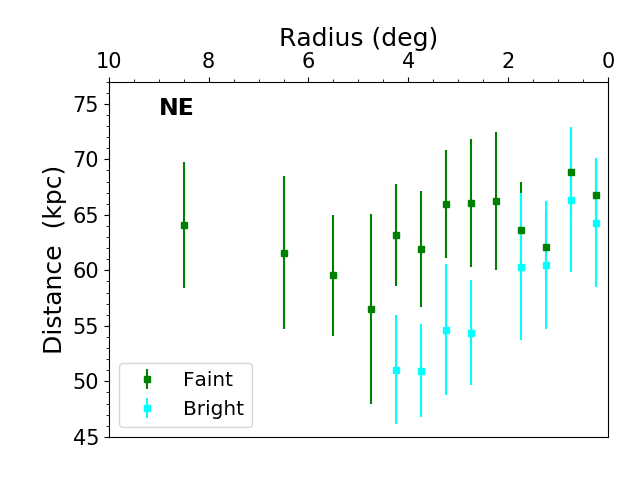}}
    \subfloat{\includegraphics[scale=0.55]{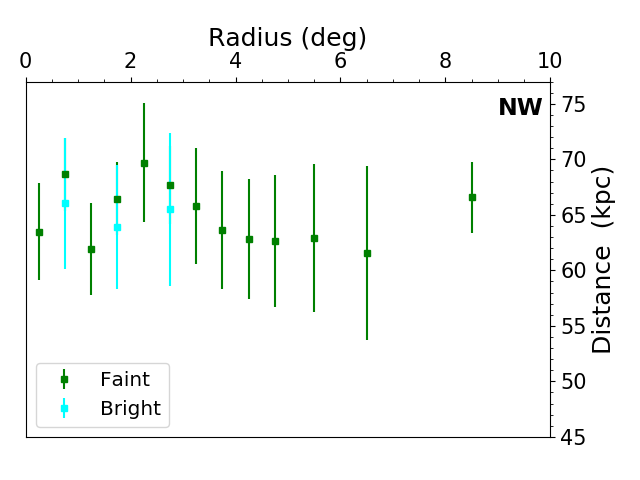}}\\
    \vspace{-0.8cm}
    \subfloat{\includegraphics[scale=0.55]{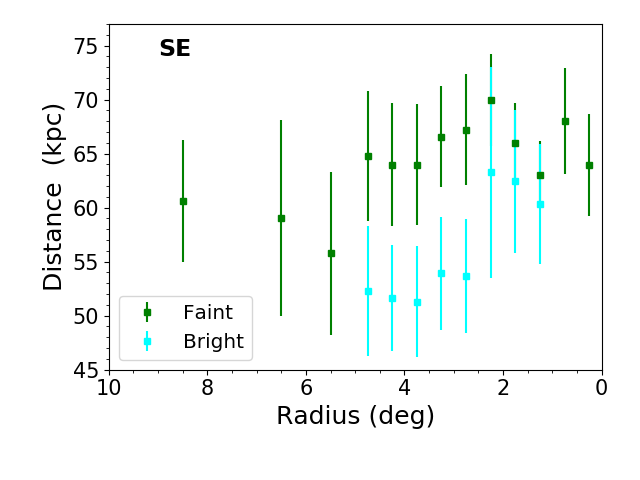}}
    \subfloat{\includegraphics[scale=0.55]{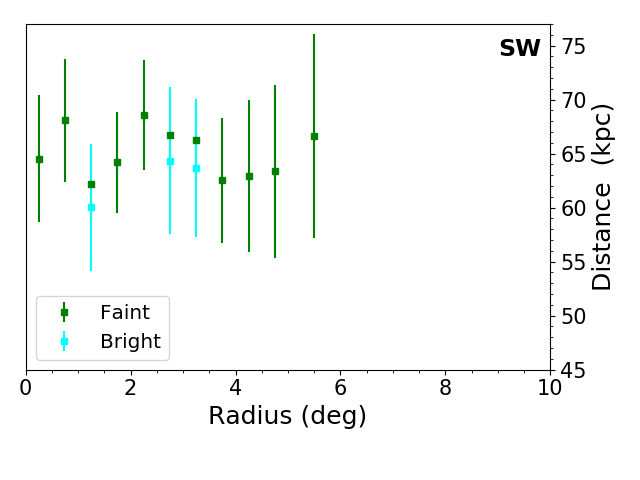}}
    \vspace{-0.6cm}
    \caption{Distance vs. radius in NE (top-left), NW (top-right), SE (bottom-left) and SW (bottom-right) regions. Cyan and green points correspond to distances of foreground (bright) and main body (faint) RC populations, respectively. The error bar is the distance corresponding to the dispersion in the magnitude distribution.}
    \label{fig:distew}
\end{figure*}
\section{3D structure}
\label{section5}
The extinction corrected peak magnitudes, ${G_0}$ of the faint and bright RC are converted to distances using the absolute magnitude in \textit{G} band ($M_{G}$) of RC stars \citep{ruiz2018}: 
\begin{equation}
    \centering
     M_{G} = 0.495 + 1.121*(G - K_{s} - 2.1 )
\end{equation}
\begin{table*}
\caption{Distances of NE, NW, SE and SW sub-regions.}
\label{tab:rad_nenw}
\small{
\begin{tabular}{|c|c|c|c|c|c|c|c|c|}
\hline
\multirow{2}{*}{\begin{tabular}[c]{@{}c@{}}Radius\\ (deg)\end{tabular}} & \multicolumn{2}{c|}{NE} & \multicolumn{2}{c|}{NE bright} & \multicolumn{2}{c|}{NW} & \multicolumn{2}{c|}{NW bright} \\ \cline{2-9} 
 & Distance (kpc) & D$_{\sigma}$ (kpc) & Distance (kpc) & D$_{\sigma}$ (kpc) & Distance (kpc) & D$_{\sigma}$ (kpc) & Distance (kpc) & D$_{\sigma}$ (kpc) \\ \hline
0--0.5 & 66.82 $\pm$ 0.43 & 3.14 & 64.29 $\pm$ 0.50 & 5.79 & 63.47 $\pm$ 0.11 & 4.38 &--&--\\ 
0.5--1 & 68.82 $\pm$ 0.32 & 4.02 & 66.36 $\pm$ 0.70 & 6.55 & 68.66 $\pm$ 0.22 & 3.29 & 66.03 $\pm$ 0.43 & 5.91 \\ 
1--1.5 & 62.10 $\pm$ 0.20 & 3.03 & 60.44 $\pm$ 0.17 & 5.77 & 61.91 $\pm$ 0.14 & 4.18 &--&--\\ 
1.5--2 & 63.61 $\pm$ 0.41 & 4.33 & 60.33 $\pm$ 1.28 & 6.64 & 66.45 $\pm$ 0.37 & 3.30 & 63.90 $\pm$ 0.77 & 5.61 \\ 
2--2.5 & 66.20 $\pm$ 0.28 & 6.21 &--&--& 69.68 $\pm$ 0.14 & 5.35 &--&--\\ 
2.5--3 & 66.06 $\pm$ 0.49 & 5.77 & 54.37 $\pm$ 0.73 & 4.72 & 67.69 $\pm$ 0.16 & 3.55 & 65.51 $\pm$ 0.30 & 6.88 \\ 
3--3.5 & 65.97 $\pm$ 1.09 & 4.85 & 54.64 $\pm$ 1.66 & 5.89 & 65.80 $\pm$ 0.25 & 5.21 &--&--\\ 
3.5--4 & 61.93 $\pm$ 0.63 & 5.21 & 50.97 $\pm$ 0.63 & 4.22 & 63.65 $\pm$ 0.25 & 5.34 &--&--\\ 
4--4.5 & 63.17 $\pm$ 0.52 & 4.62 & 51.04 $\pm$ 0.45 & 4.90 & 62.85 $\pm$ 0.17 & 5.40 &--&--\\ 
4.5--5 & 56.49 $\pm$ 0.46 & 8.58 &--&--& 62.61 $\pm$ 0.13 & 5.95 &--&--\\ 
5--6 & 59.54 $\pm$ 0.41 & 5.44 &--&--& 62.91 $\pm$ 0.40 & 6.63 &--&--\\ 
6--7 & 61.57 $\pm$ 0.55 & 6.90 &--&--& 61.55 $\pm$ 0.58 & 7.83 &--&--\\ 
7--10 & 64.08 $\pm$ 1.39 & 5.68 &--&--& 66.59 $\pm$ 0.80 & 3.21 &--&--\\ \hline
\multirow{2}{*}{\begin{tabular}[c]{@{}c@{}}Radius\\ (deg)\end{tabular}} & \multicolumn{2}{c|}{SE} & \multicolumn{2}{c|}{SE bright} & \multicolumn{2}{c|}{SW} & \multicolumn{2}{c|}{SW bright} \\ \cline{2-9} 
 & Distance (kpc) & D$_{\sigma}$ (kpc) & Distance (kpc) & D$_{\sigma}$ (kpc) & Distance (kpc) & D$_{\sigma}$ (kpc) & Distance (kpc) & D$_{\sigma}$ (kpc) \\ \hline
0--0.5 & 63.96 $\pm$ 0.11 & 4.71 &--&--& 64.53 $\pm$ 0.17 & 5.90 &--&--\\ 
0.5--1 & 67.99 $\pm$ 0.17 & 4.90 &--&--& 68.07 $\pm$ 0.10 & 5.70 &--&--\\ 
1--1.5 & 63.06 $\pm$ 0.15 & 3.10 & 60.33 $\pm$ 0.39 & 5.60 & 62.16 $\pm$ 0.17 & 3.52 & 60.02 $\pm$ 0.41 & 5.90 \\ 
1.5--2 & 65.94 $\pm$ 0.18 & 3.73 & 62.42 $\pm$ 0.69 & 6.59 & 64.21 $\pm$ 0.10 & 4.67 &--&--\\ 
2--2.5 & 69.94 $\pm$ 0.23 & 4.25 & 63.26 $\pm$ 1.19 & 9.74 & 68.57 $\pm$ 0.14 & 5.08 &--&--\\ 
2.5--3 & 67.22 $\pm$ 0.28 & 5.13 & 53.69 $\pm$ 0.69 & 5.27 & 66.70 $\pm$ 0.15 & 3.62 & 64.35 $\pm$ 0.44 & 6.79 \\ 
3--3.5 & 66.58 $\pm$ 0.58 & 4.68 & 53.92 $\pm$ 0.94 & 5.22 & 66.27 $\pm$ 0.27 & 3.78 & 63.67 $\pm$ 0.82 & 6.37 \\ 
3.5--4 & 63.96 $\pm$ 1.12 & 5.61 & 51.28 $\pm$ 1.35 & 5.15 & 62.52 $\pm$ 0.19 & 5.79 &--&--\\ 
4--4.5 & 63.96 $\pm$ 0.80 & 5.70 & 51.63 $\pm$ 0.71 & 4.88 & 62.93 $\pm$ 0.45 & 7.00 &--&--\\ 
4.5--5 & 64.79 $\pm$ 2.89 & 6.01 & 52.25 $\pm$ 2.81 & 6.04 & 63.37 $\pm$ 0.82 & 8.01 &--&--\\ 
5--6 & 55.77 $\pm$ 1.27 & 7.54 &--&--& 66.63 $\pm$ 1.76 & 9.46 &--&--\\ 
6--7 & 59.00 $\pm$ 0.39 & 9.06 &--&--&--&--&--&--\\ 
7--10 & 60.58 $\pm$ 0.58 & 5.63 &--&--&--&--&--&--\\ \hline
\end{tabular}%
}
\end{table*}

\indent We used the peak RC magnitudes (${K_{s_0}}$) provided by \cite{Subramanian2017} to calculate the $(G-K_s)$ colour and hence $M_G$. From the values listed in Table 1 of \cite{Subramanian2017}, we calculated the average $K_{s_0}$ magnitudes for the inner 0--2$^\circ$ region and 2$\rlap{.}^{\circ}$5--4$^\circ$ region from the centre. In the eastern 2$\rlap{.}^{\circ}$5--4$^\circ$ region there are faint and bright RC features and hence we calculated the average $K_{s_0}$ magnitude for the bright and faint RC. We note here that the $K_{s_0}$ magnitudes provided by \cite{Subramanian2017} are in the VISTA system and the $K_{s}$ band used by \cite{ruiz2018} is in the 2MASS system. Using the $(Y-K_s)_0$ colour of RC stars in the VISTA system (given in Fig.7 of \citealt{Subramanian2017}) and applying the  transformation equations provided by \cite{Gonzalez2018}, we converted the $K_{s_0}$ magnitudes in the VISTA system to the 2MASS system. The difference is found to be  negligible, $\sim$ 0.003 mag. The $K_{s_0}$ values obtained in the 2MASS system are 17.33 $\pm$ 0.04 mag (for the 0--2$^\circ$ region), 17.38 $\pm$ 0.05 mag (for faint RC in the 2$\rlap{.}^{\circ}$5--4$^\circ$ region) and 16.92 $\pm$ 0.05 mag (for the bright RC in the 2$\rlap{.}^{\circ}$5--4$^\circ$ region). We also calculated the average ${G_0}$ values for these regions from our estimates given in Table \ref{tab:mag_ne}. The $(G - K_{s})$ colour obtained for the RC stars is $\sim$ 1.5 mag and the absolute magnitude obtained using the colour value of 1.5 mag is $\sim$ $-0.18$ mag. Based on this value, the peak ${G_0}$ values (given in Table \ref{tab:mag_ne}) corresponding to different sub-regions are converted to distance moduli and then to distances in kpc. Table \ref{tab:rad_nenw} gives the distance and the associated errors in the northern and southern sub-regions respectively. The error in the distance value is the distance corresponding to the error in the peak RC magnitude. The distance corresponding to the dispersion of the magnitude distribution is also shown in Table \ref{tab:rad_nenw} as D$_\sigma$.\\ \indent Different panels of Fig. \ref{fig:distew} show the estimated distances in the NE, SE, NW and SW sub-regions, as a function of radius. The plots clearly show that the eastern sub-regions at 2$\rlap{.}^{\circ}$5--5$^\circ$ radius of the SMC have two populations of RC at different distances along the line of sight. The fainter RC is at the distance of the main body of the SMC and the brighter RC is in the foreground of the SMC. The relative distance between these populations is $\sim$ 12 kpc. The mean distance, corresponding to a single RC, of the SE 5--6$^\circ$ region is similar to the foreground (bright) RC population. This suggests that the foreground RC population  probably extends till SE 5--6$^\circ$, in agreement with a recent study (El Youssoufi et al. submitted) based on NIR data from the VISTA Hemisphere Survey (VHS).\\
\indent The error bars shown in Fig. \ref{fig:distew} are the distances corresponding to the dispersion of the respective Gaussian components of the best fit to the RC magnitude distribution. Hence it represents the extent of the structure (corresponding to each Gaussian component) along the line of sight. As the dispersion of the magnitude distribution includes contributions from internal extinction and photometric errors, the error bars in Fig. \ref{fig:distew} provide an upper limit to the depth along the line of sight. As shown in other studies (\citealt{subramanian2012} and Tatton et al. 2020, submitted), the main body of the SMC (as traced by the fainter RC clump) is extended along the line of sight with a depth of $\sim$ 5--8 kpc (corresponding to $\sim\pm$1$\sigma$ of the Gaussian component corresponding to the faint RC). However, the foreground RC population (corresponding to the bright RC) identified in the regions between 2$\rlap{.}^{\circ}$5--4$\rlap{.}^{\circ}$5 in the NE and 2$\rlap{.}^{\circ}$5--5$^\circ$ in the SE, is distinct and well separated from the main body of the SMC. As we can appreciate from Fig. \ref{fig:distew}, the errors bars on the mean distances to the foreground and main body RC populations in these eastern sub-regions are not overlapping. Thus the foreground RC stars in the eastern sub-regions between 2$\rlap{.}^{\circ}$5--5$^\circ$ from the SMC centre are more likely part of a separate stellar sub-structure in front of the SMC, which could be connected to the main body only at low density levels.\\
\indent As discussed in Section \ref{section3.1}, stars in the outer regions are corrected for interstellar extinction using the extinction values of regions at 3$^\circ$ from the SMC centre. This may lead to an over-subtraction of extinction in outer regions and can affect the distance estimation. Though it can affect the absolute distance estimation, relative distance between the 
 foreground stellar structure and the main body population in the same region of the sky will not get affected. This is because we apply a constant extinction value to all the stars in a sub-region. However, if there is significant dust in between the two RC populations, the background population will have more extinction than the foreground population. This effect is found to be negligible based on our analysis on the colour of the two RC populations in \ref{section4}. We also note that the variation of RC population effects across the SMC could also affect the estimation of the $(G - K_{s})$ colour and hence the $M_G$ value and distance estimation. However, our final results based on the relative distance between the main body (faint) and foreground (bright) RC will not be affected significantly by these effects.\\
\begin{table*}
\centering
\caption{Proper motion values for NE, NW, SE and SW sub-regions.}
\label{tab:pm ne nw}
\resizebox{\textwidth}{!}{%
\begin{tabular}{|c|c|c|c|c|c|c|c|c|c|c|c|c|}
\hline
\multirow{2}{*}{\begin{tabular}[c]{@{}c@{}}Radius \\  (deg)\end{tabular}} & \multicolumn{2}{c|}{NORTH EAST} & \multicolumn{2}{c|}{NORTH EAST (BRIGHT)} & \multicolumn{2}{c|}{NORTH WEST} \\ \cline{2-7} 
 & {$\mu_\alpha$ (mas yr$^{-1}$)} &{$\mu_\delta$ (mas yr$^{-1}$)} & {$\mu_\alpha$ (mas yr$^{-1}$)} & {$\mu_\delta$ (mas yr$^{-1}$)} & {$\mu_\alpha$ (mas yr$^{-1}$)} & {$\mu_\delta$ (mas yr$^{-1}$)} \\ \hline
0--0.5 & 0.672 $\pm$ 0.013 & $-$1.080 $\pm$ 0.011 &-- &-- & 0.779 $\pm$ 0.012 &$-$1.051 $\pm$ 0.010 \\ 
0.5--1 & 0.704 $\pm$ 0.008 &$-$1.081 $\pm$ 0.006 &-- &-- & 0.692 $\pm$ 0.008 &$-$1.130 $\pm$ 0.006 \\ 
1--1.5 & 0.789 $\pm$ 0.007 &$-$1.102 $\pm$ 0.005 &-- &-- & 0.607 $\pm$ 0.008 &$-$1.187 $\pm$ 0.006 \\ 
1.5--2 & 0.786 $\pm$ 0.007 &$-$1.144 $\pm$ 0.005 &-- &-- & 0.589 $\pm$ 0.010 &$-$1.191 $\pm$ 0.008 \\ 
2--2.5 & 0.815 $\pm$ 0.009 &$-$1.134 $\pm$ 0.007 &-- &-- & 0.574 $\pm$ 0.012 &$-$1.192 $\pm$ 0.009 \\ 
2.5--3 & 0.843 $\pm$ 0.012 &$-$1.150 $\pm$ 0.009 & 0.991 $\pm$ 0.012 &$-$1.225 $\pm$ 0.008 & 0.538 $\pm$ 0.014 &$-$1.172 $\pm$ 0.011 \\ 
3--3.5 & 0.863 $\pm$ 0.016 &$-$1.110 $\pm$ 0.012 & 1.009 $\pm$ 0.011 &$-$1.195 $\pm$ 0.008 & 0.558 $\pm$ 0.017 &$-$1.165 $\pm$ 0.013 \\ 
3.5--4 & 0.931 $\pm$ 0.016 &$-$1.108 $\pm$ 0.012 & 1.122 $\pm$ 0.012 &$-$1.213 $\pm$ 0.009 & 0.580 $\pm$ 0.018 &$-$1.188 $\pm$ 0.014 \\ 
4--4.5 & 0.899 $\pm$ 0.021 &$-$1.099 $\pm$ 0.014 & 1.138 $\pm$ 0.013 &$-$1.185 $\pm$ 0.010 & 0.543 $\pm$ 0.020 &$-$1.180 $\pm$ 0.016 \\ 
4.5--5 & 1.039 $\pm$ 0.014 &$-$1.146 $\pm$ 0.011 &--&--& 0.607 $\pm$ 0.019 &$-$1.185 $\pm$ 0.018 \\ 
5--6 & 0.943 $\pm$ 0.019 &$-$1.109 $\pm$ 0.014 &-- &-- & 0.620 $\pm$ 0.022 &$-$1.181 $\pm$ 0.018 \\ 
6--7 & 1.017 $\pm$ 0.031 &$-$1.054 $\pm$ 0.030 &-- &-- & 0.776 $\pm$ 0.037 &$-$1.240 $\pm$ 0.031 \\ 
7--10 & 1.103 $\pm$ 0.036 &$-$0.985 $\pm$ 0.033 &-- &-- & 0.825 $\pm$ 0.045 &$-$1.287 $\pm$ 0.041 \\ \hline 		
\multirow{2}{*}{\begin{tabular}[c]{@{}c@{}}Radius \\  (deg)\end{tabular}} & \multicolumn{2}{c|}{SOUTH EAST} & \multicolumn{2}{c|}{SOUTH EAST (BRIGHT)} & \multicolumn{2}{c|}{SOUTH WEST} \\ \cline{2-7} 
 & {$\mu_\alpha$ (mas yr$^{-1}$)} & {$\mu_\delta$ (mas yr$^{-1}$)} & {$\mu_\alpha$ (mas yr$^{-1}$)} & {$\mu_\delta$ (mas yr$^{-1}$)} & {$\mu_\alpha$ (mas yr$^{-1}$)} & {$\mu_\delta$ (mas yr$^{-1}$)} \\ \hline
0--0.5 & 0.647 $\pm$ 0.015 &$-$1.039 $\pm$ 0.011 &-- &-- & 0.552 $\pm$ 0.016 &$-$1.043 $\pm$ 0.012 \\ 
0.5--1 & 0.710 $\pm$ 0.008 &$-$1.063 $\pm$ 0.006 &-- &-- & 0.644 $\pm$ 0.008 &$-$1.067 $\pm$ 0.006 \\ 
1--1.5 & 0.674 $\pm$ 0.008 &$-$1.120 $\pm$ 0.005 &-- &-- & 0.627 $\pm$ 0.006 &$-$1.133 $\pm$ 0.005 \\ 			
1.5--2 & 0.791 $\pm$ 0.008 &$-$1.112 $\pm$ 0.006 &-- &-- & 0.534 $\pm$ 0.007 &$-$1.148 $\pm$ 0.005 \\ 
2--2.5 & 0.775 $\pm$ 0.012 &$-$1.132 $\pm$ 0.008 &-- &-- & 0.502 $\pm$ 0.008 &$-$1.175 $\pm$ 0.006 \\ 
2.5--3 & 0.832 $\pm$ 0.015 &$-$1.117 $\pm$ 0.010 & 1.042 $\pm$ 0.011 &$-$1.265 $\pm$ 0.008 & 0.498 $\pm$ 0.011 &$-$1.171 $\pm$ 0.007 \\ 
3--3.5 & 0.845 $\pm$ 0.018 &$-$1.128 $\pm$ 0.012 & 1.067 $\pm$ 0.012 &$-$1.258 $\pm$ 0.009 & 0.488 $\pm$ 0.015 &$-$1.208 $\pm$ 0.010 \\ 
3.5--4 & 0.968 $\pm$ 0.020 &$-$1.102 $\pm$ 0.014 & 1.157 $\pm$ 0.013 &$-$1.252 $\pm$ 0.010 & 0.495 $\pm$ 0.020 &$-$1.222 $\pm$ 0.014 \\ 
4--4.5 & 1.029 $\pm$ 0.022 &$-$1.080 $\pm$ 0.015 & 1.216 $\pm$ 0.015 &$-$1.218 $\pm$ 0.011 & 0.517 $\pm$ 0.029 &$-$1.239 $\pm$ 0.020 \\   
4.5--5 & 1.053 $\pm$ 0.027 &$-$1.020 $\pm$ 0.019 & 1.228 $\pm$ 0.017 &$-$1.181 $\pm$ 0.014 & 0.554 $\pm$ 0.037 &$-$1.252 $\pm$ 0.026 \\ 
5--6 & 1.190 $\pm$ 0.017 &$-$1.091 $\pm$ 0.014 &-- &-- & 0.705 $\pm$ 0.044 &$-$1.304 $\pm$ 0.033 \\ 
6--7 & 1.215 $\pm$ 0.024 &$-$1.003 $\pm$ 0.023 &-- &-- &--&--\\ 
7--10 & 1.431 $\pm$ 0.025 &$-$0.682 $\pm$ 0.029 &-- &-- &--&--\\ \hline
\end{tabular}%
}
\end{table*}

 \section{Proper Motion of the foreground and main body RC populations}
\label{section6}
In this study, we find that there are two populations of RC stars which are located at a relative distance of $\sim$ 12 kpc along the line of sight in the eastern sub-regions (at 2$\rlap{.}^{\circ}$5--5$^\circ$ radius). Using \textit{Gaia} DR2 proper motion measurements ($\mu_\alpha$, $\mu_\delta$) we analyse in this section the kinematics of these two populations and also compare them with RC stars within other sub-regions.\\
\begin{figure}
     \centering
     \hspace*{-0.8em}
     \includegraphics[scale = 0.57]{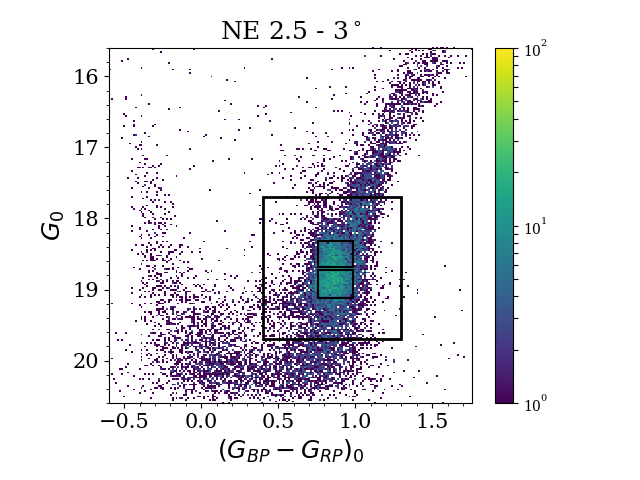}
     \caption{CMD showing the selection of RC stars. The outer black box shows our initial selection of RC stars and the inner ones represents the 1$\sigma$ ranges in selection with peak values of 0.870 $\pm$ 0.113 mag in colour and 18.922 $\pm$ 0.189 mag (faint); 18.499 $\pm$ 0.19 mag (bright) in magnitude. The colour bar from blue to yellow indicates the increase in stellar density.}
     \label{fig:clumpselec}
 \end{figure}
 \begin{figure*}
     \hspace{-0.5cm}
     \includegraphics[width=0.41\textwidth]{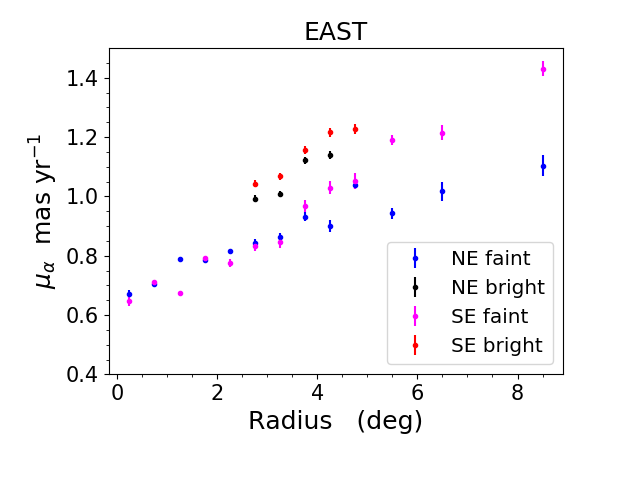}
     \hspace{-0.5cm}
     \includegraphics[width=0.41\textwidth]{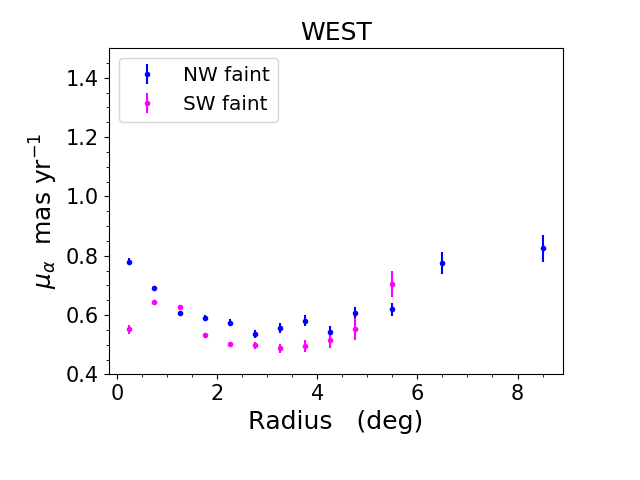}\\
     \vspace{-0.5cm}
     \hspace{-0.5cm}
     \includegraphics[width=0.41\textwidth]{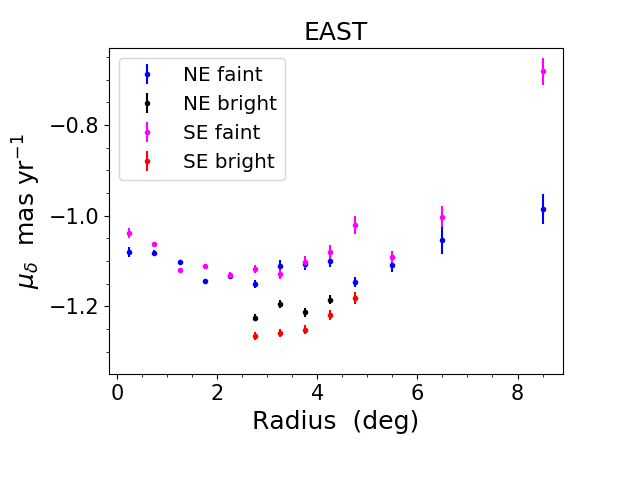}
     \hspace{-0.5cm}
     \includegraphics[width=0.41\textwidth]{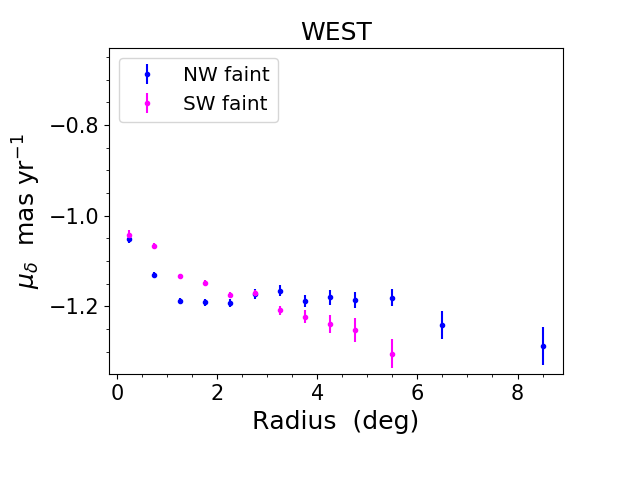}
      \vspace{-0.5cm}
     \caption{Proper motion components $\mu_\alpha$ (top) and $\mu_\delta$ (bottom) as a function of radius for the NE and SE sub-regions (left) and for the NW and SW sub-regions (right). The error bars represent the standard errors in the respective sub-regions. Blue (magenta) points in the figures represent the values corresponding to the main body (faint RC) of the SMC in NE (SE) and NW (SW) sub-regions. Black and red points show the foreground (bright) RC population in the NE and SE, respectively.}
     \label{fig:rad pm}
 \end{figure*}
 \begin{figure*}
    \hspace{-0.5cm}
    \vspace{-0.5cm}
    \includegraphics[width=0.41\textwidth]{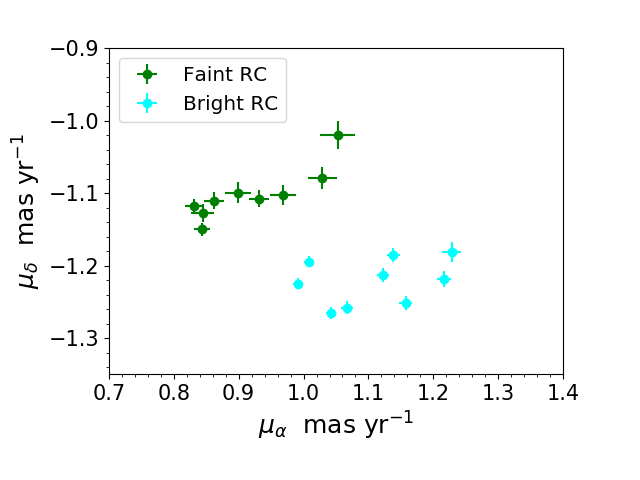}
    \hspace{-0.5cm}
    \includegraphics[width=0.41\textwidth]{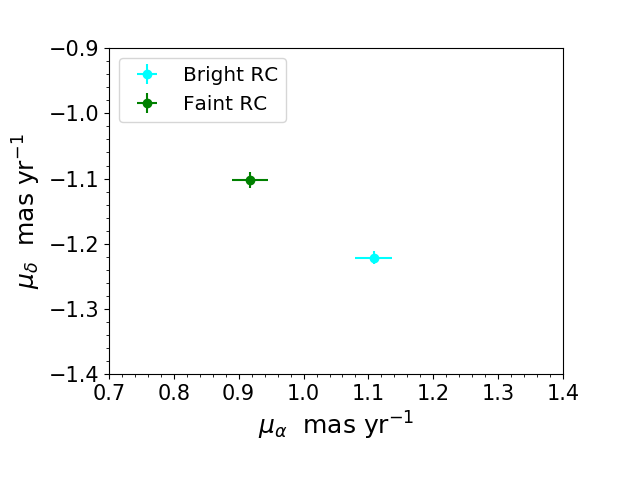}\\
    \hspace{-0.5cm}
    \includegraphics[width=0.41\textwidth]{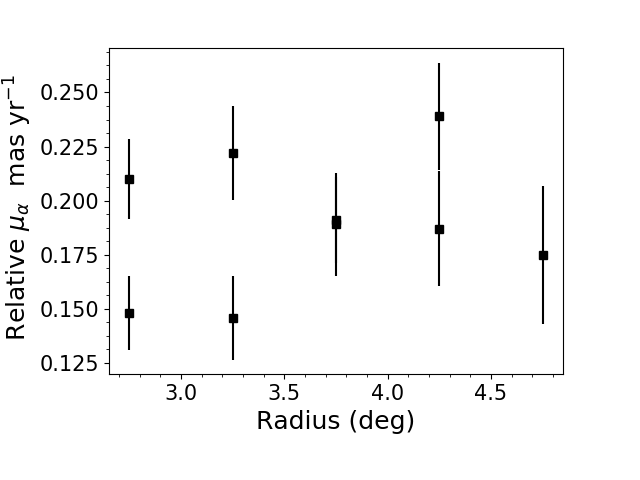}
    \hspace{-0.5cm}
    \includegraphics[width=0.41\textwidth]{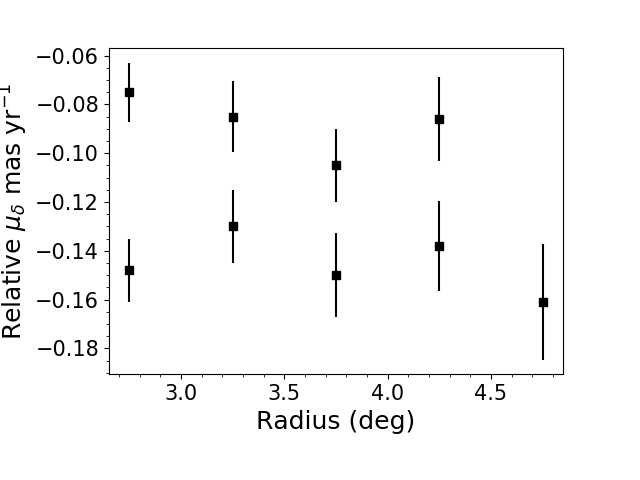}\\
    \caption{Mean proper motion for bright and faint RC stars (top-left) and their mean (top-right). Relative $\mu_{\alpha}$ (bottom-left) and $\mu_{\delta}$ (bottom-right) as a function of radius within the eastern sub-regions at 2$\rlap{.}^{\circ}$5--5$^\circ$ from the centre of the SMC.}
    \label{fig:pm mean}
 \end{figure*}
   \begin{figure*}
    \hspace{-0.5cm}
    \vspace{-0.5cm}
    \includegraphics[width=0.41\textwidth]{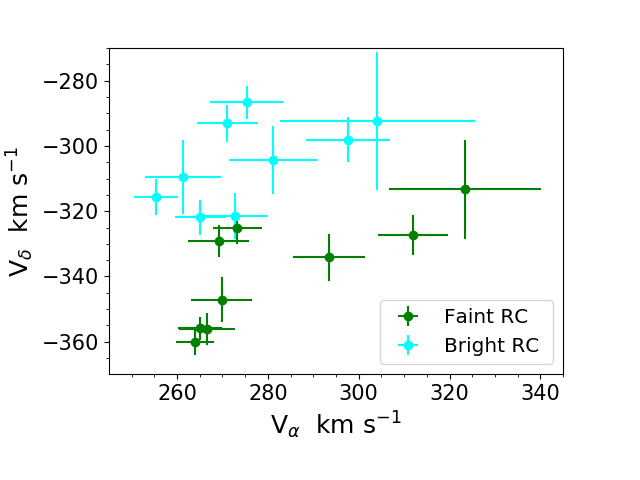}
    \hspace{-0.5cm}
    \includegraphics[width=0.41\textwidth]{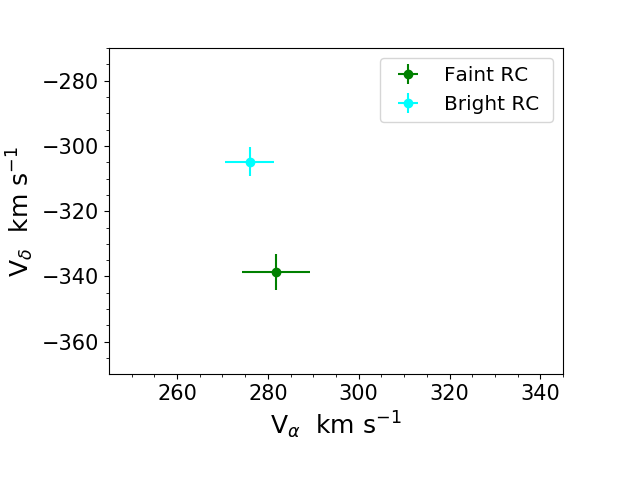}
    \hspace{-0.5cm}
    \includegraphics[width=0.41\textwidth]{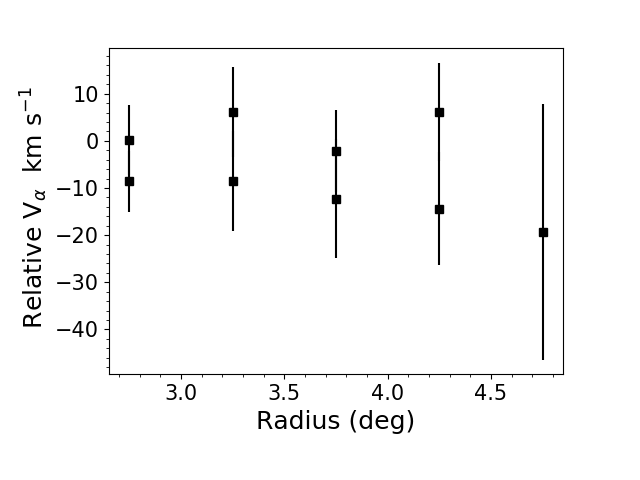}
    \hspace{-0.5cm}
    \includegraphics[width=0.41\textwidth]{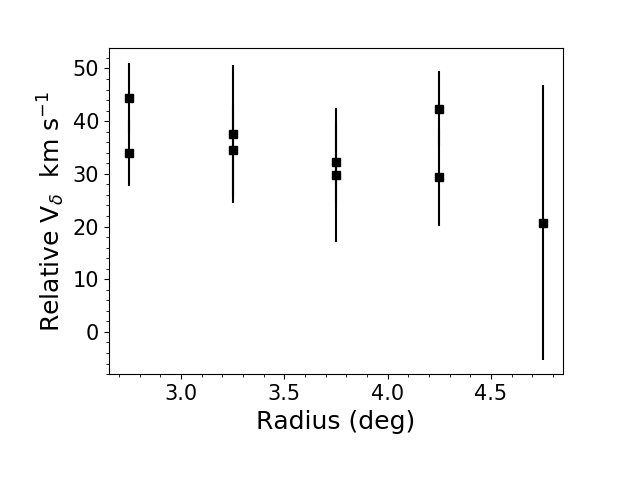}
    \vspace{-0.5cm}
    \caption{Mean tangential velocity components for bright and faint RC stars (top-left) and their mean (top-right). Relative V$_{\alpha}$ (bottom-left) and V$_{\delta}$ (bottom-right) as a function of radius within the eastern sub-regions at 2$\rlap{.}^{\circ}$5--5$^\circ$ from the centre of the SMC.} 
    \label{fig:velocity}
 \end{figure*}
 
\begin{table*}
\caption{Velocity components and relative velocity of bright RC stars with respect to the faint ones in the NE and SE sub-regions.}
\label{tab:vel_ne_rel}
\resizebox{\textwidth}{!}{%
\begin{tabular}{|c|c|c|c|c|c|c|c|c|c|}
\hline
\multirow{2}{*}{\begin{tabular}[c]{@{}c@{}}RADIUS  \\  (deg)\end{tabular}} & \multicolumn{3}{c|}{NORTH EAST Faint RC} & \multicolumn{3}{c|}{NORTH EAST   Bright RC} & \multicolumn{3}{c|}{Relative velocity} \\ \cline{2-10} 
 & \begin{tabular}[c]{@{}c@{}}V$_{\alpha}$ \\ km s$^{-1}$\end{tabular} & \begin{tabular}[c]{@{}c@{}}V$_{\delta}$\\  km s$^{-1}$\end{tabular} & \begin{tabular}[c]{@{}c@{}}V$_{t}$ \\ km s$^{-1}$\end{tabular} & \begin{tabular}[c]{@{}c@{}}V$_{\alpha}$ \\ km s$^{-1}$\end{tabular} & \begin{tabular}[c]{@{}c@{}}V$_{\delta}$ \\ km s$^{-1}$\end{tabular} & \begin{tabular}[c]{@{}c@{}}V$_{t}$ \\ km s$^{-1}$\end{tabular} & \begin{tabular}[c]{@{}c@{}}V$_{\alpha}$ \\ km s$^{-1}$\end{tabular} & \begin{tabular}[c]{@{}c@{}}V$_{\delta}$ \\ km s$^{-1}$\end{tabular} & \begin{tabular}[c]{@{}c@{}}V$_{t}$ \\ km s$^{-1}$\end{tabular} \\ \hline
0--0.5 & 212.84 $\pm$ 4.34 & $-$342.06 $\pm$ 4.12 & 402.88 $\pm$ 5.98 &--& -- & -- & -- & -- & -- \\ 
0.5--1 & 229.65 $\pm$ 2.82 & $-$352.63 $\pm$ 2.55 & 420.82 $\pm$ 3.80 & -- & -- & -- & -- & -- & -- \\ 
1--1.5 & 232.25 $\pm$ 2.19 & $-$324.38 $\pm$ 1.80 & 398.95 $\pm$ 2.84 & -- & -- & -- & -- & -- & -- \\ 
1.5--2 & 236.99 $\pm$ 2.61 & $-$344.93 $\pm$ 2.69 & 418.50 $\pm$ 3.74 & -- & -- & -- & -- & -- & -- \\ 
2--2.5 & 255.74 $\pm$ 3.02 & $-$355.84 $\pm$ 2.66 & 438.20 $\pm$ 4.03 & -- & -- & -- & -- & -- & -- \\ 
2.5--3 & 263.96 $\pm$ 4.24 & $-$360.09 $\pm$ 3.88 & 446.48 $\pm$ 5.75 & 255.39 $\pm$ 4.94 & $-$315.70 $\pm$ 5.44 & 406.07 $\pm$ 7.34 & $-$8.57 $\pm$ 6.51 & 44.39 $\pm$ 6.68 & 45.21 $\pm$ 9.33 \\ 
3--3.5 & 269.86 $\pm$ 6.70 & $-$347.09 $\pm$ 6.85 & 439.66 $\pm$ 9.59 & 261.32 $\pm$ 8.29 & $-$309.5 $\pm$ 11.21 & 405.07 $\pm$ 13.95 & $-$8.54 $\pm$ 10.66 & 37.59 $\pm$ 13.14 & 38.55 $\pm$ 16.92 \\ 
3.5--4 & 273.29 $\pm$ 5.46 & $-$325.25 $\pm$ 4.83 & 424.83 $\pm$ 7.29 & 271.07 $\pm$ 6.75 & $-$293.06 $\pm$ 5.66 & 399.20 $\pm$ 8.81 & $-$2.22 $\pm$ 8.68 & 32.19 $\pm$ 7.44 & 32.27 $\pm$ 11.43 \\ 
4--4.5 & 269.18 $\pm$ 6.67 & $-$329.07 $\pm$ 4.99 & 425.14 $\pm$ 8.33 & 275.32 $\pm$ 8.00 & $-$286.69 $\pm$ 5.03 & 397.48 $\pm$ 9.45 & 6.14 $\pm$ 10.42 & 42.38 $\pm$ 7.09 & 42.82 $\pm$ 12.60 \\ 
4.5--5 & 278.21 $\pm$ 4.38 & $-$306.86 $\pm$ 3.86 & 414.20 $\pm$ 5.84 & -- & -- & -- & -- & -- & -- \\ 
5--6 & 266.13 $\pm$ 5.67 & $-$312.98 $\pm$ 4.50 & 410.83 $\pm$ 7.24 & -- & -- & -- & -- & -- & -- \\ 
6--7 & 296.80 $\pm$ 9.43 & $-$307.60 $\pm$ 9.18 & 427.45 $\pm$ 13.16 & -- & -- & -- & -- & -- & -- \\ 
7--10 & 335.02 $\pm$ 13.13 & $-$299.18 $\pm$ 11.94 & 449.17 $\pm$ 17.75 & -- & -- & -- & -- & -- & -- \\ \hline
\multirow{2}{*}{\begin{tabular}[c]{@{}c@{}}RADIUS \\   (deg)\end{tabular}} & \multicolumn{3}{c|}{SOUTH EAST Faint RC} & \multicolumn{3}{c|}{SOUTH EAST   Bright RC} & \multicolumn{3}{c|}{Relative velocity} \\ \cline{2-10} 
 & \begin{tabular}[c]{@{}c@{}}V$_{\alpha}$\\ km s$^{-1}$\end{tabular} & \begin{tabular}[c]{@{}c@{}}V$_{\delta}$\\  km s$^{-1}$\end{tabular} & \begin{tabular}[c]{@{}c@{}}V$_{t}$ \\ km s$^{-1}$\end{tabular} & \begin{tabular}[c]{@{}c@{}}V$_{\alpha}$\\  km s$^{-1}$\end{tabular} & \begin{tabular}[c]{@{}c@{}}V$_{\delta}$ \\ km s$^{-1}$\end{tabular} & \begin{tabular}[c]{@{}c@{}}V$_{t}$\\  km s$^{-1}$\end{tabular} & \begin{tabular}[c]{@{}c@{}}V$_{\alpha}$\\  km s$^{-1}$\end{tabular} & \begin{tabular}[c]{@{}c@{}}V$_{\delta}$\\  km s$^{-1}$\end{tabular} & \begin{tabular}[c]{@{}c@{}}V$_{t}$ \\ km s$^{-1}$\end{tabular} \\ \hline
0--0.5 & 196.15 $\pm$ 4.56 & $-$314.99 $\pm$ 3.38 & 371.07 $\pm$ 5.68 & -- & -- & -- & --& -- & --\\ 
0.5--1 & 228.81 $\pm$ 2.64 & $-$342.58 $\pm$ 2.11 & 411.96 $\pm$ 3.38 & -- & -- & -- & -- & -- & -- \\ 
1--1.5 & 201.46 $\pm$ 2.44 & $-$334.77 $\pm$ 1.69 & 390.72 $\pm$ 2.97 & -- & -- & -- & -- & -- & -- \\ 
1.5--2 & 247.23 $\pm$ 2.59 & $-$347.56 $\pm$ 2.10 & 426.52 $\pm$ 3.34 & -- & -- & -- & -- & -- & -- \\ 
2--2.5 & 256.92 $\pm$ 4.07 & $-$375.28 $\pm$ 2.93 & 454.80 $\pm$ 5.01 & -- & -- & -- & -- & -- & -- \\ 
2.5--3 & 265.09 $\pm$ 4.91 & $-$355.90 $\pm$ 3.51 & 443.78 $\pm$ 6.03 & 265.18 $\pm$ 5.56 & $-$321.93 $\pm$ 5.22 & 417.08 $\pm$ 7.63 & 0.09 $\pm$ 7.42 & 33.97 $\pm$ 6.29 & 33.97 $\pm$ 9.73 \\ 
3--3.5 & 266.67 $\pm$ 6.14 & $-$355.98 $\pm$ 4.89 & 444.79 $\pm$ 7.85 & 272.70 $\pm$ 7.31 & $-$321.52 $\pm$ 7.14 & 421.60 $\pm$ 10.22 & 6.03 $\pm$ 9.55 & 34.46 $\pm$ 8.65 & 34.98 $\pm$ 12.89 \\ 
3.5--4 & 293.47 $\pm$ 7.95 & $-$334.09 $\pm$ 7.23 & 444.68 $\pm$ 10.74 & 281.23 $\pm$ 9.75 & $-$304.32 $\pm$ 10.37 & 414.37 $\pm$ 14.23 & $-$12.24 $\pm$ 12.58 & 29.77 $\pm$ 12.64 & 32.19 $\pm$ 17.83 \\ 
4--4.5 & 311.96 $\pm$ 7.73 & $-$327.42 $\pm$ 6.12 & 452.25 $\pm$ 9.86 & 297.59 $\pm$ 9.23 & $-$298.08 $\pm$ 6.92 & 421.20 $\pm$ 11.54 & $-$14.37 $\pm$ 12.04 & 29.34 $\pm$ 9.24 & 32.67 $\pm$ 15.18 \\ 
4.5--5 & 323.38 $\pm$ 16.64 & $-$313.25 $\pm$ 15.14 & 450.22 $\pm$ 22.50 & 304.13 $\pm$ 21.47 & $-$292.49 $\pm$ 21.15 & 421.96 $\pm$ 30.14 & $-$19.25 $\pm$ 27.16 & 20.76 $\pm$ 26.01 & 28.31 $\pm$ 37.61 \\ 
5--6 & 314.58 $\pm$ 8.46 & $-$288.41 $\pm$ 7.54 & 426.77 $\pm$ 11.33 & -- & -- & -- & -- & -- & -- \\ 
6--7 & 339.79 $\pm$ 7.08 & $-$280.50 $\pm$ 6.69 & 440.61 $\pm$ 9.74 & -- & -- & -- & -- & -- & -- \\ 
7--10 & 410.91 $\pm$ 8.19 & $-$195.84 $\pm$ 8.54 & 455.19 $\pm$ 11.83 & -- & -- & -- & -- & -- & -- \\ \hline
\end{tabular}%
}
\end{table*}
\indent As mentioned in Section \ref{section3.3}, our initial selection of RC stars is contaminated by RGB stars. In principle, RGB stars in the same region are expected to have similar kinematics as RC stars and should not have any impact on the estimation of the average proper motion values of the sample. But as some of the sub-regions have dual RC populations, which are located at two different distances along the line of sight, the presence of RGB stars in the selected sample can affect the estimates. This is mainly because RGB stars corresponding to the main body (faint) RC population can be present in the location of the foreground (bright) RC in the CMD and vice versa.  To obtain a cleaner sample of RC stars we re-defined the RC selection box. The colour range is defined as the peak value of $(G_{BP} - G_{RP})_{0}$ $\pm$ $\sigma_{(G_{BP} - G_{RP})_{0}}$ and the magnitude range is defined as the peak value of $G_{0}$ $\pm$ $\sigma_{G_{0}}$, where the peak and $\sigma$ values in colour and magnitude are the best fit values for observed magnitude and colour distributions of RC stars discussed in Section \ref{section3.3} and given in Tables \ref{tab:mag_ne} and \ref{tab:color}. \\ 
\indent  For those sub-regions where a dual (faint and bright) RC population is found, the magnitude selection range corresponds to the best fit values obtained for the faint and bright Gaussian components. The dual RC populations considered here are only those in the eastern sub-regions at 2$\rlap{.}^{\circ}$5--5$^\circ$ radius. Fig. \ref{fig:clumpselec} shows, as an example, the CMD of the NE region (2$\rlap{.}^{\circ}$5--3$^\circ$ from the centre) with the selection boxes. The outer box is the initial RC selection box. The inner small boxes correspond to the selection of faint and bright RC stars based on the best fit Gaussian parameters. In the inner boxes the contribution from the RC stars are expected to dominate that of the RGB stars. For other sub-regions with dual RC populations (central regions and western regions), the peak magnitudes of the faint and bright RC are not well separated and are within the dispersion of the two Gaussian components. So for these sub-regions we considered only the single component values corresponding to the narrow Gaussian component for the selection of the RC stars.\\
 \indent Using the re-defined selection criteria, we selected the RC stars in all the sub-regions and calculated the mean proper motion values. For the dual RC populations, the mean proper motion values corresponding to the faint and bright RC populations are estimated separately. To obtain robust results of the mean proper motions and their associated uncertainties we employed a combination of Gaussian Mixture Model (GMM) fitting and bootstrapping. We started by creating 5,000 bootstrapping samples for each RC in the different sub-regions. These samples are generated by resampling the original data set with replacement. Then, we fitted a two dimensional, single-component GMM to the un-binned proper motions in the RA and Dec directions. This was done for each bootstrap sample. As the mean proper motions of the RC and the uncertainties we used the simple mean and standard deviation of the expectation values obtained from the GMMs for all bootstrap samples. The mean $\mu_\alpha$ and $\mu_\delta$ along with their standard errors for each of the sub-regions are presented in Table \ref{tab:pm ne nw} (NE, NW, SE and SW sub-regions).

\indent The estimated proper motion values have a large range. To decipher the variation in proper motion values, we plotted the mean $\mu_\alpha$ and $\mu_\delta$ values as a function of radius (Fig. \ref{fig:rad pm}) for all the sub-regions. The plots indicate that the proper motion components of the foreground RC population are distinct from those of the main body population at the same location in the plane of the sky. The plots also suggest that there is a radial variation in proper motion components of both the RC populations and variation between the eastern and western regions. These observed variations in $\mu_{\alpha}$ and  $\mu_{\delta}$ could be due to internal kinematics, geometry of the system and/or effects of tidal interactions of the MCs (\citealp{zivick2018,zivick2019,niederhofer2018}, 2020; Schmidt et al.2020). 
A proper modelling of the kinematics is required to understand these effects and is beyond the scope of this work. Here, we are mainly interested in the relative kinematic variation of the stars in the foreground stellar structure and the main body of the SMC.\\
\indent To reveal the relative kinematic difference of the dual RC population in more detail, in the top-left panel of Fig. \ref{fig:pm mean} we plotted the mean $\mu_\alpha$ vs. the mean $\mu_\delta$ of the eastern sub-regions (2$\rlap{.}^{\circ}$5--4$\rlap{.}^{\circ}$5 in NE and 2$\rlap{.}^{\circ}$5--5$^{\circ}$ in SE from the centre where distinct dual RC populations are observed). The faint and bright RC populations are clearly separated in this plot. The top-right panel of Fig. \ref{fig:pm mean} shows the mean values of the proper motion components corresponding to the main body (faint) and foreground (bright) RC, with standard errors as error bars. The difference in mean $\mu_\alpha$ between the bright RC (foreground population) and the faint RC (main body population) is $\sim$ 0.19$\pm$0.04 mas yr$^{-1}$ and the corresponding difference in mean $\mu_\delta$ is $\sim$ 0.12$\pm$0.02 mas yr$^{-1}$. The relative difference in $\mu_\alpha$ and $\mu_\delta$ as a function of radius in the eastern sub-regions (where distinct dual population is seen) is shown in the bottom-left and bottom-right panels of Fig. \ref{fig:pm mean} respectively. The mean relative difference (with standard errors) of bright and faint RC stars in $\mu_\alpha$ is $\sim$ 0.19$\pm$0.01 mas yr$^{-1}$ and $\mu_\delta$ is $\sim$ 0.12$\pm$0.01 mas yr$^{-1}$. One can note that the errors are slightly less as compared with the values from mean difference.  The proper motion values of the bright RC is significantly larger than that of the faint RC. This is expected if  the bright RC is at a closer distance. Thus the observed proper motion values also support that the two RC populations are at two different distances. \\ 
\indent In order to check whether there is any true kinematic variation between the two RC populations, we calculated the tangential velocity (V$_t$ in km s$^{-1}$) and the velocity components (V$_{\alpha}$ and V$_{\delta}$ in km s$^{-1}$) of the RC stars in the eastern sub-regions by incorporating the estimated proper motion values ($\mu_{\alpha}$ and $\mu_{\delta}$ in mas yr$^{-1}$) and distances (D in kpc) in the following equations:\\
\vspace{-0.5cm}
 \begin{equation}
    V_\alpha = 4.74 \times \mu_\alpha \times D   ;\\ V_\delta = 4.74 \times \mu_\delta \times D \end{equation}
    \vspace{-0.6cm}
    \begin{equation}
        V_t = (V_{\alpha}^2 + V_{\delta}^2)^{0.5}
    \end{equation}
The calculated values are tabulated in Table \ref{tab:vel_ne_rel}. We note that the V$_t$ value of RC stars in the SE sub-regions (5--6$^\circ$ from the centre) is  similar to the foreground (bright) RC population. This supports the result from photometric analysis that the foreground stellar structure in front of the SMC probably extends till SE 5--6$^\circ$.\\
\indent The velocity estimates in Table \ref{tab:vel_ne_rel} suggest that the two RC populations are indeed kinematically distinct and the foreground (bright) RC population is 34.4$\pm$3.8 km s$^{-1}$ slower than the main body (faint) RC population. This is illustrated in Fig. \ref{fig:velocity} where the top-left panel shows the two RC populations (between 2$\rlap{.}^{\circ}$5--5$^\circ$) in the V$_\alpha$ - V$_\delta$ plane. The top-right panel shows the mean values of the tangential velocity components corresponding to the faint (main body) and foreground (bright) RC with standard errors as the error bars. The relative difference in V$_\alpha$ (bottom-left) and V$_\delta$ (bottom-right) is calculated for the eastern sub-regions between 2$\rlap{.}^{\circ}$5--5$^\circ$ and plotted as a function of radius (Fig. \ref{fig:velocity}).  The relative difference in V$_\delta$ (34$\pm$2 km s$^{-1}$ towards North) is more significant than in V$_\alpha$ (6$\pm$3 km s$^{-1}$ towards West). This suggests that the foreground (bright) RC population is moving to the NW relative to the main body (faint) RC population.\\
\indent Recent studies which analysed the internal proper motion structure of the SMC (e.g. \citealp{Oey2018,zivick2018,deleo2020}); Niederhofer et al. 2020) 
 found that stars east of the main body of the galaxy move preferentially away from the SMC, towards the East. This motion has been interpreted as a signature of tidal stripping of the outer parts of the SMC. The results from the mentioned studies, however, are based on the assumption that all stars are at the same distance and the measured proper motions directly reflect tangential velocities. 

\section{Comparison with simulations and Discussion}
\label{section7}
\begin{figure}
    \subfloat{\includegraphics[scale=0.33]{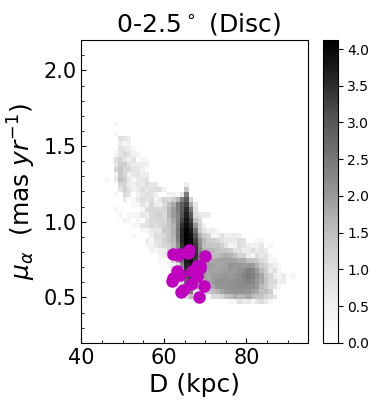}}
    \subfloat{\includegraphics[scale=0.33]{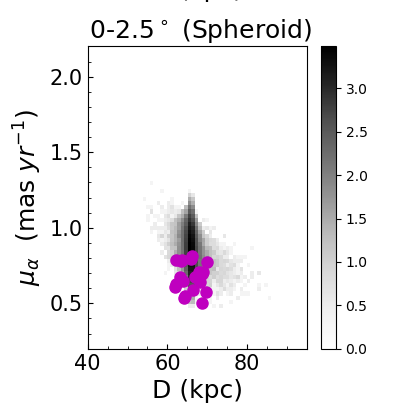}}\\
    \subfloat{\includegraphics[scale=0.3]{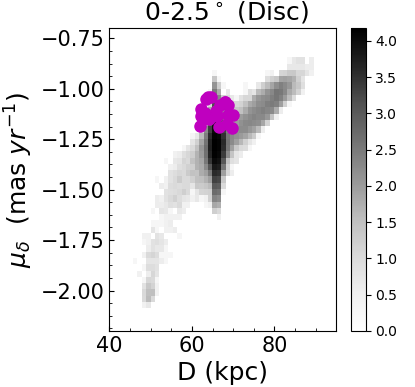}}
    \subfloat{\includegraphics[scale=0.3]{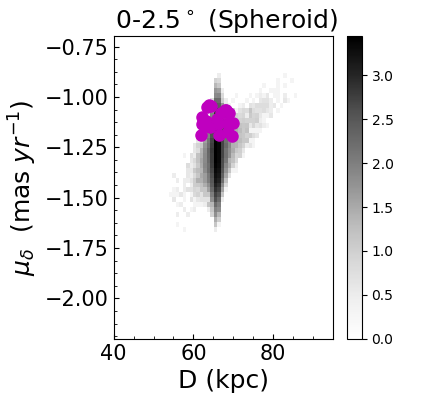}}
    \caption{Distribution of mean proper motions as function of distance for the NE, NW, SE and SW sub-regions within 0--2$\rlap{.}^{\circ}$5 from the SMC centre compared with simulations. Top (Bottom) panels refer to the $\mu_{\alpha}$ ($\mu_{\delta}$) direction for the disc and spheroid. The density of simulated points are shown in grey-scale and magenta points indicate the observed values.}
    \label{fig:0-2.5new}
\end{figure}

\begin{figure*}
    \captionsetup[subfigure]{labelformat=empty}
    \centering
    \vspace{-0.6cm}
    \subfloat[]{\includegraphics[scale=0.32]{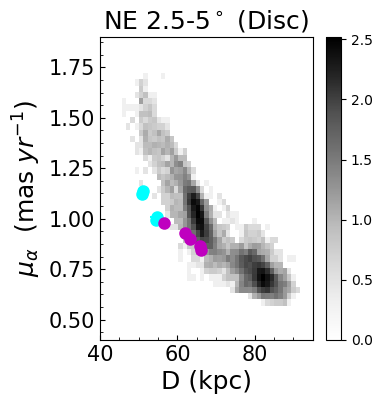}}
    \subfloat[]{\includegraphics[scale=0.32]{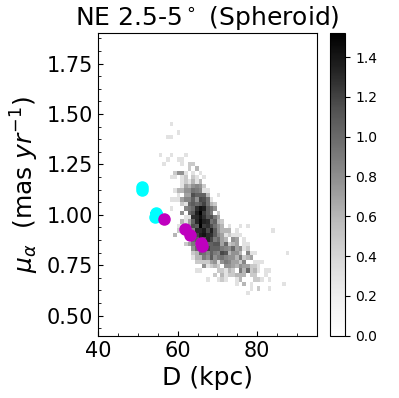}}
    \subfloat[]{\includegraphics[scale=0.32]{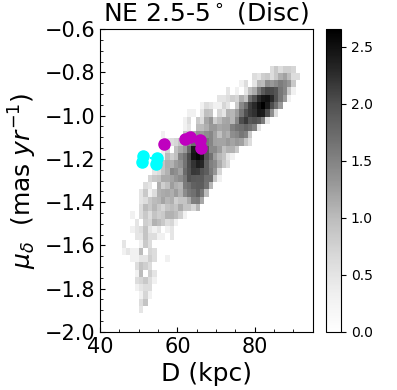}}
    \subfloat[]{\includegraphics[scale=0.32]{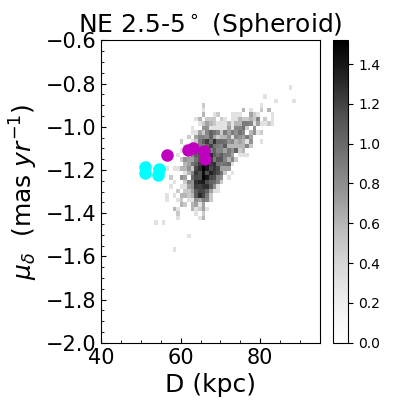}}\\
    \vspace{-0.6cm}
    \subfloat[]{\includegraphics[scale=0.3]{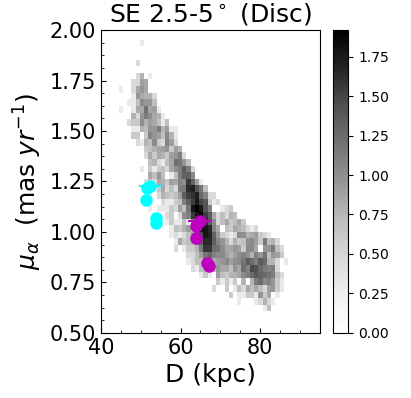}}
    \subfloat[]{\includegraphics[scale=0.3]{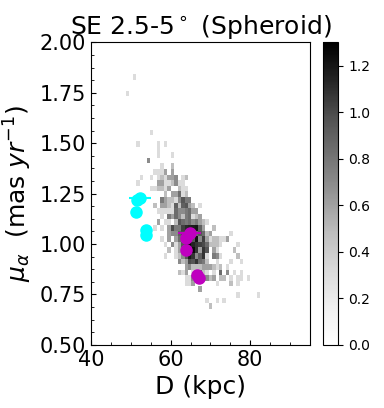}}
    \subfloat[]{\includegraphics[scale=0.3]{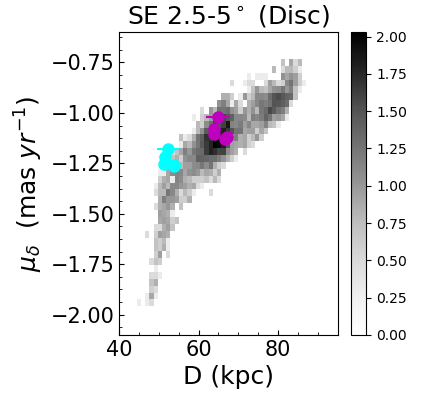}}
    \subfloat[]{\includegraphics[scale=0.3]{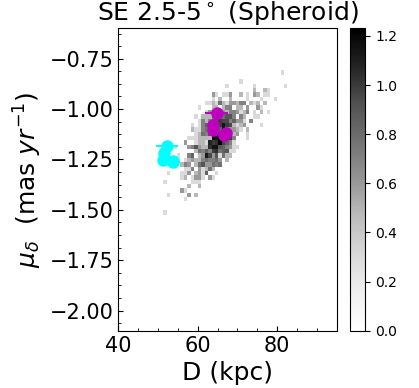}}\\
    \vspace{-0.6cm}
    \subfloat[]{\includegraphics[scale=0.3]{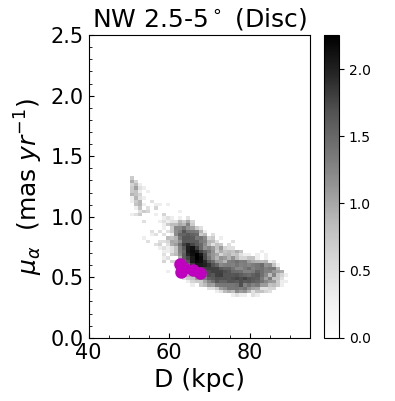}}
    \subfloat[]{\includegraphics[scale=0.3]{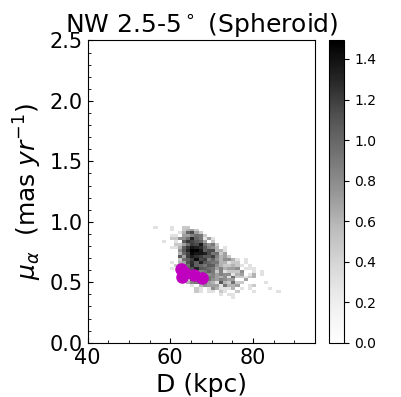}}
    \subfloat[]{\includegraphics[scale=0.3]{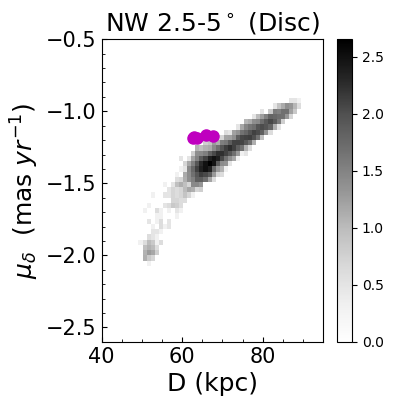}}
    \subfloat[]{\includegraphics[scale=0.3]{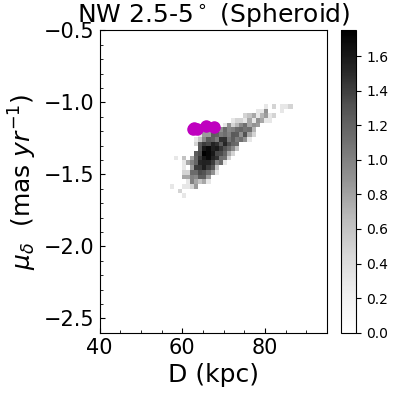}}\\
    \vspace{-0.6cm}
    \subfloat[]{\includegraphics[scale=0.3]{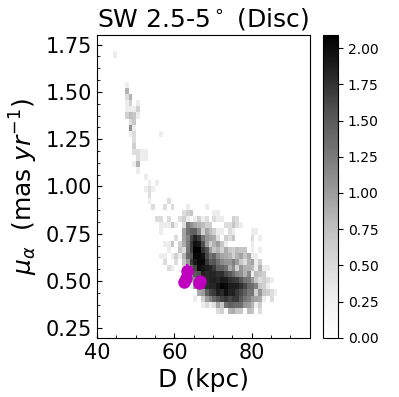}}
    \subfloat[]{\includegraphics[scale=0.3]{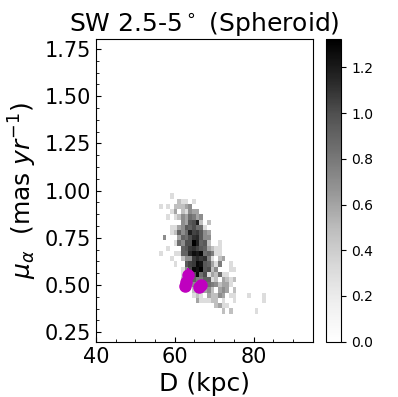}}
    \subfloat[]{\includegraphics[scale=0.3]{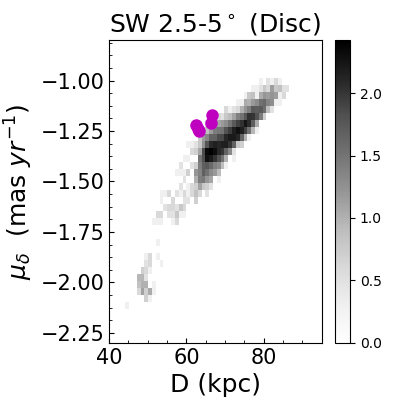}}
    \subfloat[]{\includegraphics[scale=0.3]{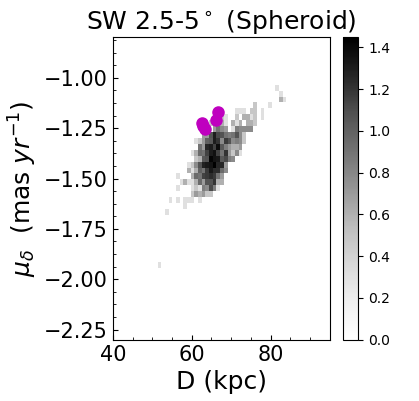}}
    \vspace{-0.6cm}
    \caption{Same as Fig. \ref{fig:0-2.5new} but for the NE (first row), SE (second row), NW (third row) and SW (fourth row) within 2$\rlap{.}^{\circ}$5--5$^\circ$ from the centre. Cyan (first two rows) and magenta points in all rows correspond to the bright and faint RC population, respectively.}
    \label{fig:sim2-2.5}
\end{figure*}

\begin{figure*}
    \captionsetup[subfigure]{labelformat=empty}
    \centering
    \subfloat[]{\includegraphics[scale=0.3]{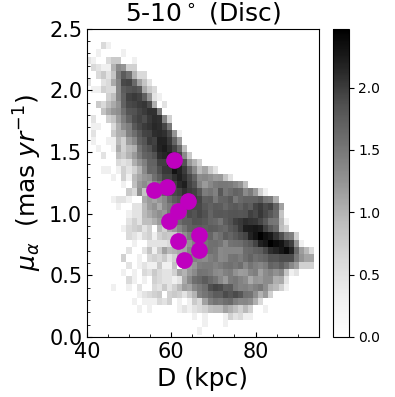}}
    \subfloat[]{\includegraphics[scale=0.3]{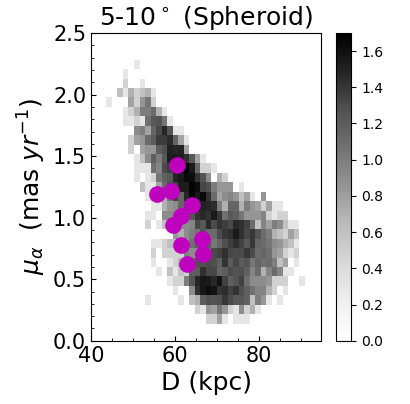}}
    \subfloat[]{\includegraphics[scale=0.3]{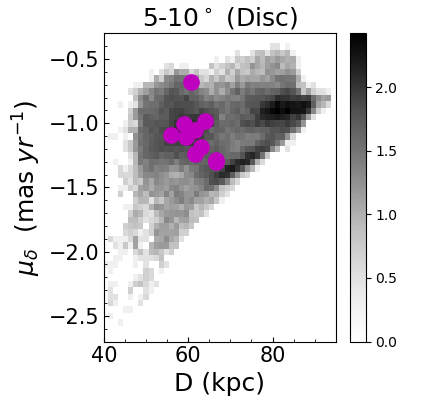}}
    \subfloat[]{\includegraphics[scale=0.3]{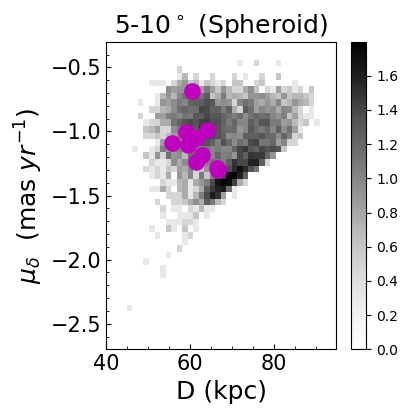}}
    \vspace{-0.6cm}
    \caption{Same as Fig. \ref{fig:0-2.5new} but for the sub-regions within 5--10$^{\circ}$ from the SMC centre.}
    \label{fig:sim5-10}
\end{figure*}

In the N-body simulation of the Magellanic System by \cite{Diaz2012} the LMC is treated as a point mass and the SMC is represented as a multi-component system composed of an exponential disc (truncation radius = 5 kpc and disc scale-length = 1 kpc), a central spheroid and a dark matter halo. The authors consider three models for the central spheroidal component of the SMC and found that an extended spheroid (truncation radius = 7.5 kpc and scale-length = 1.5 kpc) best reproduces the observed features. The assumed total mass of the LMC and the SMC are  10$^{10}$ M$_{\sun}$ and 3$\times$10$^{9}$ M$_{\sun}$ respectively. The disc and the dark matter halo of the SMC have equal mass, 1.36 $\times$ 10$^9$ M$_{\sun}$. The mass ratio of the spheroid to the disc is taken as 0.2. The Milky Way is represented by a realistic potential having a bulge, disk and a Navarro-Frenk-White (NFW) dark matter halo, with a total mass of 1.73 $\times$ 10$^{12}$ within r = 300 kpc.\\ 
\indent The assumption in the simulation is that the SMC spheroidal component mainly contains old stars and the disc (mainly in the outer 3--5 kpc) contains gas. The inner part of the disc (up to 2--3 kpc) contains both stars and gas. During the tidal interaction of the MCs $\sim$ 260 Myr ago, particles from both the disc and spheroid components of the SMC were stripped to create the MB. The gas stripped from the disc is responsible for the gaseous bridge. Stellar particles from the disc and spheroid were also stripped during the interaction, predicting stellar structures similar to gaseous features. In this section we compare the observed properties of RC stars in our study with the predictions from the simulation. We note that the simulation by \cite{Diaz2012} is mainly based on the gravitational effects and ignore the effect of drag forces induced by the hot halo of the Milky Way. Hence no offsets are expected between the stars and the gas from the disc.\\ 
\indent Figs. \ref{fig:0-2.5new} to \ref{fig:sim5-10} show the density distribution of the simulated particles of the present day SMC, from both disc and spheroid components, in the $\mu_\alpha$ vs. distance and $\mu_\delta$ vs. distance for different sub-regions and compare with the observed values. In order to make the comparison between the observations and simulations meaningful we chose the centre of the SMC in the simulation similar to the optical centre used in our study. \cite{Diaz2012} assumed the present day distance to the centre of mass of the N-body system of the SMC as 61.6 kpc. The mean distance to the SMC obtained by us from the main body RC population in the sub-regions between 0--2$\rlap{.}^{\circ}$5 (using only a narrow component) is 65.8 kpc. We applied this difference in distance as a systematic offset to the distances of the particles in the simulations and accordingly re-scaled the proper motion values provided in the simulations.\\

\begin{figure*}
    \centering
    \vspace{-0.6cm}
    \subfloat{\includegraphics[scale=0.28]{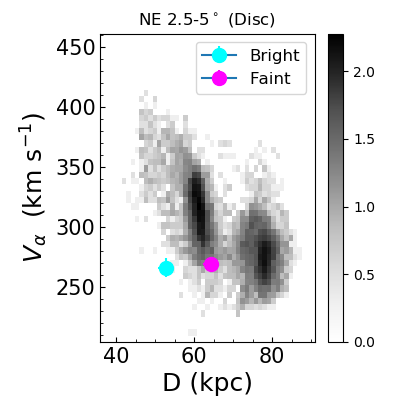}}
    \subfloat{\includegraphics[scale=0.28]{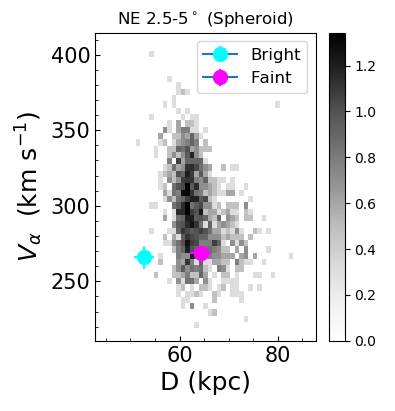}}
    \subfloat{\includegraphics[scale=0.28]{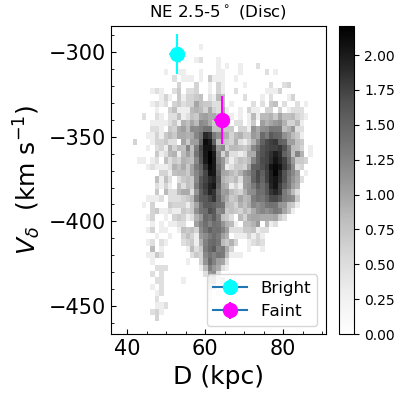}}
    \subfloat{\includegraphics[scale=0.28]{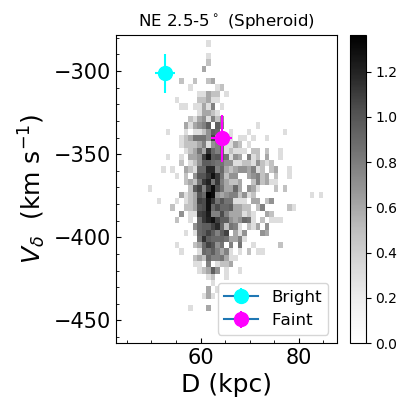}}\\
    \vspace{-0.5cm}
    \subfloat{\includegraphics[scale=0.28]{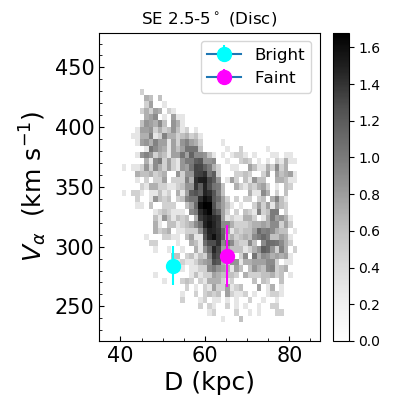}}
    \subfloat{\includegraphics[scale=0.28]{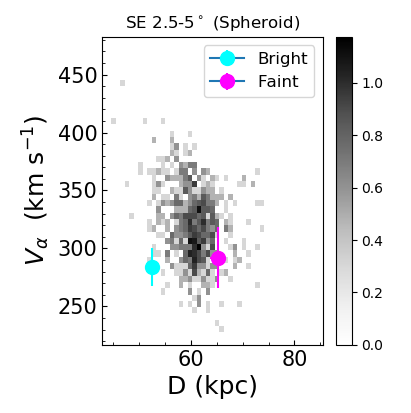}}
    \subfloat{\includegraphics[scale=0.28]{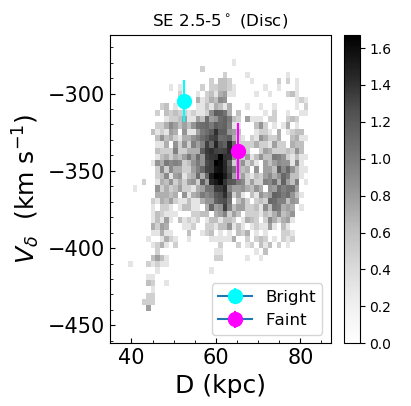}}
    \subfloat{\includegraphics[scale=0.28]{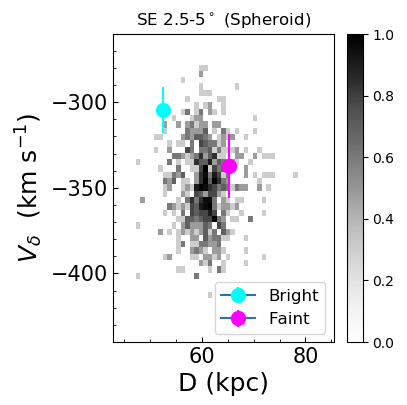}}\\
    \vspace{-0.3cm}
    \caption{From left to right the first two panels refer to the V$_{\alpha}$ against distance (disc and spheroid) while the other two panels refer to the V$_{\delta}$ against distance (disc and spheroid) for the 2$\rlap{.}^{\circ}$5--5$^\circ$ NE (first row) and SE (second row), respectively. Grey points indicate simulations and cyan/magenta points are observed values.}
    \label{fig:sim_v}
\end{figure*}
\indent Fig. \ref{fig:0-2.5new} shows the comparison for the 0--2$\rlap{.}^{\circ}$5 region. The grey-scale image represents the density of the simulated particles whereas the magenta points indicate the observed values. The majority of the simulated data points are distributed at the distance of the SMC for both the disc and spheroid components. However the disc component has particles distributed at closer distances as well as at farther distances. The density of particles at farther distances is more prominent, which \cite{Diaz2012} suggest is a Counter Bridge population. Though the observed proper motion values are within the range of the SMC main body proper motion values of the simulated particles, the measured motions (both $\mu_{\alpha}$ and $\mu_\delta$) are smaller than the mean values of the simulation. \\
\indent For the 2$\rlap{.}^{\circ}$5--5$^\circ$ sub-regions from the SMC centre we compare the observations with simulations for the NE, SE, NW and SW regions separately which are shown in Fig. \ref{fig:sim2-2.5}. Magenta and cyan points in all the plots represent the observed values corresponding to the SMC main body and to the foreground stellar sub-structure respectively. The disc sub-structures at closer distances are prominent in eastern regions compared to western regions. Whereas the disc sub-structure at a farther distance, the Counter Bridge, is most prominent in the NE region, though the feature is visible in all other regions as well. Similar sub-structures of spheroidal component are at a low density and they appear less extended along the line of sight than as seen in the disc component. The foreground stellar sub-structure observed in our study in the eastern SMC (shown with cyan points) more or less overlaps with the sub-structures at closer distances. But as seen in Fig. \ref{fig:0-2.5new}, the observed $\mu_{\alpha}$ and $\mu_{\delta}$ values are smaller compared to the simulated particles. However, we did not find any signature of the Counter Bridge in our study. The comparison between the observational results from this study and simulations suggests that the observed foreground stellar sub-structure in the eastern SMC has properties similar to that predicted by simulations for the sub-structures formed during the interaction between the MCs, with more similarity to that of the disc model (see first and second rows of Fig. \ref{fig:sim2-2.5}). Fig. \ref{fig:sim_v} shows the comparison between observations and simulations in the distance vs. tangential velocity components plane, for the sub-regions in the east (between 2$\rlap{.}^{\circ}$5--5$\rlap{.}^{\circ}$0 from the SMC centre) where we find a foreground stellar sub-structure. This figure also suggests that the properties of foreground stellar structure agrees more with the predictions from the disc model, although the values of the velocity components are not matching. The differences between the observations and simulations could be due to effect of hydro-dynamical effect of the Milky Way halo on gas (eg: ram-pressure effects), which the simulation has not considered. \\ 
\indent Beyond 5$\degree$ we identified only single populations of RC in all the sub-regions which are more or less at the distance of the main body of the SMC. Fig. \ref{fig:sim5-10} is the same as Fig. \ref{fig:0-2.5new} but for the sub-regions located 5--10$^\circ$ from the centre. Simulated particles in these sub-regions also show that there are extended structures towards both closer and farther distances. However, the sub-structures are less prominent in the spheroid model than in the disc model suggesting that in the regions 5--10$^\circ$ from the SMC centre observations of stellar features map the predictions from simulations adopting a spheroidal model. This is contrary to the observations in the 2$\rlap{.}^{\circ}$5--5$^\circ$ eastern region which map better the predictions from the disc model. The observed features in the entire region up to 10$^\circ$ from the SMC centre can be explained if the foreground stars are those which got tidally stripped from the inner disc region. This is  possible as the disc assumed in the simulations contains stars in the inner regions while the outer regions are dominated by gas.\\
 \indent \cite{muraveva2018} used VMC data of RR Lyrae stars and found an ellipsoidal distribution. They did not find distance bimodality in the 2$\rlap{.}^{\circ}$5--4$^\circ$ eastern regions of the SMC as observed in the RC distribution. RR Lyrae stars are older (age $\gtrapprox$ 10 Gyr) than RC stars and are expected to be distributed in the spheroidal component of the galaxy. Simulations suggest that the spheroidal component has relatively less sub-structure than the disc component. If the foreground RC population is tidally stripped from the inner disc of the SMC and the RR Lyrae stars are in the spheroidal component then the observed difference in the distribution of RC stars and RR Lyrae stars is naturally explained.\\
\indent Tatton et al. 2020, submitted suggested that ram-pressure effects, along with tidal effects, are required to explain the presence of bimodality in RC stars and the absence of this feature in RR Lyrae stars. They speculated that the foreground RC stars might have formed due to star formation in ram-pressure stripped gas. However, recent hydro-dynamical simulations by \cite{Wang2019} did not show a foreground population of intermediate-age stars in the eastern SMC.\\
\indent Classical Cepheids, which are younger (age $\sim$ 100--300 Myr) than RC stars, are used as tracers (\citealt{Haschke2012,Subramanian2015,Scowcroft2016,Jacyszyn-Dobrzeniecka2016} and \citealt{Ripepi2017}) to analyse the 3D structure of the SMC. All these studies found that the 3D distribution of Cepheids in the SMC is highly elongated ($\sim$ 20--30 kpc in the NE--SW direction). \cite{Haschke2012} and \cite{Subramanian2015} interpreted this elongated 3D structure as a highly inclined disc plane. However \cite{Scowcroft2016}, \cite{Jacyszyn-Dobrzeniecka2016} and \cite{Ripepi2017} suggested that the 3D distribution of Cepheids describes the SMC as a very disturbed galaxy which cannot be well described by a plane. \cite{Scowcroft2016} suggested that the structure traced by Cepheids traces a cylindrical shape, which we are viewing from one end. The 3D structure of the SMC depicted by the intermediate-age RC stars described in Section \ref{section5} is different from that implied by the younger Cepheids. This is expected since the distribution of Cepheids might have been driven by star formation during and/or after the recent interaction of the MCs, $\sim$ 150--300 Myr ago. Also, hydro-dynamical effects on gas can play a significant role in star formation during and/or after the interactions and shaping the 3D distribution of stars younger than $\sim$ 150--300 Myr.\\
\indent A detailed chemical and 3D kinematic study of the RC stars along with improved theoretical simulations, including both tidal and hydro-dynamical effects, are required to confirm the nature and origin of this foreground intermediate-age stellar structure of the SMC. So, in our future work we plan to analyse in detail the three-dimensional velocities and the chemical composition of the RC stars.

\section{Summary and Conclusions}
\label{section8}
In this study we used data from \textit{Gaia} DR2 to study a stellar sub-structure in front of the SMC. We obtained the following results.
\begin{itemize}
    \item{We traced the presence of a dual RC feature which corresponds to two populations at different distances along the line of sight. The brighter one corresponds to a foreground population (at a closer distance from us) and the fainter one corresponds to the main body of the SMC. The foreground population is located $\sim$12 kpc in front of the main body of the SMC. \textit{Gaia} data trace this feature from 2$\rlap{.}^{\circ}$5 to $\sim$5--6$^\circ$ from the centre of the SMC in the eastern regions.}
    \item{Beyond 6$\degree$ only a single RC population is identified even in the eastern regions. The distances corresponding to the single peaks of the RC magnitude distributions beyond 6$\degree$ are similar to that of the main body of the SMC. This suggests that the foreground stellar structure is not present beyond 6$^\circ$ from the SMC centre and hence does not fully overlap with the gaseous MB.}
    \item{From the \textit{Gaia} proper motion measurements, we found that the foreground stellar structure is kinematically distinct from the main body population with $\sim$ 35 km s$^{-1}$ slower tangential velocity. The  foreground RC population is moving to the NW relative to the main body population. The relative difference in V$_\delta$ (34$\pm$2 km s$^{-1}$ towards North) is more significant than in V$_\alpha$ (6$\pm$3 km s$^{-1}$ towards West).}
    \item{The observed properties of the RC stars are compared with numerical simulations to understand the origin of the foreground structure. Though the observed properties are not fully consistent with the simulations, a comparison indicates that the foreground stellar structure is most likely to be the tidally stripped stellar counterpart of the gaseous MB and might have formed from the inner disc of the SMC. }
    \item {A detailed chemical and 3D kinematic study of the RC stars along with improved theoretical simulations, including both tidal and hydro-dynamical effects, are required to better understand the nature and origin of this foreground stellar structure of the SMC.}
\end{itemize}  
{\bf Acknowledgements}
    AOO acknowledges support from the Indian Institute of Astrophysics through the Visiting Students Programme. SS acknowledges support from the Science and Engineering Research Board of India through a Ramanujan Fellowship and support from the Australia-India Council/Department of Foreign Affairs and Trade (via grant AIC2018-067) which funded a visit to the International Centre for Radio Astronomy Research (ICRAR), University of Western Australia during the period of this research. MRC, FN and DEY acknowledge support from the European Research Council (ERC) under the European Horizon 2020 research and innovation programme (grant agreement no. 682115). We thank Annapurni Subramaniam for comments on an earlier version of the manuscript. This work has made use of data from the European Space Agency (ESA) space mission Gaia (\url{https://www.cosmos.esa.int/gaia}). Gaia data are being processed by the Gaia Data Processing and Analysis Consortium (DPAC). Funding for the DPAC is provided by national institutions, in particular the institutions participating in the Gaia MultiLateral Agreement (MLA). This research made use of numpy \citep{numpy}, scipy \citep{scipy2020}, matplotlib \citep{matplotlib} and  astropy,\footnote{http://www.astropy.org} a community-developed core Python package for Astronomy \citep{astropy:2013, astropy:2018}. Finally, it is our pleasure to thank the referee for constructive suggestions.\\
\newline
{\bf Data Availability}   
The mean magnitudes, colours, distances, proper motions and tangential velocities of RC stars in  different sub-regions are provided in various tables in the respective sections of the article. The Gaia data used to derive these parameters were released as part of Gaia data release 2 (DR2) and is available in Gaia Archive at \url{https://archives.esac.esa.int/gaia}. 

\appendix
\section{Hess diagrams for all sub-regions}
\begin{figure*}
    \captionsetup[subfigure]{labelformat=empty}
    \centering
    \hspace*{-1.1em}
    \subfloat[]{\includegraphics[width=0.2\textwidth]{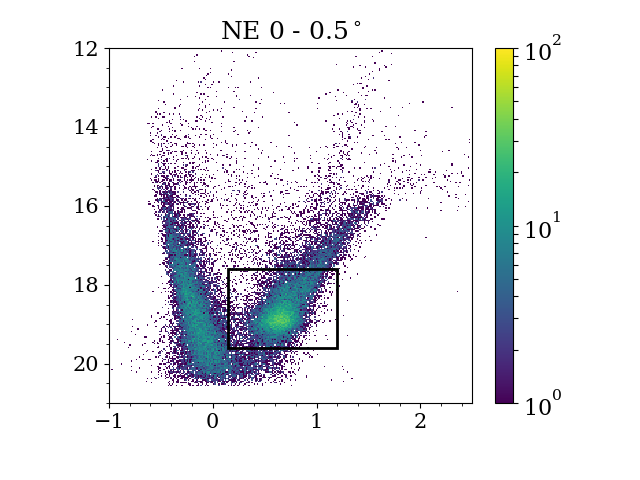}}
    \hspace*{-1.1em}
    \subfloat[]{\includegraphics[width=0.2\textwidth]{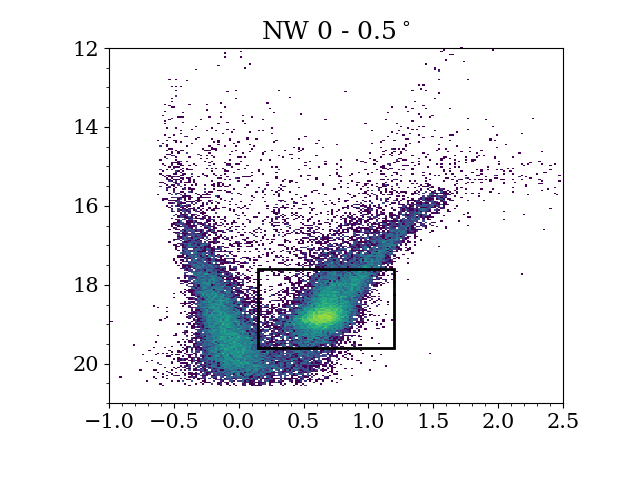}}
    \hspace*{-1.1em}
    \subfloat[]{\includegraphics[width=0.2\textwidth]{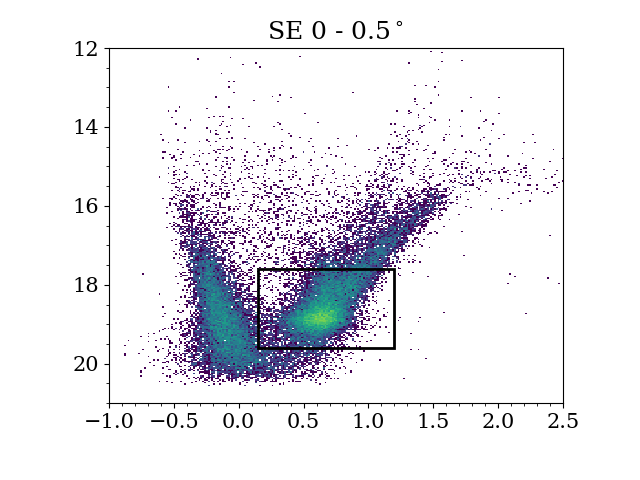}}
    \hspace*{-1.1em}
    \subfloat[]{\includegraphics[width=0.2\textwidth]{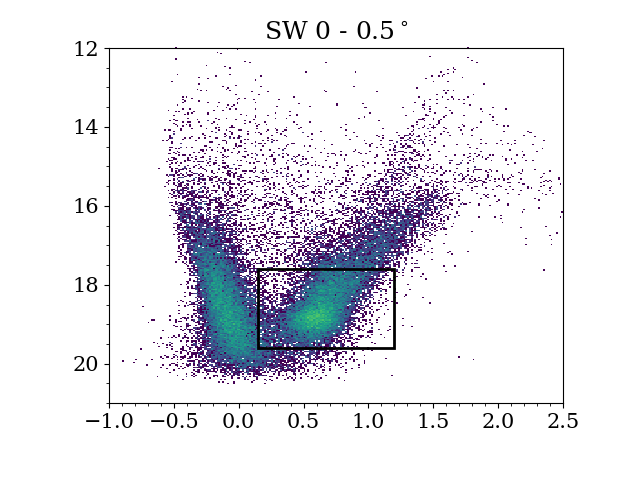}}\\
    \vspace{-1.1cm}
    \hspace*{-1.1em}
    \subfloat[]{\includegraphics[width=0.2\textwidth]{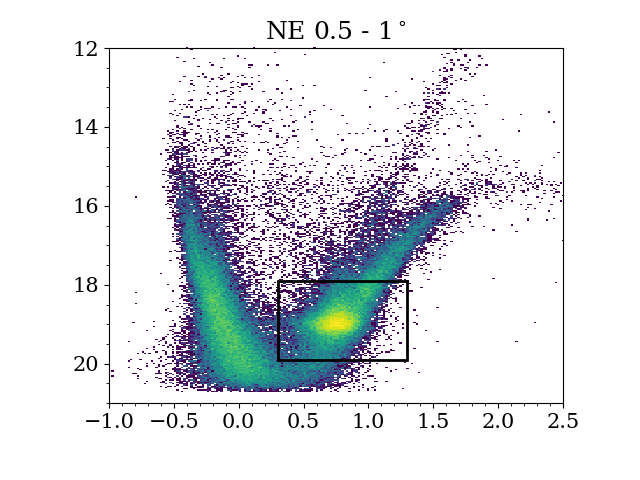}}
    \hspace*{-1.1em}
    \subfloat[]{\includegraphics[width=0.2\textwidth]{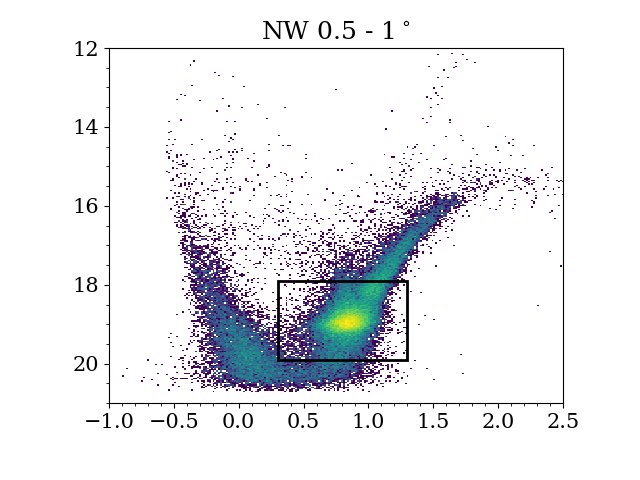}}
    \hspace*{-1.1em}
    \subfloat[]{\includegraphics[width=0.2\textwidth]{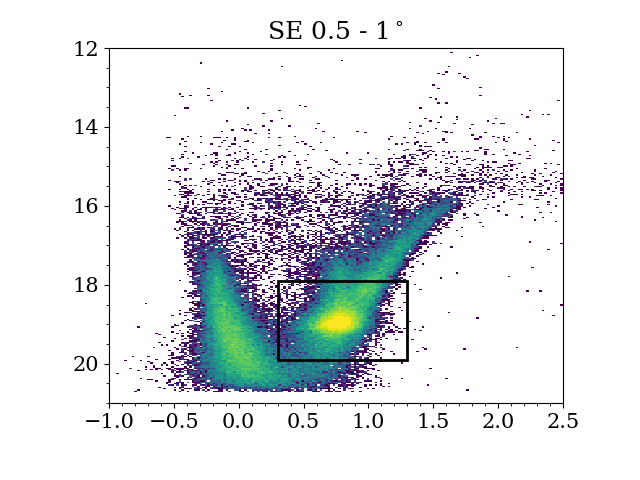}}
    \hspace*{-1.1em}
    \subfloat[]{\includegraphics[width=0.2\textwidth]{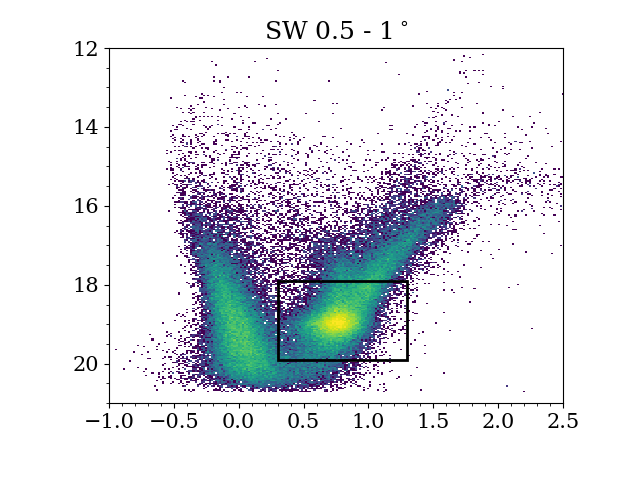}}\\
    \vspace{-1.1cm}
    \hspace*{-1.1em}
    \subfloat[]{\includegraphics[width=0.2\textwidth]{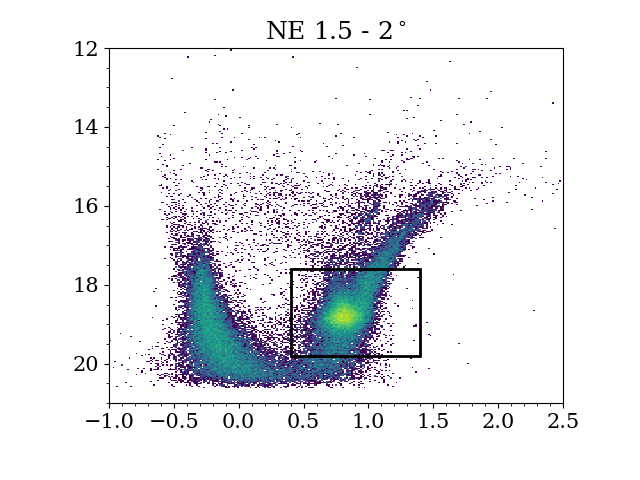}}
    \hspace*{-1.1em}
    \subfloat[]{\includegraphics[width=0.2\textwidth]{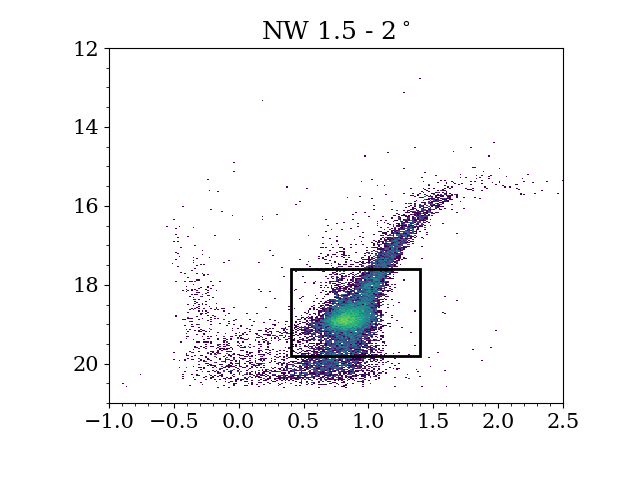}}
    \hspace*{-1.1em}
    \subfloat[]{\includegraphics[width=0.2\textwidth]{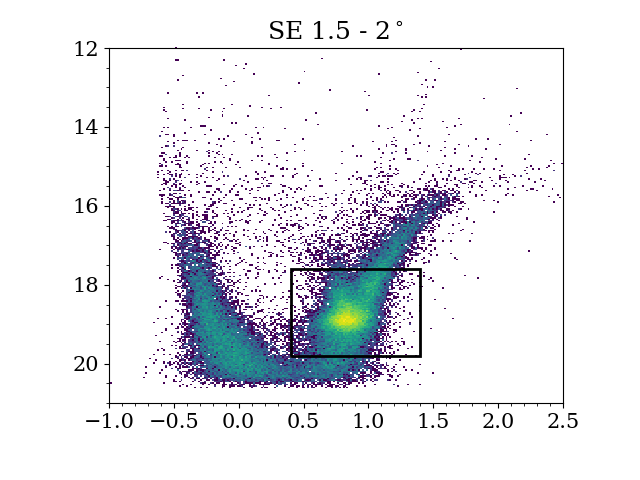}}
    \hspace*{-1.1em}
    \subfloat[]{\includegraphics[width=0.2\textwidth]{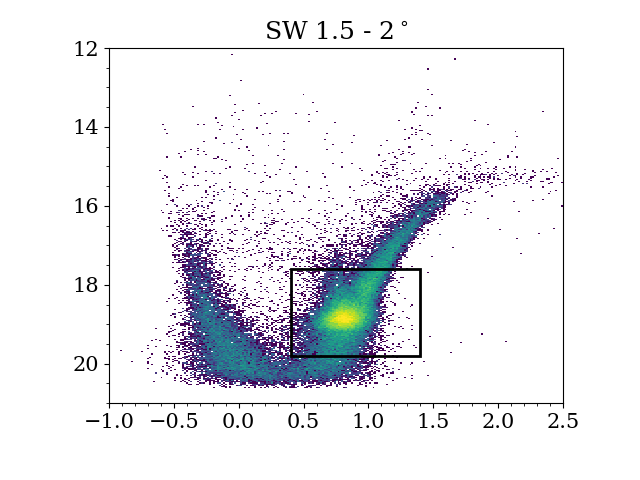}}\\
    \vspace{-1.1cm}
    \hspace*{-1.1em}
    \subfloat[]{\includegraphics[width=0.2\textwidth]{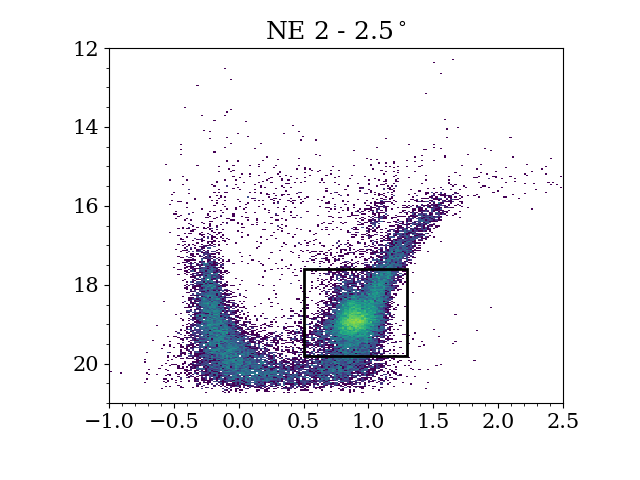}}
    \hspace*{-1.1em}
    \subfloat[]{\includegraphics[width=0.2\textwidth]{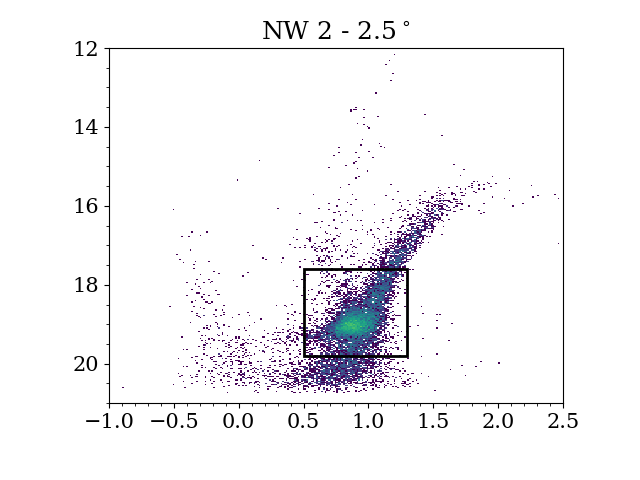}}
    \hspace*{-1.1em}
    \subfloat[]{\includegraphics[width=0.2\textwidth]{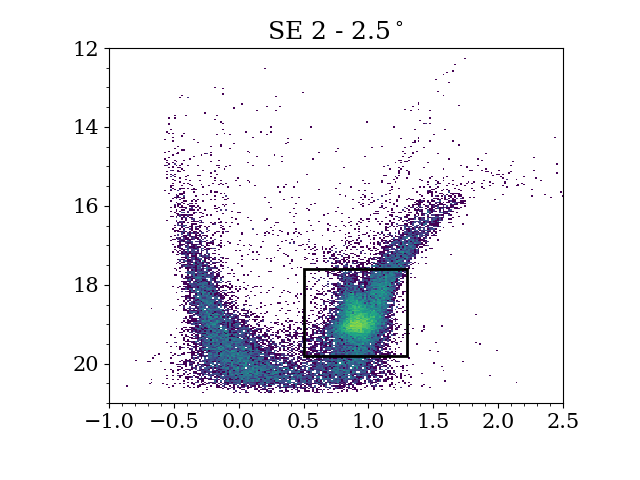}}
    \hspace*{-1.1em}
    \subfloat[]{\includegraphics[width=0.2\textwidth]{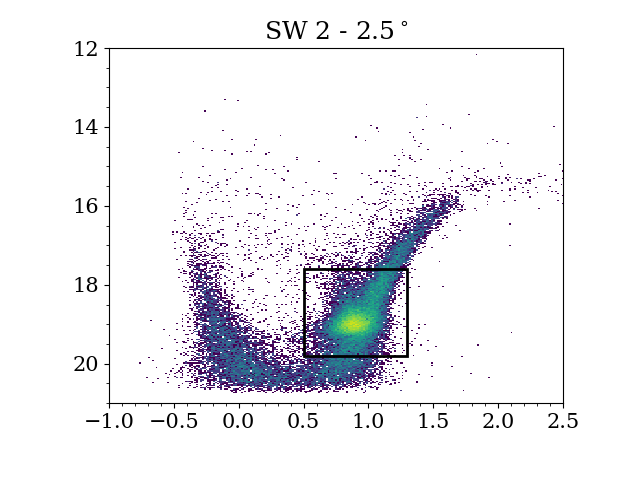}}\\
    \vspace*{-1.5em}
    \caption{Hess diagrams representing the stellar density in the observed CMD of \textit{Gaia} DR2 sources enclosed in sub-regions within the 0--2$\rlap{.}^{\circ}$5 (except 1--1$\rlap{.}^{\circ}$5) radial region from the SMC centre. The colour bar from blue to yellow represents the increase in stellar density. In black are the rectangular boxes to select RC stars within each sub-region. The axis labels and colorbars are the same and shown only for the top left panel.}
    \label{fig:CMD0-2.5}
\end{figure*}
\begin{figure*}
    \captionsetup[subfigure]{labelformat=empty}
    \centering
    \hspace*{-1.1em}
    \subfloat[]{\includegraphics[width=0.2\textwidth]{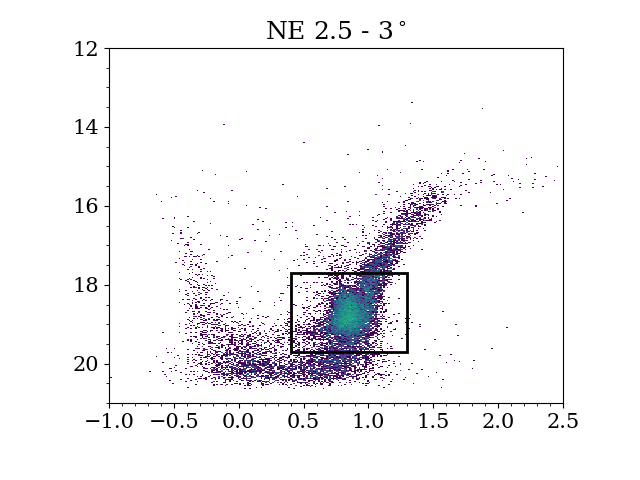}}
    \hspace*{-1.1em}
    \subfloat[]{\includegraphics[width=0.2\textwidth]{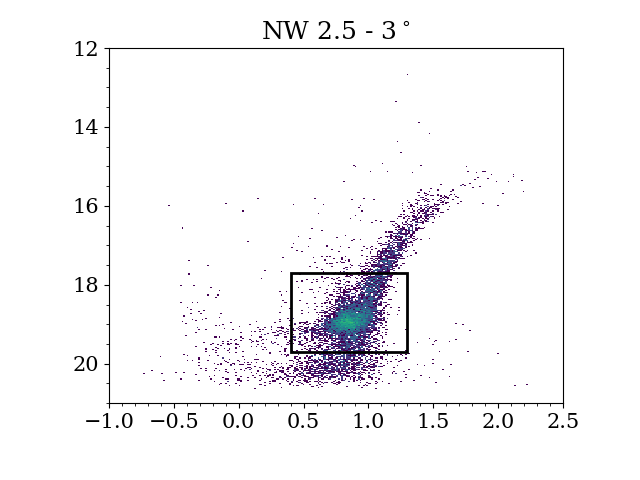}}
    \hspace*{-1.1em}
    \subfloat[]{\includegraphics[width=0.2\textwidth]{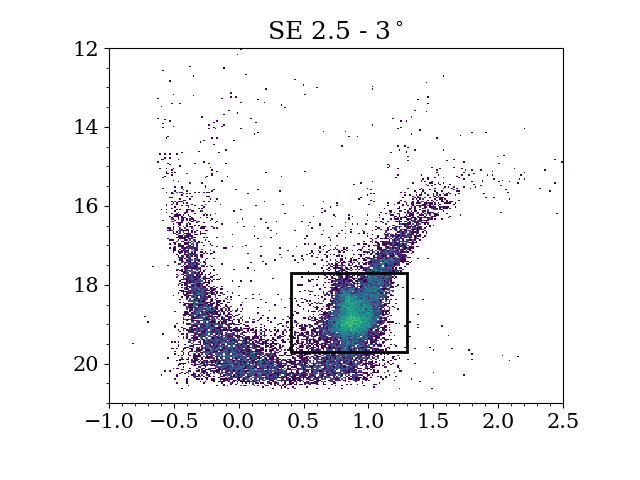}}
    \hspace*{-1.1em}
    \subfloat[]{\includegraphics[width=0.2\textwidth]{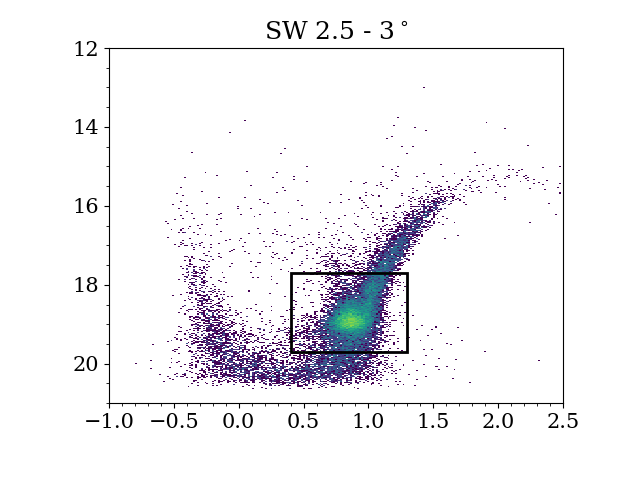}}\\
    \vspace{-0.9cm}
    \hspace*{-1.1em}
    \subfloat[]{\includegraphics[width=0.2\textwidth]{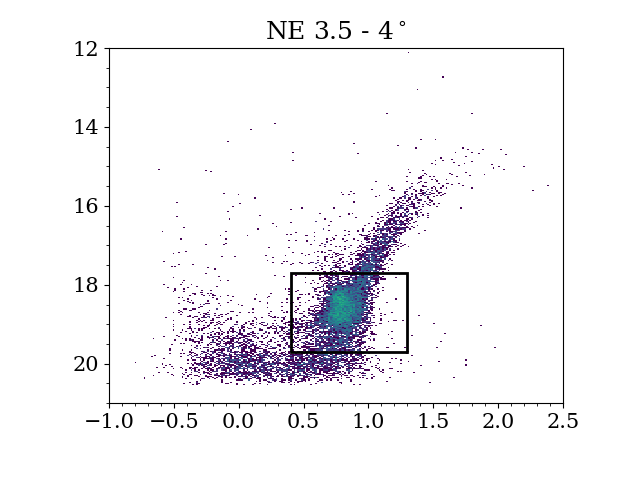}}
    \hspace*{-1.1em}
    \subfloat[]{\includegraphics[width=0.2\textwidth]{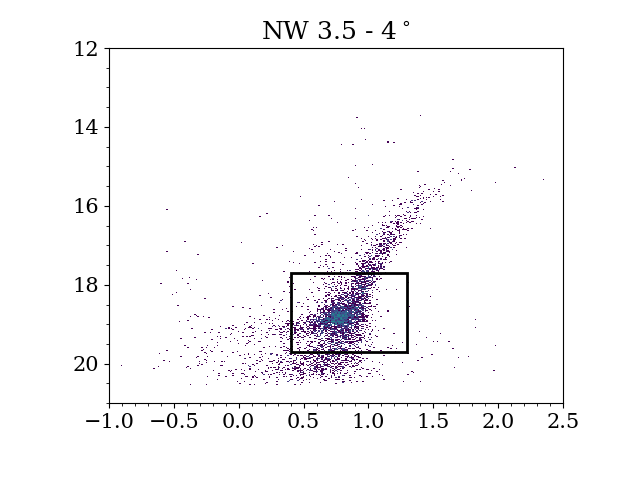}}
    \hspace*{-1.1em}
    \subfloat[]{\includegraphics[width=0.2\textwidth]{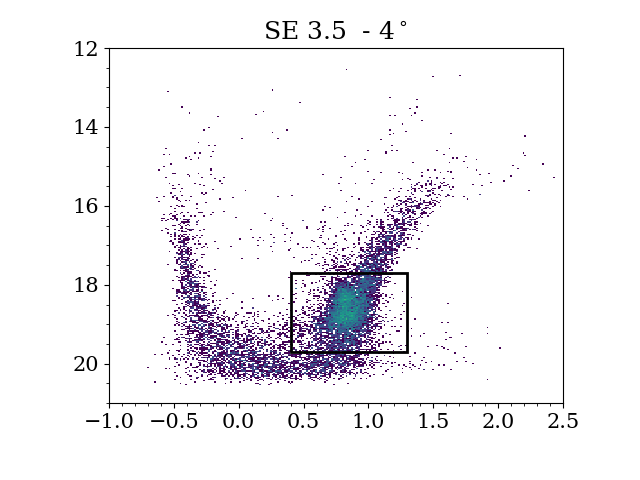}}
    \hspace*{-1.1em}
    \subfloat[]{\includegraphics[width=0.2\textwidth]{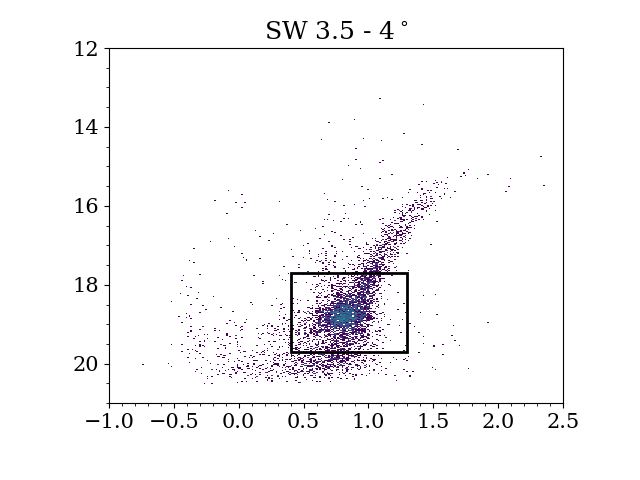}}\\
    \vspace{-1.1cm}
    \hspace*{-1.1em}
    \subfloat[]{\includegraphics[width=0.2\textwidth]{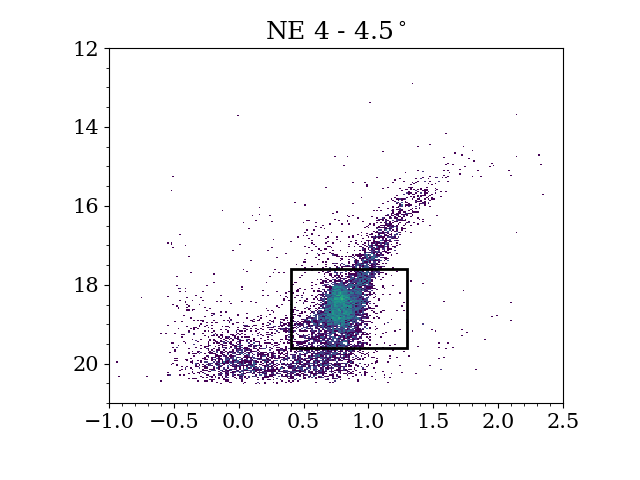}}
    \hspace*{-1.1em}
    \subfloat[]{\includegraphics[width=0.2\textwidth]{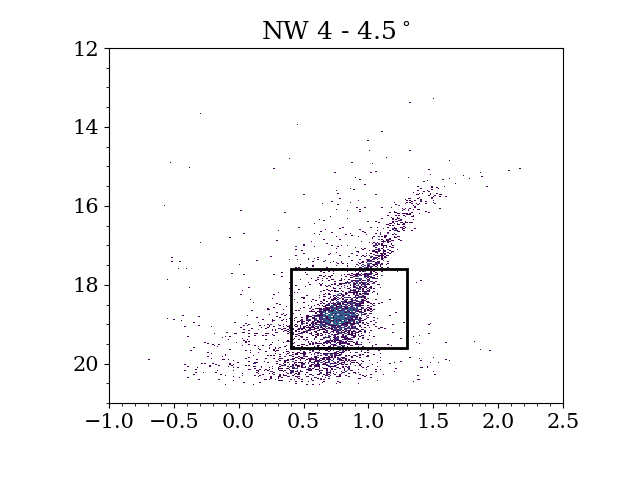}}
    \hspace*{-1.1em}
    \subfloat[]{\includegraphics[width=0.2\textwidth]{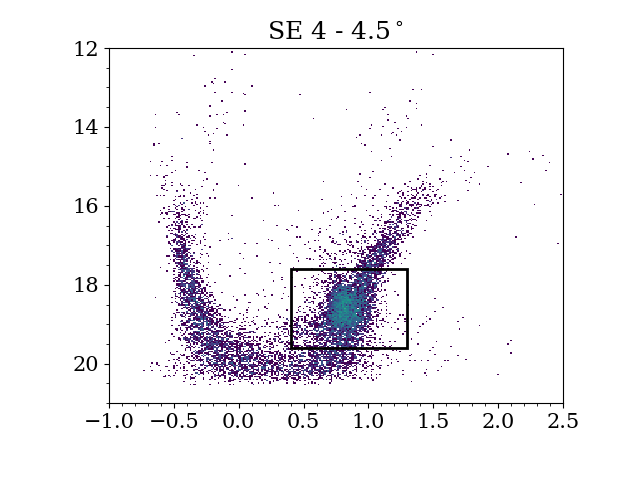}}
    \hspace*{-1.1em}
    \subfloat[]{\includegraphics[width=0.2\textwidth]{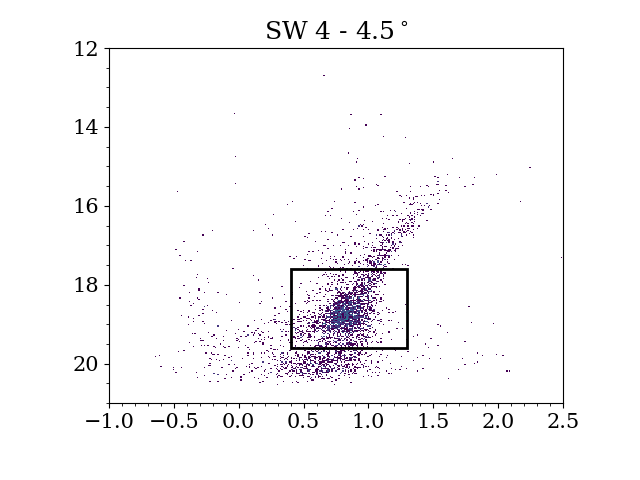}}\\
    \vspace{-1.1cm}
    \hspace*{-1.1em}
    \subfloat[]{\includegraphics[width=0.2\textwidth]{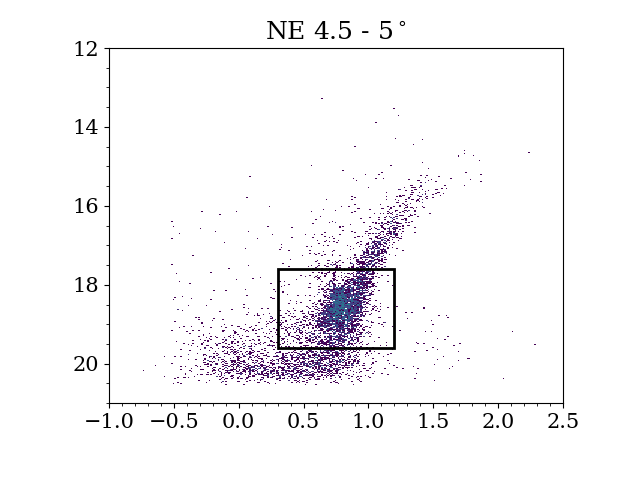}}
    \hspace*{-1.1em}
    \subfloat[]{\includegraphics[width=0.2\textwidth]{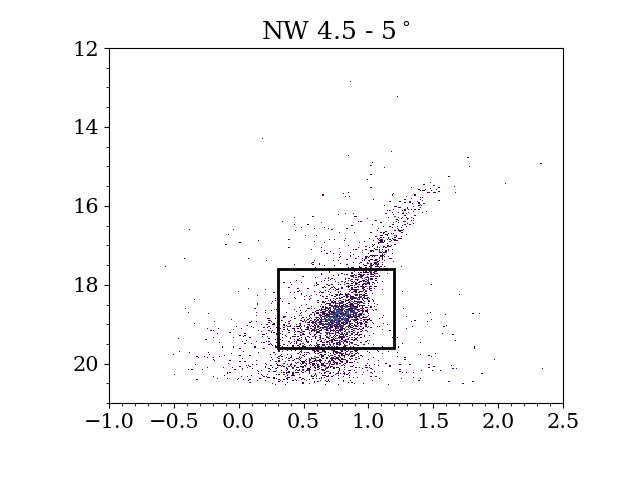}}
    \hspace*{-1.1em}
    \subfloat[]{\includegraphics[width=0.2\textwidth]{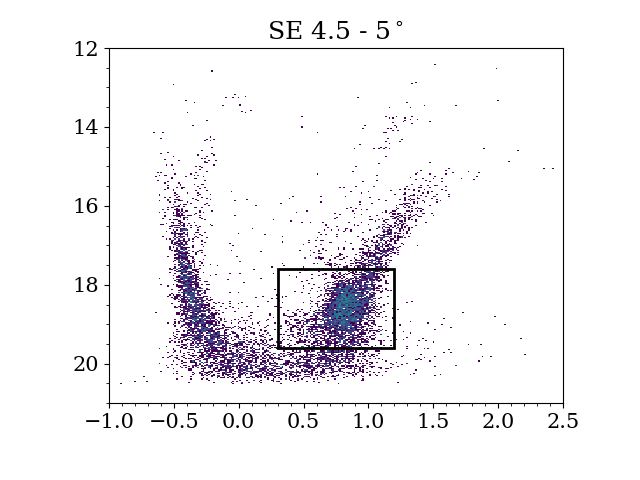}}
    \hspace*{-1.1em}
    \subfloat[]{\includegraphics[width=0.2\textwidth]{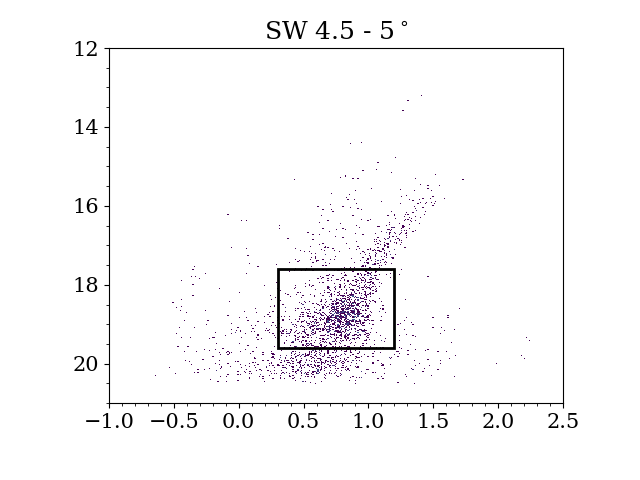}}\\
    \vspace*{-1.3em}
    \caption{Same as Fig.\ref{fig:CMD0-2.5} but for the 2$\rlap{.}^{\circ}$5--5$^\circ$ sub-regions from the SMC centre.}
    \label{fig:CMD2.5-5}
\end{figure*}
\begin{figure*}
    \captionsetup[subfigure]{labelformat=empty}
    \centering
    \hspace*{-1.1em}
    \subfloat[]{\includegraphics[width=0.2\textwidth]{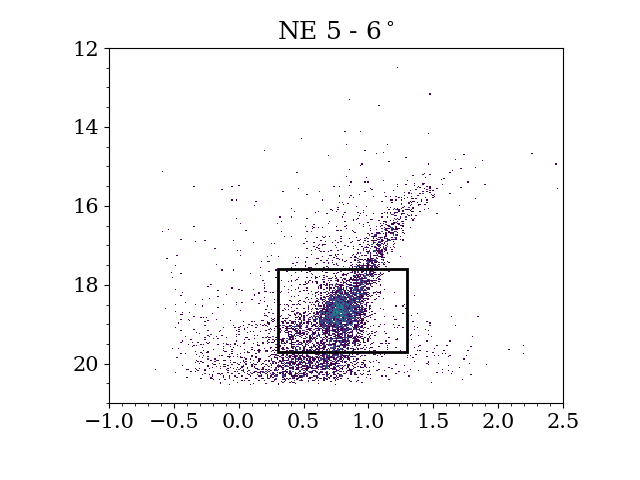}}
    \hspace*{-1.1em}
    \subfloat[]{\includegraphics[width=0.2\textwidth]{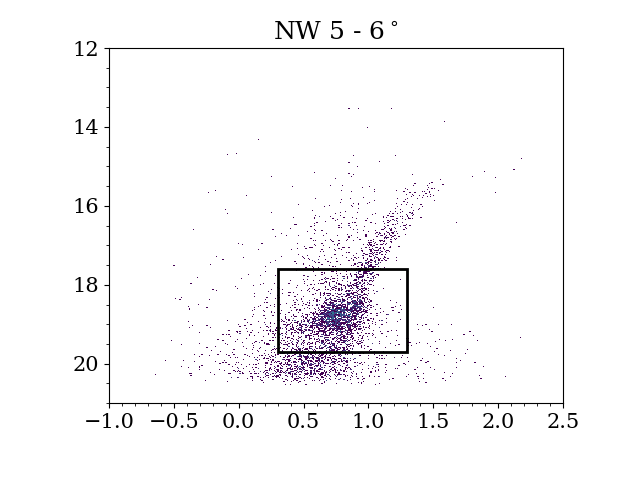}}
    \hspace*{-1.1em}
    \subfloat[]{\includegraphics[width=0.2\textwidth]{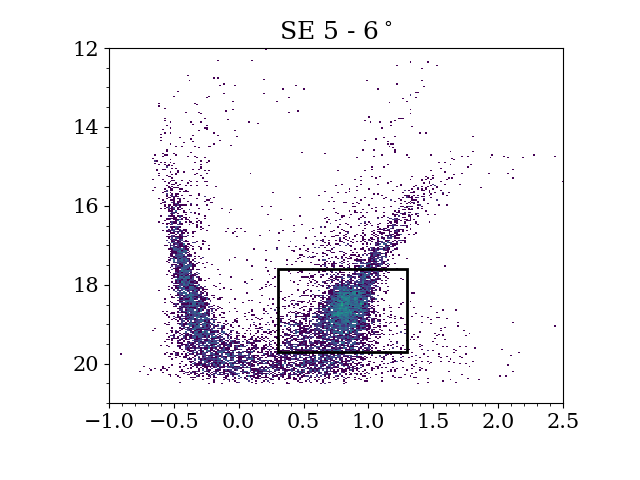}}
    \hspace*{-1.1em}
    \subfloat[]{\includegraphics[width=0.2\textwidth]{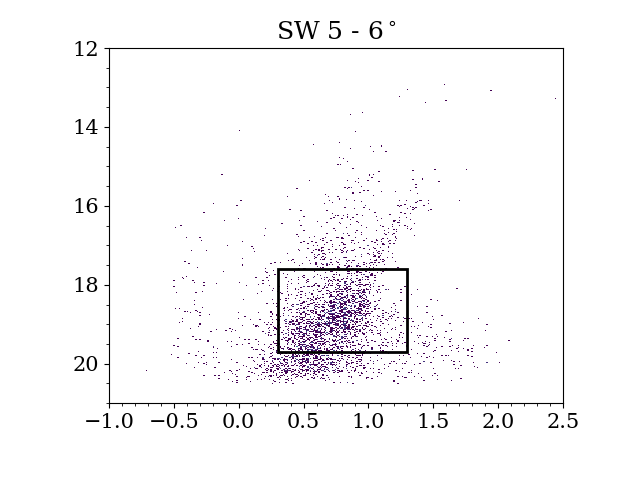}}\\
    \vspace{-0.9cm}
    \hspace*{-1.1em}
    \subfloat[]{\includegraphics[width=0.2\textwidth]{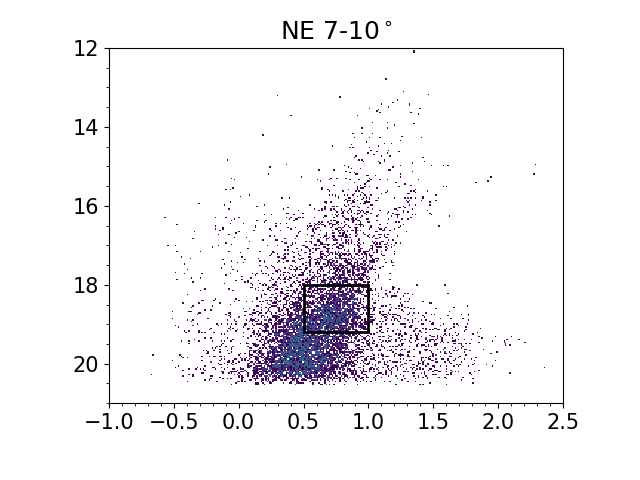}}
    \hspace*{-1.1em}
    \subfloat[]{\includegraphics[width=0.2\textwidth]{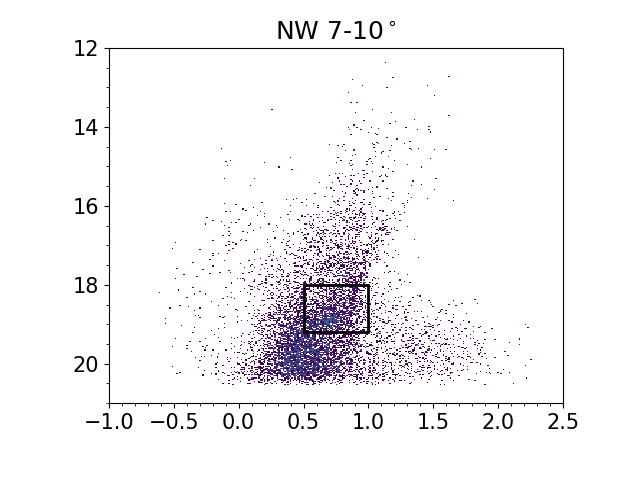}}
    \hspace*{-1.1em}
    \subfloat[]{\includegraphics[width=0.2\textwidth]{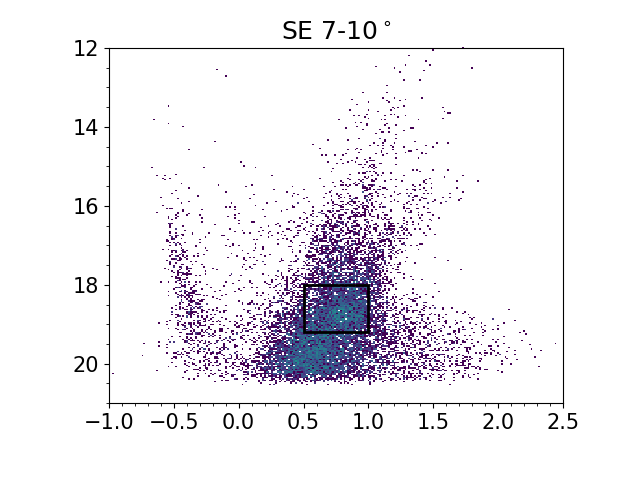}}
    \hspace*{-1.1em}
    \subfloat[]{\includegraphics[width=0.2\textwidth]{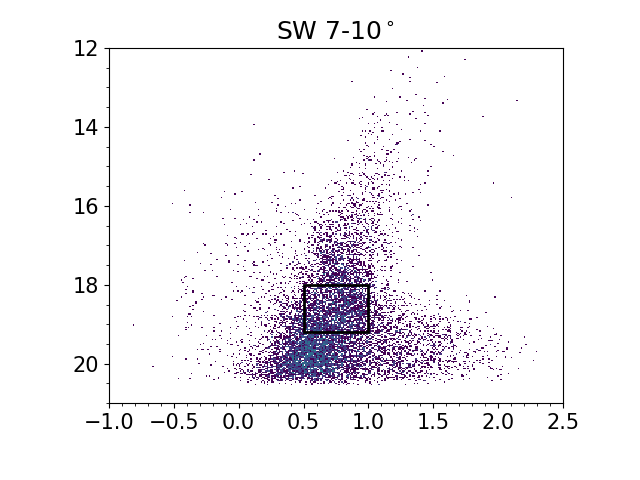}}\\
    \vspace*{-1.3em}
    \caption{Same as Fig.\ref{fig:CMD0-2.5} but for the 5--10$^\circ$ sub-regions from the SMC centre.}
    \label{fig:CMD5-10}
\end{figure*}

\section{Magnitude distributions of RC stars}
\begin{figure*}
    \captionsetup[subfigure]{labelformat=empty}
    \centering
    \hspace*{-1.6em}
    \vspace{-0.8cm}
    \subfloat[]{\includegraphics[width=0.23\textwidth]{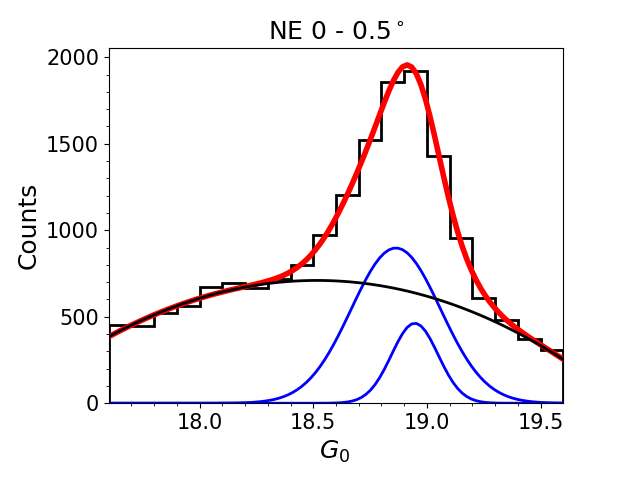}}
    \hspace*{-1.6em}
    \subfloat[]{\includegraphics[width=0.23\textwidth]{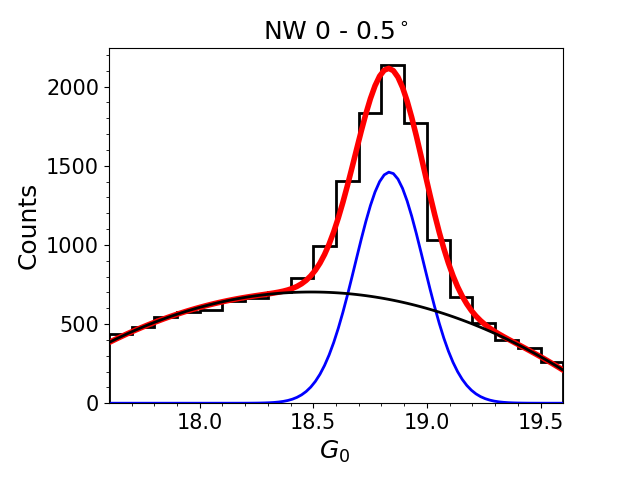}}
    \hspace*{-1.6em}
    \subfloat[]{\includegraphics[width=0.23\textwidth]{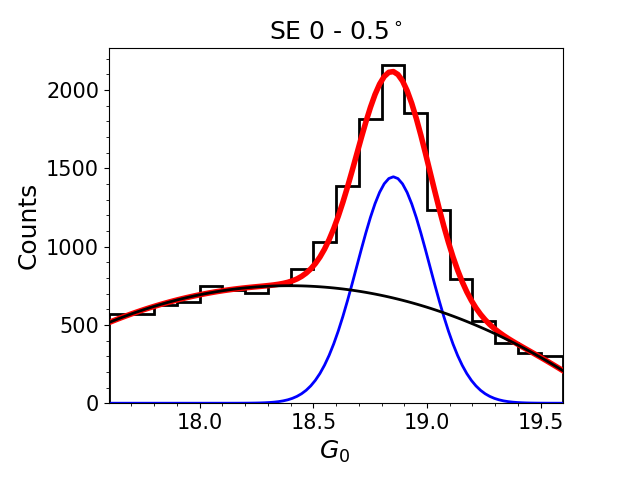}}
    \hspace*{-1.6em}
    \subfloat[]{\includegraphics[width=0.23\textwidth]{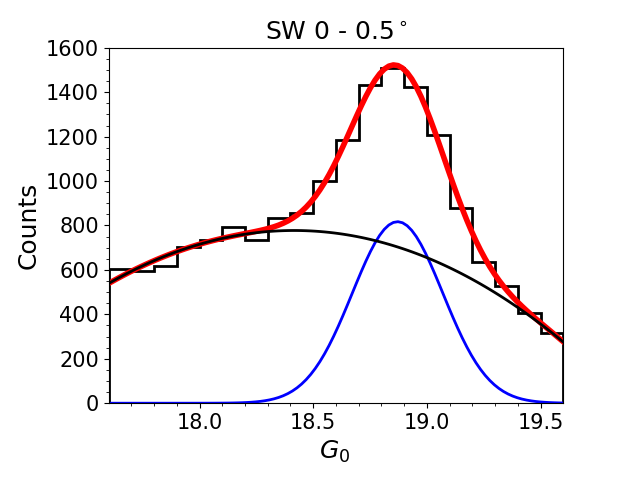}}\\
    \hspace*{-1.6em}
    \vspace{-0.8cm}
    \subfloat[]{\includegraphics[width=0.23\textwidth]{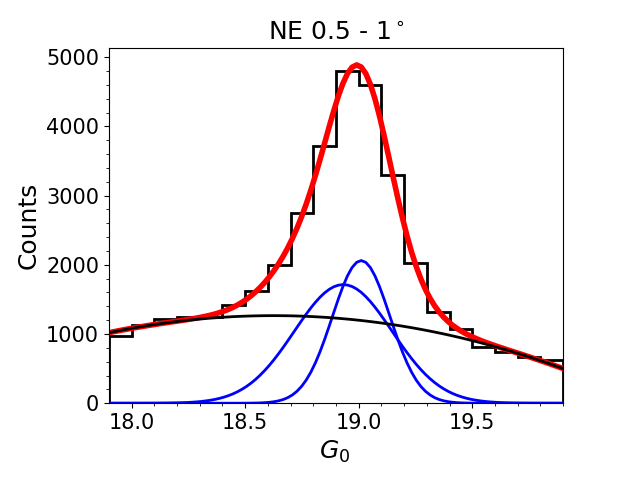}}
    \hspace*{-1.6em}
    \subfloat[]{\includegraphics[width=0.23\textwidth]{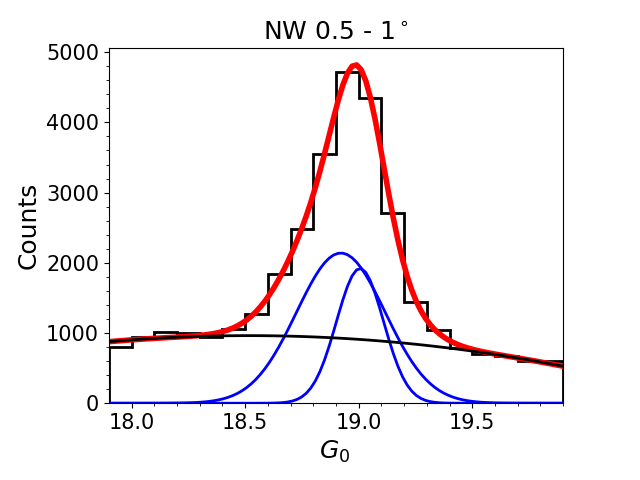}}
    \hspace*{-1.6em}
    \subfloat[]{\includegraphics[width=0.23\textwidth]{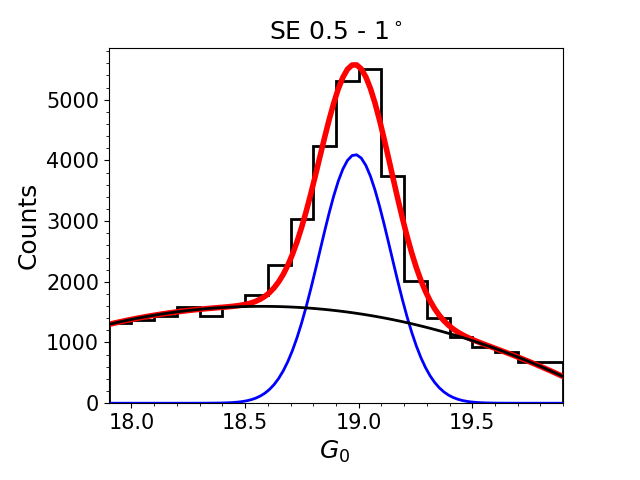}}
    \hspace*{-1.6em}
    \subfloat[]{\includegraphics[width=0.23\textwidth]{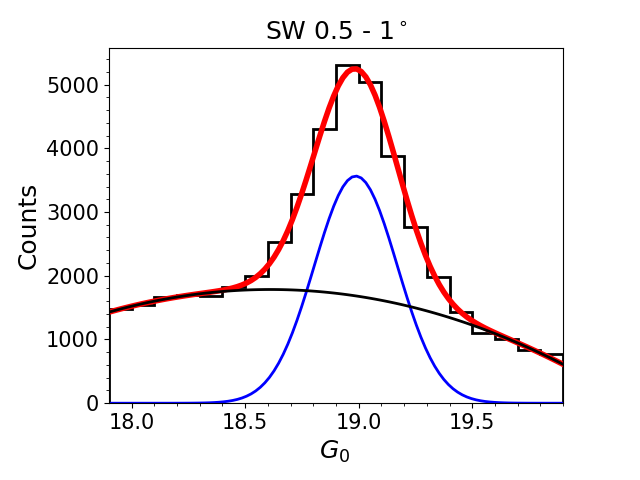}}\\
    \hspace*{-1.6em}
    \vspace{-0.8cm}
    \subfloat[]{\includegraphics[width=0.23\textwidth]{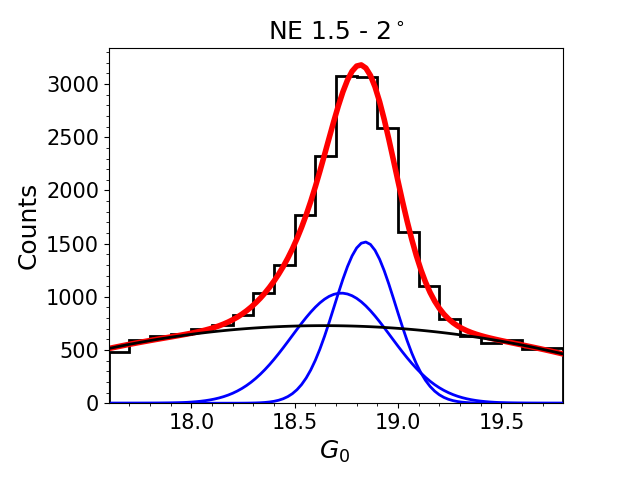}}
    \hspace*{-1.6em}
    \subfloat[]{\includegraphics[width=0.23\textwidth]{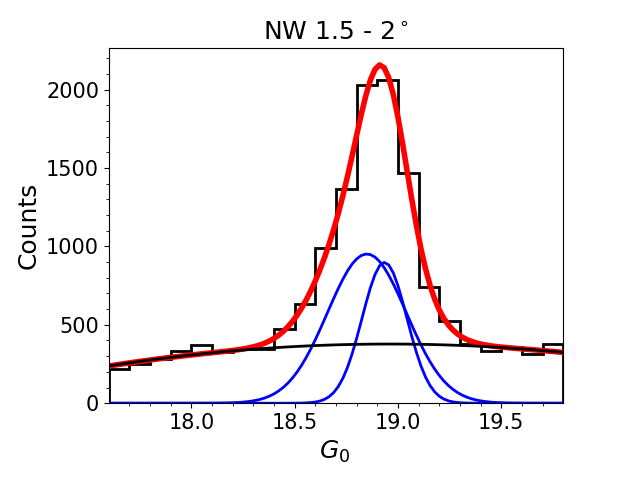}}
    \hspace*{-1.6em}
    \subfloat[]{\includegraphics[width=0.23\textwidth]{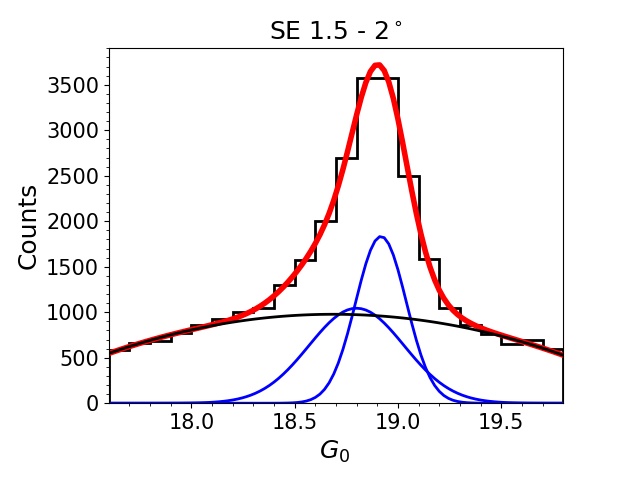}}
    \hspace*{-1.6em}
    \subfloat[]{\includegraphics[width=0.23\textwidth]{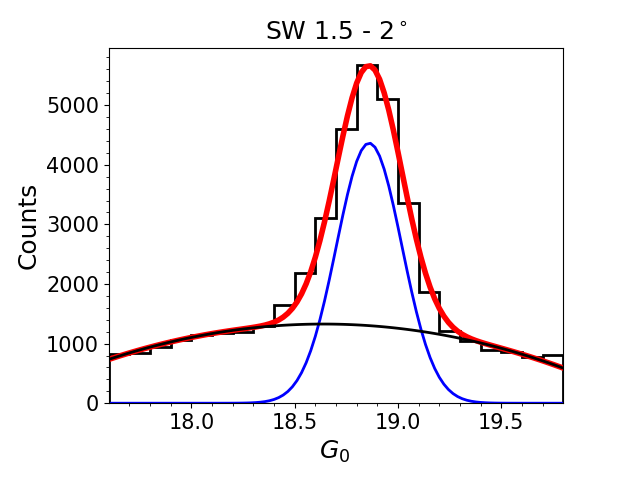}}\\
    \hspace*{-1.6em}
    \subfloat[]{\includegraphics[width=0.23\textwidth]{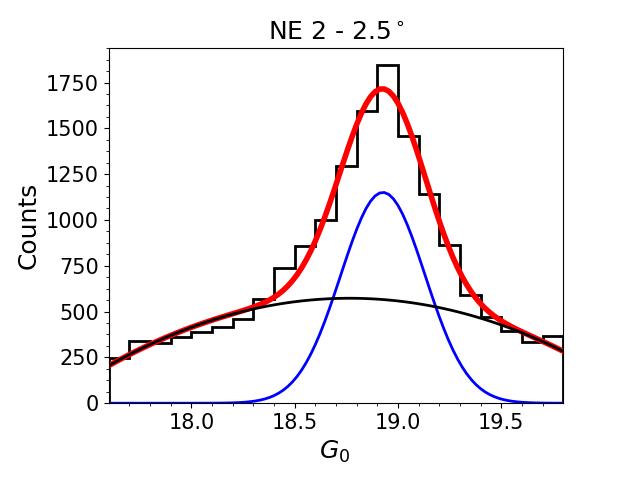}}
    \hspace*{-1.6em}
    \subfloat[]{\includegraphics[width=0.23\textwidth]{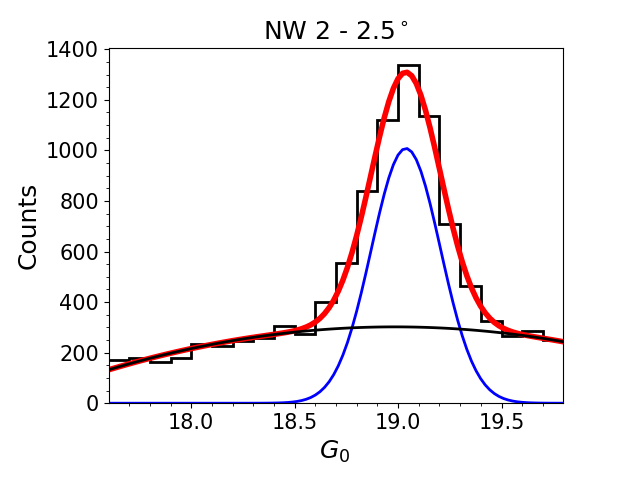}}
    \hspace*{-1.6em}
    \subfloat[]{\includegraphics[width=0.23\textwidth]{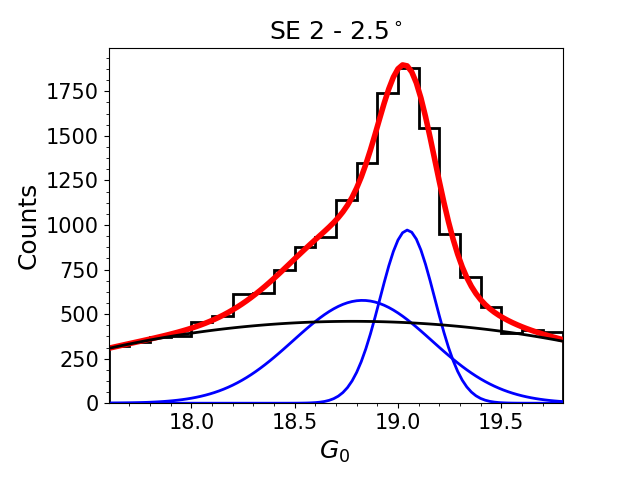}}
    \hspace*{-1.6em}
    \subfloat[]{\includegraphics[width=0.23\textwidth]{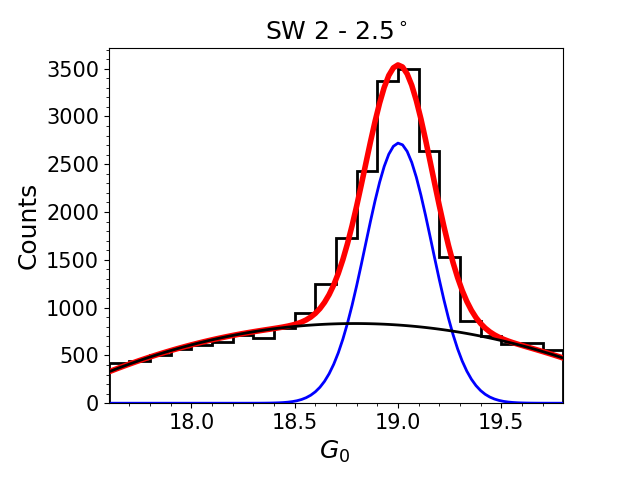}}\\
    \vspace*{-1.3em}
    \caption{Magnitude distributions of RC stars in the 0--2$\rlap{.}^{\circ}$5 sub-regions and their best fits. Blue, black and red lines indicate the Gaussian function, the quadratic polynomial and the total fit respectively.}
    \label{fig:magfit0-2.5}
\end{figure*}

\begin{figure*}
    \captionsetup[subfigure]{labelformat=empty}
    \centering
    \hspace*{-1.6em}
    \vspace{-0.8cm}
    \subfloat[]{\includegraphics[width=0.23\textwidth]{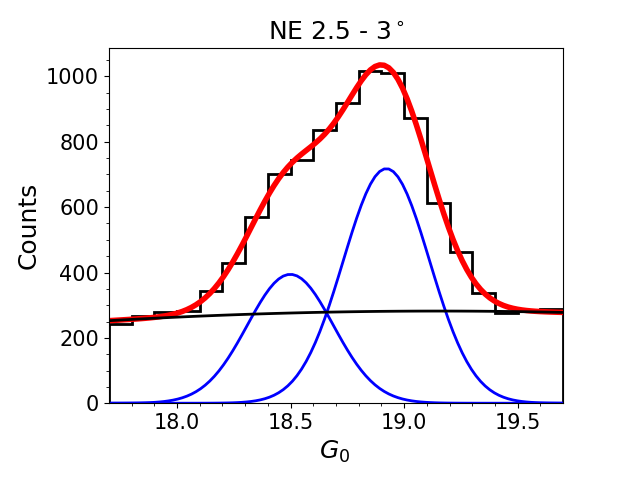}}
    \hspace*{-1.6em}
    \subfloat[]{\includegraphics[width=0.23\textwidth]{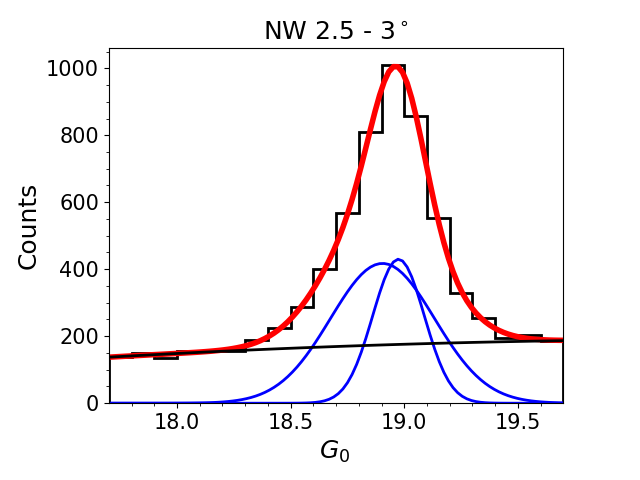}}
    \hspace*{-1.6em}
    \subfloat[]{\includegraphics[width=0.23\textwidth]{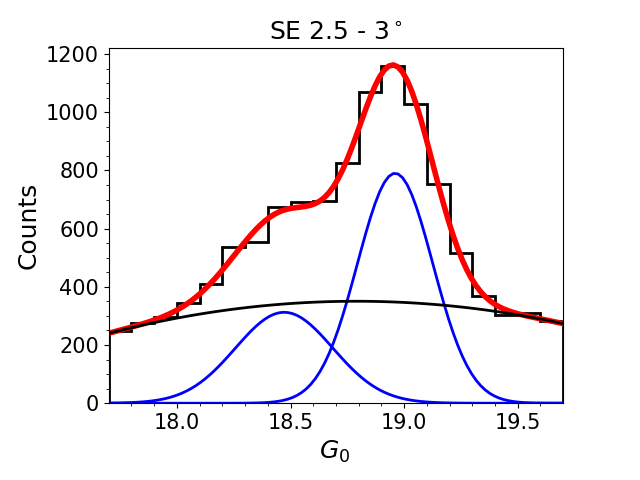}}
    \hspace*{-1.6em}
    \subfloat[]{\includegraphics[width=0.23\textwidth]{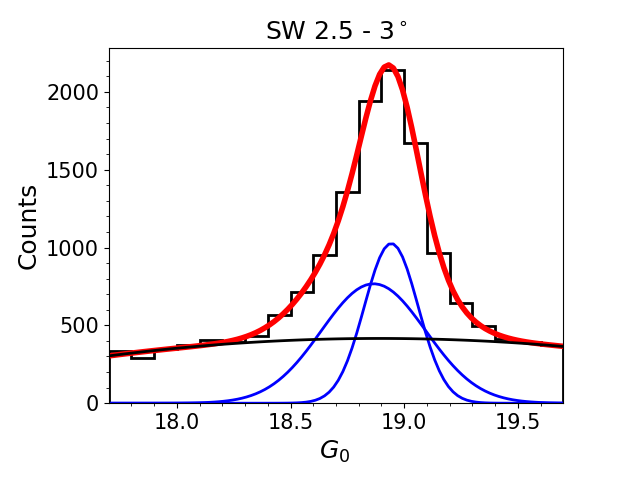}}\\
    \hspace*{-1.6em}
    \vspace{-0.8cm}
    \subfloat[]{\includegraphics[width=0.23\textwidth]{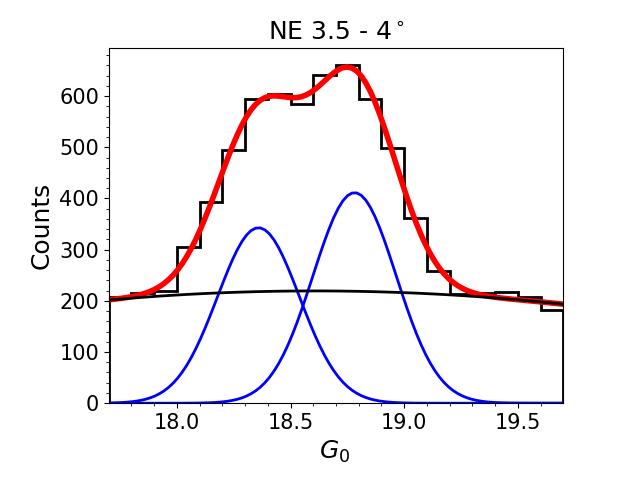}}
    \hspace*{-1.6em}
    \subfloat[]{\includegraphics[width=0.23\textwidth]{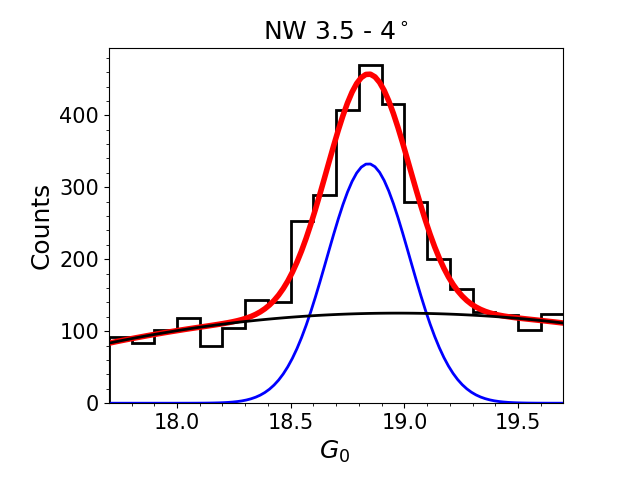}}
    \hspace*{-1.6em}
    \subfloat[]{\includegraphics[width=0.23\textwidth]{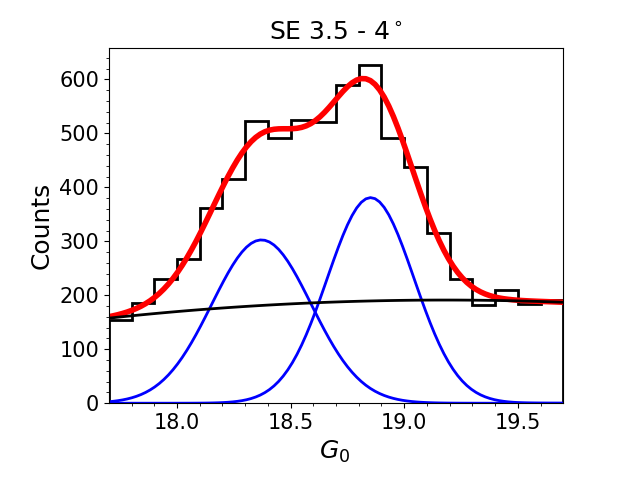}}
    \hspace*{-1.6em}
    \subfloat[]{\includegraphics[width=0.23\textwidth]{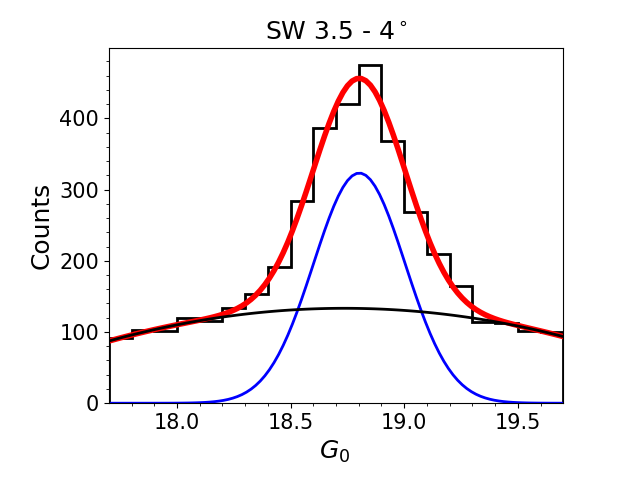}}\\
    \hspace*{-1.6em}
    \vspace{-0.8cm}
    \subfloat[]{\includegraphics[width=0.23\textwidth]{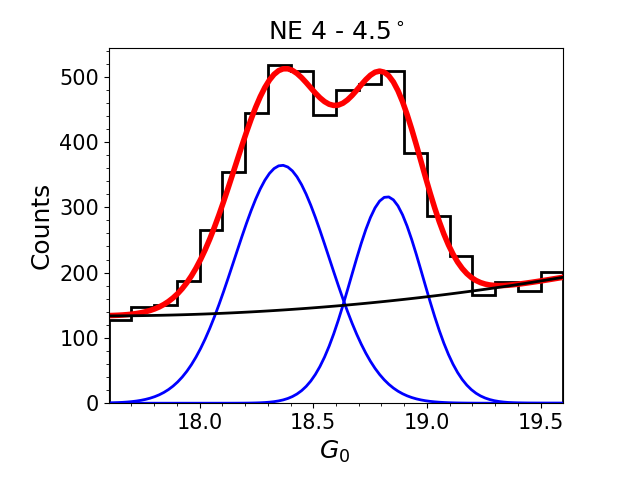}}
    \hspace*{-1.6em}
    \subfloat[]{\includegraphics[width=0.23\textwidth]{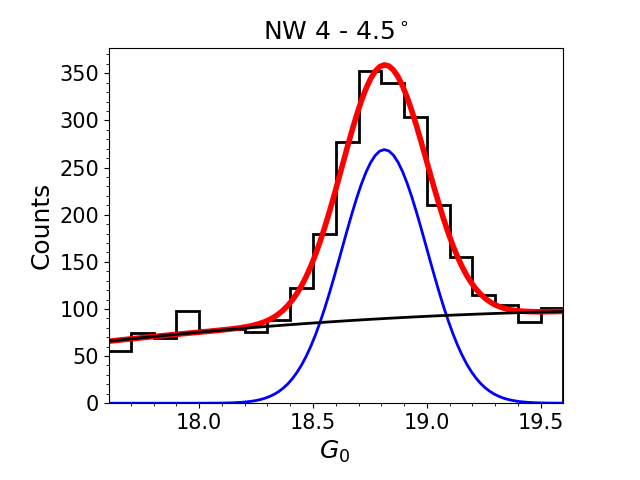}}
    \hspace*{-1.6em}
    \subfloat[]{\includegraphics[width=0.23\textwidth]{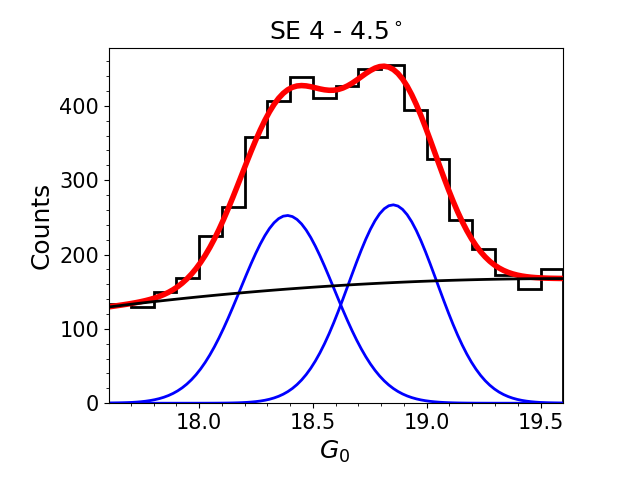}}
    \hspace*{-1.6em}
    \subfloat[]{\includegraphics[width=0.23\textwidth]{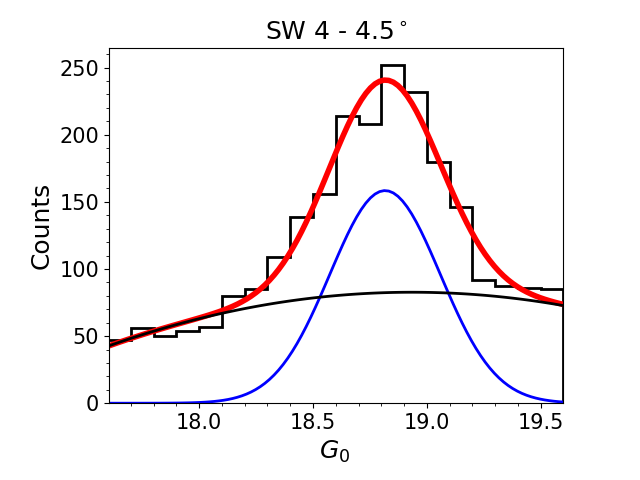}}\\
    \hspace*{-1.6em}
    \subfloat[]{\includegraphics[width=0.23\textwidth]{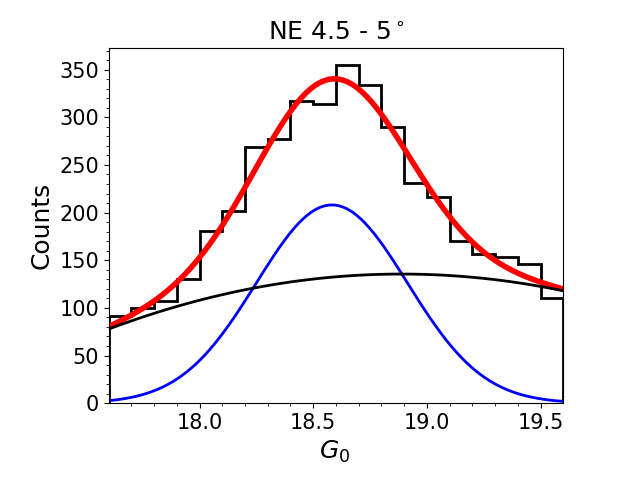}}
    \hspace*{-1.6em}
    \subfloat[]{\includegraphics[width=0.23\textwidth]{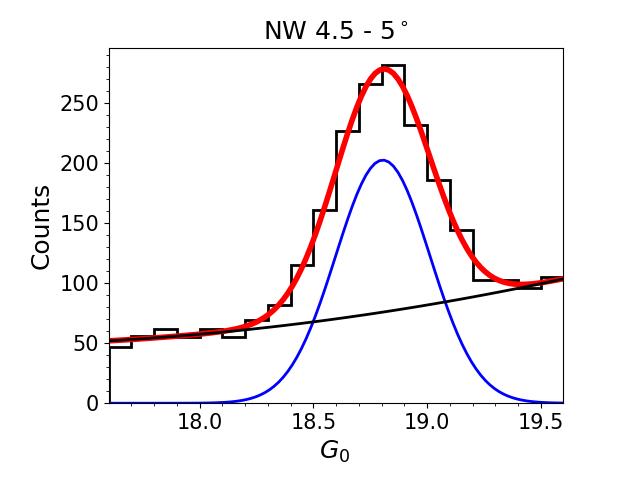}}
    \hspace*{-1.6em}
    \subfloat[]{\includegraphics[width=0.23\textwidth]{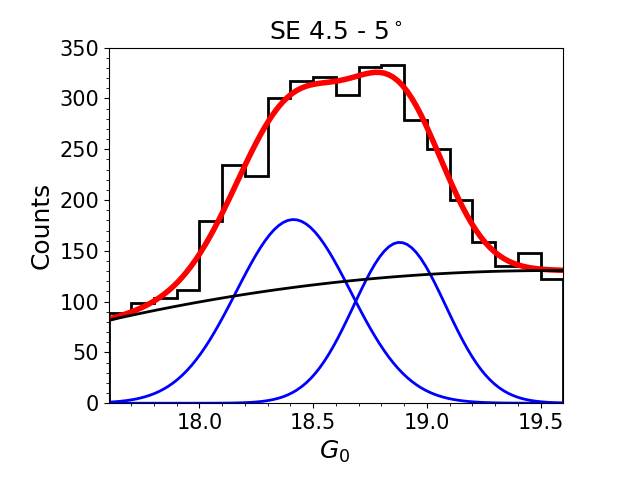}}
    \hspace*{-1.6em}
    \subfloat[]{\includegraphics[width=0.23\textwidth]{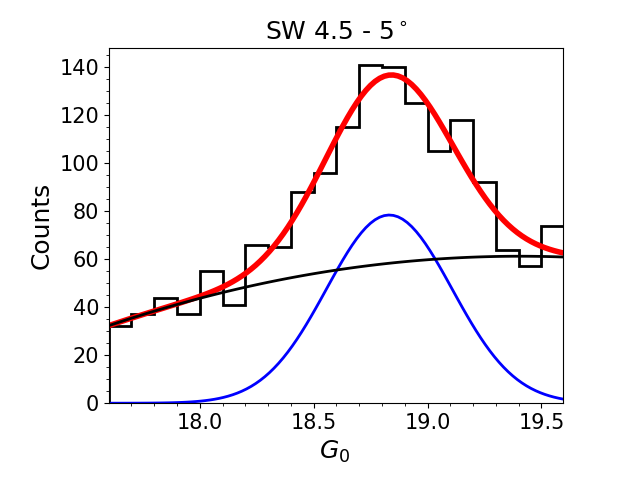}}\\
    \vspace*{-1.4em}
    \caption{Same as Fig. \ref{fig:magfit0-2.5} but now for the 2$\rlap{.}^{\circ}$5--5$^\circ$ sub-regions.}
    \label{fig:magfit2.5-5}
\end{figure*}
\begin{figure*}
    \captionsetup[subfigure]{labelformat=empty}
    \centering
    \hspace*{-1.1em}
    \vspace{-1cm}
    \subfloat[]{\includegraphics[width=0.23\textwidth]{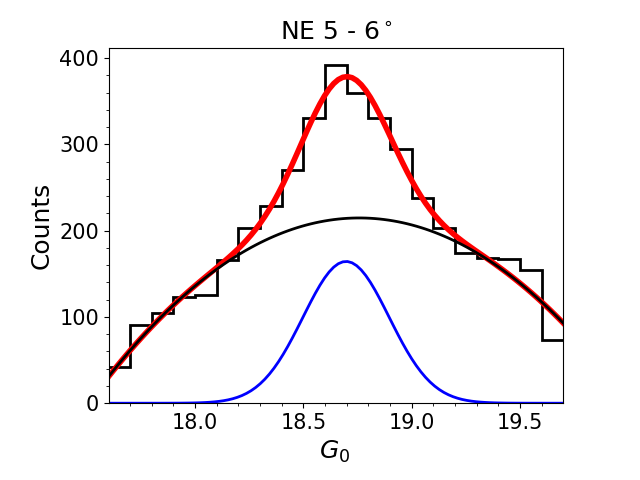}}
    \hspace*{-1.1em}
    \subfloat[]{\includegraphics[width=0.23\textwidth]{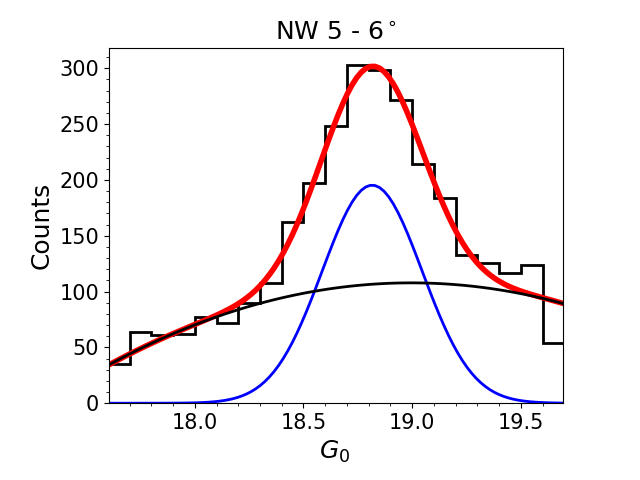}}
    \hspace*{-1.1em}
    \subfloat[]{\includegraphics[width=0.23\textwidth]{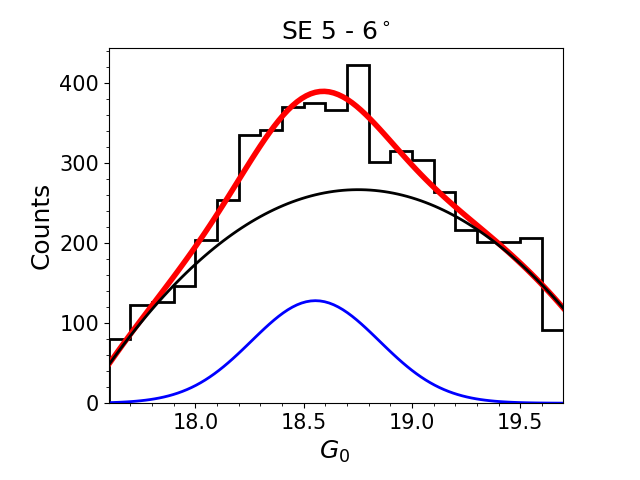}}
    \hspace*{-1.1em}
    \subfloat[]{\includegraphics[width=0.23\textwidth]{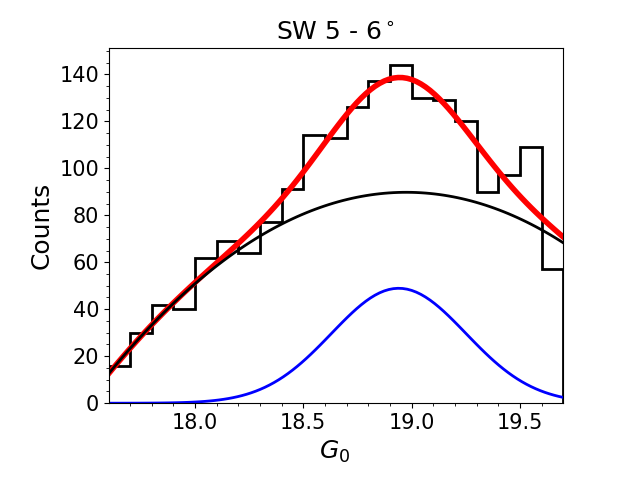}}\\
    \hspace*{-1.1em}
    \subfloat[]{\includegraphics[width=0.26\textwidth]{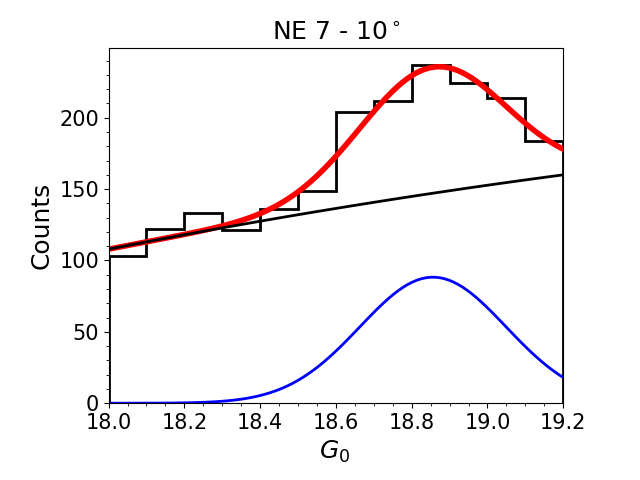}}
    \hspace*{-1.1em}
    \subfloat[]{\includegraphics[width=0.26\textwidth]{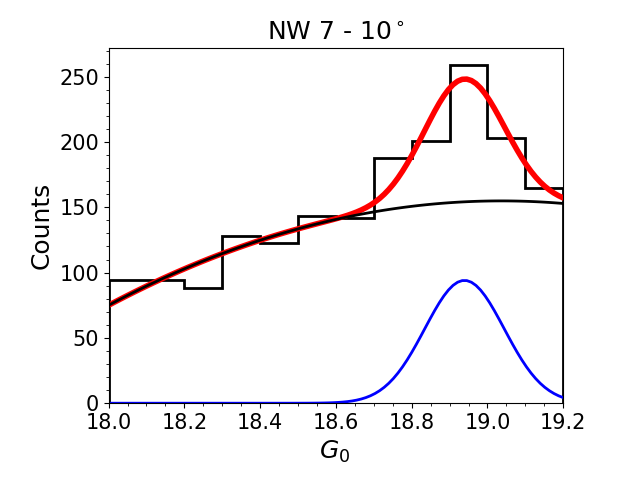}}
    \hspace*{-1.1em}
    \subfloat[]{\includegraphics[width=0.26\textwidth]{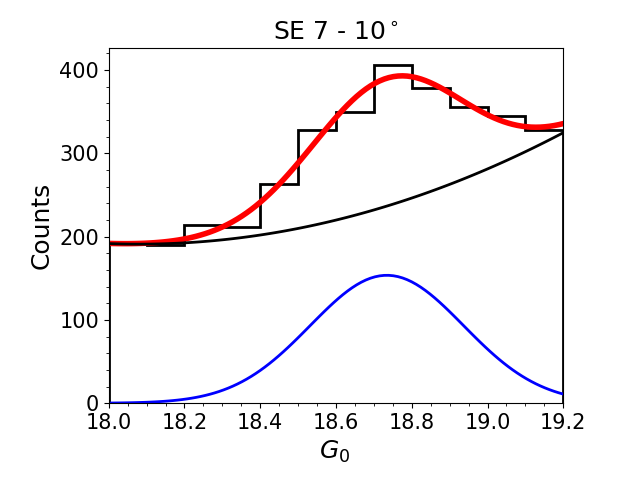}}
    \vspace*{-1.4em}
    \caption{Same as Fig. \ref{fig:magfit0-2.5} but for the 5--10$^\circ$ sub-regions.}
    \label{fig:magfit5-10}
\end{figure*}

\bibliographystyle{mnras}
\bibliography{refer} 


\bsp	
\label{lastpage}
\end{document}